\def\be{\begin{equation}}
\def\ee{\end{equation}}
\def\te{\end{equation}}
\def\bea{\begin{eqnarray}}
\def\ba{\begin{eqnarray}}

\def\ta{\end{eqnarray}}
\def\tea{\end{eqnarray}}

\def\ben{\begin{enumerate}}
\def\een{\end{enumerate}}

\def\ha{{1\over 2}}

\def\c{\raisebox{.4ex}{$\chi$}}

\def\m{\mu}

\documentclass[11pt,a4paper]{article}
\usepackage{jheppub}
\usepackage{latexsym, graphicx}
\usepackage{pbox,ulem}
\usepackage{amssymb,comment}

\allowdisplaybreaks

\title{Quantum Entanglement at High Temperatures? II.  Bosonic Systems in Nonequilibrium Steady State}
\author[a]{Jen-Tsung Hsiang,}
\author[a,b]{B. L. Hu}

\affiliation[a]{Department of Physics and Center for Field Theory and Particle Physics,  \\ Fudan University, Shanghai 200433, China}
\affiliation[b]{Joint Quantum Institute and Maryland Center for Fundamental Physics, \\ University of Maryland, College Park, Maryland 20742, USA}
\emailAdd{cosmology@gmail.com}
\emailAdd{blhu@umd.edu}
\date{March 10, 2015}

\abstract{This is the second of a series of three papers examining how viable it is for entanglement to be sustained at high temperatures  for quantum systems in thermal equilibrium (Case A), in nonequilibrium  (Case B) and in nonequilibrium steady state conditions (Case C). The system we analyze here consists of two coupled quantum harmonic oscillators each interacting with its own bath described by a scalar field, set at temperatures $T_1 > T_2$.  For  \textit{constant bilinear inter-oscillator coupling} studied here (Case C1) owing to the Gaussian nature,  the problem can be solved exactly at arbitrary temperatures even for strong coupling. We find that the valid entanglement criterion  in general is not a function of the bath temperature difference, in contrast to thermal transport in the same NESS setting~\cite{HHNESS}.   Thus lowering the temperature of one of the thermal baths does not necessarily help to safeguard the entanglement between the oscillators.  Indeed, quantum entanglement will disappear if any one of the thermal baths has a temperature higher than the critical temperature $T_c$.  With the Langevin equations derived we give a full display of how entanglement dynamics in this system depends on  $T_{1}$,  $T_{2}$ , the inter-oscillator coupling and the system-bath coupling strengths.  For weak oscillator-bath coupling the critical temperature $T_c$ is about the order of the inverse oscillator frequency, but for strong oscillator-bath coupling it will depend on the bath cutoff frequency.  We conclude that in most realistic circumstances, for bosonic systems in NESS with constant bilinear coupling, `hot entanglement' is largely a fiction.  In Paper III we will  examine the case (C2) of \textit{time-dependent driven coupling } which  contains the parametric pumping type described in~\cite{GalvePRL} wherein entanglement was  first shown to sustain at high temperatures.}


\arxivnumber{1503.xxxx}

\begin{document}

\maketitle

\section{Introduction}

Recently Galve et al  \cite{GalvePRL} (see also \cite{EP}) pointed out the possibility of keeping quantum entanglement alive in a system at high temperatures by driving the system of two oscillators with a time-dependent interaction term.  This has generated a great deal of interest in understanding the underlying issues and the basic mechanisms of obtaining the so-called `hot entanglement' \cite{Vedral}).   The word `hot'  conveys three layers of meaning in three different contexts, referring to quantum systems  A)  kept in thermal \textit{equilibrium} at all times,  B) in a \textit{nonequilibrium }condition and evolving and  C) in a \textit{nonequilibrium steady state} at late times.  Thus before making sweeping statements one needs to discern and analyze systems under at least these three separate situations for the behavior of quantum entanglement therein.

We have analyzed Case B) described above in some detail in our first paper,  obtaining the parameter ranges for entanglement to survive at a finite temperature and comparing with the results for \textbf{Case A} obtained earlier in e.g., \cite{AndWin,Anders}. Our results indicate that, generically, when two coupled oscillators separated at a fixed distance evolve under the influence of a shared thermal bath, their dynamics is usually highly non-Markovian. The asymptotic correlation /entanglement between the oscillators tends to survive better under 1) stronger inter-oscillator coupling, 2) weaker oscillator-bath interaction and at 3) a shorter distance between them. In the case of weak oscillator-bath coupling, the critical temperature is still bounded by the inverse oscillator's natural frequency, but tends to be lower than that the critical temperature in {\bf Case C}, due to the finite separation between the oscillators. The largest separation before the entanglement drops significantly is of the order of the inverse cutoff frequency inherited in the thermal bath, and the distance will decrease with higher bath temperature. In this limit, the results are similar to \textbf{Case A}. This is not unexpected since it is known~\cite{FRH,FH} that in the weak coupling limit, both configurations will yield similar results; furthermore, the non-Markovian mutual interaction between the oscillators is minimal in the weak oscillator-bath coupling regime. For stronger oscillator-bath interaction, the mutual interaction can sustain over a very long history in the evolution of both oscillators. Deviation in results between \textbf{Case A} and \textbf{Case B} will emerge. Nonetheless a strong oscillator-bath interaction can likely induce dynamical instability in the oscillators, a case worthy of closer analysis later.  

In this paper we analyze condition C)  where the system can maintain a nonequilibrium steady state (NESS) at late times.  Since NESS is a distinctly generic state, playing an important role for nonequilibrium systems as fundamental as the equilibrium state in quantum statistical mechanics, it is important to clarify the behavior of high temperature quantum entanglement under such conditions.  We illustrate these two conditions with two generic models:   \textbf{Case B)} is exemplified by  a quantum system made of at least two harmonic oscillators (HO) interacting with \textit{one common thermal bath}; \textbf{Case C)} is exemplified by a quantum system composed of two coupled harmonic oscillators each interacting with its own (\textit{private) thermal bath}.  We wish to  inquire about how entanglement initially present between the two quantum oscillators evolves in time, and calculate at what  temperature (approaching from below) it begins to die out.


To identify  the root cause  of quantum entanglement existing at high temperatures, if it does at all, one needs to identify the determining factors. Coupling in the system is certainly an important factor. Intuitively the stronger the coupling in the system, the weaker the coupling of the system to the baths, the better preserved the entanglement will be. If the coupling can be tuned to ``cruise alongside" how entanglement evolves in time, to even amplify it  along the way,  the better the chance of keeping the entanglement alive.  To see these effects more clearly we further divide the nonequilibrium steady state cases into two subcases,  C1 and C2.  \textbf{Case C1} is for  \textit{time-independent} inter-oscillator coupling,  and\textbf{ Case C2 }for \textit{time-dependent} inter-oscillator coupling.   Before one can  bring these cases under the same roof of nonequilibrium steady state condiiton one needs to prove or demonstrate that indeed a steady state exists at  late times in these setups.  We have so far shown the existence of NESS only for Case C1 in \cite{HHNESS}.
\footnote{It naturally behooves upon advocates of  hot entanglement \cite{Vedral} under NESS, namely, those with time-dependent coupling as exemplified by \cite{GalvePRL}  to prove or show the existence of a NESS under those conditions. It may not be a straightforward task. In fact,  for lack of a proof that systems with time-dependent coupling can approach NESS it is probably more prudent  to call this setup Class D, and only after such a proof shall one reinstate it into Class C  for systems which admit NESS. } .

Before we treat the Case C1 scenario in full which is the main goal of this paper,  we first give a brief description of a Case C2 model to mark the differences so the results of our work can be placed in perspective.

As a model for Case C2 the system is made up of two quantum oscillators interacting with each other via a time-dependent (sinusoidal) coupling.   Unlike Case C1 where the temperatures of the two baths are different, here they could be the same. In fact the temperature of the thermal bath and how strong the oscillators are coupled to the baths are not important. The nonequilibrium condition is provided by the external driving agent. Driving leads to production of entanglement even at very high  temperatures. For instance, even with a weak environmental coupling, a strong driving amplitude still provides a  higher critical temperature. 

The physics for these two cases albeit both in NESS is also very different. As explained in \cite{GalvePRL} , it is the squeezing of the system provided by the external agent and the parametric amplification (pumping) which  can offset the thermalization /equilibration process naturally expected for the systems interacting with a bath and dominant at high temperatures. Parametric driving is what sustains the entanglement in the system.  We will study this case in our sequel paper.

\subsection{Time-independent bilinear inter-oscillator coupling }

In the case of a chain of quantum harmonic oscillators coupled bilinearly with each other and with the baths the dynamics of the total system admits a complete solution, by virtue of its Gaussian nature, for all temperatures and for strong coupling within the system and with the baths.  This model has been studied by many authors \cite{AEPW02,Adesso,Ludwig,NQBM,HZH}. In our recent work \cite{HHNESS} functional methods are used to provide an explicit demonstration of the existence of a nonequilibrium steady state. Here we apply the results obtained therein to a study of quantum entanglement in NESS, with the aim of quantifying the claims made in the literature \cite{Vedral},  alluding to the possibility of entanglement survival at high temperatures for systems in NESS.  Note the present setup of bilinear coupling is different from that of Galve et al  \cite{GalvePRL} where the interaction between the two oscillators is via parametric pumping.  For this setup a recent paper closest to our intent is that by Ghesquire, Sinayskiy \& Petruccione \cite{GSP}.


\subsection{Comments on Claims by Other Authors}

We make a brief summary of what GSP have done and what claims they made below.



For the same model as mentioned above,  namely,  two bilinearly coupled quantum harmonic oscillators each interacting with its own bath  
GSP derived a perturbative `pre-Lindblad'  master equation without invoking the rotating wave approximation (non-rotating-wave, or NRW) \cite{nRWA}.  They consider two situations: For the study of entanglement they consider the high temperature regime in their Eq. (3) valid for both strong and weak interaction strength with the baths.  For the consideration of entropy dynamics related to equilibration issues they take the weak system-bath coupling limit and arrived at their Eq. (4). We will only be concerned with the entanglement issue here.   GSP made the following claims:

a) Entanglement persists for longer times at lower temperatures.

b) In the weak system-bath coupling limit, the late time steady state developed is independent of the initial conditions. 

c) For the equilibrium case, there exists a critical temperature which is consistent with the result of \cite{AndWin} in the limit.

We limit to two comments  regarding their method and claims here. The major differences will become clear in our results with quantitative representation via graphs found in later sections.

1) Regarding the method and approximations:  A perturbative `pre-Lindblad' master equation, even without the RWA, does not in general satisfy the complete positivity condition. Although it works better for strong coupling to the environment the results obtained under these approximations have unphysical behavior at low temperatures. For example, Ludwig et al \cite{Ludwig} pointed out the effect from the environment cutoff has to be handled with care.

2) The claim statements are too general -- they may not hold for specific conditions.  They need be qualified more carefully by specifying the range of (in)validity of the approximations introduced. E.g., Point a) above is sort of expected, but does it also imply that entanglement can be generated and be sustained if the temperature of both baths are sufficiently low,  even though the system state is initially separable?  Point b) regarding the existence of a NESS -- it  has been demonstrated for arbitrary strength in bilinear inter-oscillator coupling and for  arbitrary temperatures of the two baths \cite{HHNESS}. Point c) There is a distinction between  i) a system of two coupled oscillators each with its private bath under NESS studied here,  setting the two baths to be at the same temperature (presumably what their 'equilibrium' condition entails) and ii) the system in one common thermal bath (what we call Case A).  The situation is a lot more complex -- see discussions in the last section of this paper.

The above questions and a broader set of issues will be addressed in a fuller treatment of this generic (bilinear coupling) NESS model in the sections below. 

\subsection{Our Method and Key Findings}

The model we use in this work to describe entanglement dynamics at high temperature, namely, two coupled oscillators each interacting with its private bath at different temperatures, has been treated in full in our earlier paper \cite{HHNESS}, where one can find more technical details of the whole framework.  Entanglement in a harmonic chain is also a well-explored subject . The Gaussian nature of this model allows us to obtain exact solutions for arbitrary coupling strengths and temperatures . The central quantity to calculate is the covariance matrix at finite temperature and at late times, where it has been shown that the system approaches NESS. The Peres-Horodecki-Simon entanglement criterion \cite{Peres,Horod,PPTSimon} can be calculated without approximation. This approach has been shown to be totally equivalent to that of directly deriving the reduced density operator of the {system~\cite{HHNESS,CRV}}. A short way to report on our findings is that quantum entanglement will disappear when the bath temperatures become higher than a critical temperature ($T_c=1/\beta_c$). Also not surprisingly, asymptotic entanglement is easier to sustain for stronger inter-oscillator coupling and weaker oscillator-bath coupling.  The true gain of this investigation is a full display via the Langevin equations we derived of  the dependence of entanglement dynamics on the three parameters in this model, temperatures ($T_1,T_2$) of the baths,  the intra-system (inter-oscillator) coupling $\sigma$ and the system-bath coupling strengths $\gamma$. Their interplay is presented in the plots, where the critical temperature dependence on different coupling strengths can be easily seen.  For the special case when both baths have the same temperature, we show that the critical temperature, above which the system becomes separable, satisfies $\beta_{c}\omega\sim2\bigl(1+4\sigma/3\omega^{2}\bigr)^{-1}$ for \textit{weak} oscillator-bath coupling, $\omega$ being the oscillator natural frequency.  It is consistent with the general expectation that $\beta_{c}\omega\sim\mathcal{O}(1)$ in the \textit{vanishing} inter-oscillator coupling $\sigma$ limit. In the opposite limit, when the oscillator-bath coupling is \textit{strong}, correction terms with  bath cutoff frequency dependence will show up. This is a noteworthy point in a lesser-explored regime, namely, one needs to be mindful of the choice of the environment cut-off frequency in the treatment of strong system-environment coupling.

A cautionary remark is in place here about entanglement measures used for quantum systems at finite temperature:  Although the Peres-Horodecki-Simon (PHS)  criterion is totally valid to identify the existence of entanglement in a quantum system,  it does not serve as  a quantifiable measure. We find from explicit calculations that at finite temperature it does not necessarily vary monotonically with the parameters in our system,  namely, the temperature or coupling constants. One should exercise caution in using  the PHS criterion  for a physical understanding of thermal entanglement.  In contrast negativity is a valid measure to quantify the dependence of quantum entanglement on these physical parameters.

\subsection{Differences from the common bath case}

To highlight the qualitative features in the behavior of the separability criterion it is useful to contrast  the private bath case (Case C1) studied here and the shared bath case (Case B) studied in Paper I.  A more detailed description can be found in the last section:
\begin{enumerate}
	\item The initial Gaussian conditions will be irrelevant in the private bath case, but remain significant in the shared bath case, so the state of entanglement is sensitive to the initial conditions.
	\item At late times the entanglement measure for the private bath case is time-independent, but it continues  to oscillate in time.
   \item The inter-oscillator coupling $(\sigma>0)$  plays a more important role in the private bath case  than in the shared bath case.
 	\item In the private bath case, entanglement is easier to survive for stronger inter-oscillator coupling and weaker oscillator-bath coupling, but in the shared bath case, both factors seem to be overshadowed by the intrinsic  quantum dynamics of the system which depends on the initial conditions of the oscillators.
\end{enumerate}

This paper is organized as follows: In Sec. 2, we briefly discuss the dynamics of the reduced system in the NESS configuration, and introduce the separability/entanglement criterion. In particular we pay attention to the covariance matrix, which constitutes the building blocks of the separability criterion. In Sec. 3 and 4, we highlight the calculations of the covariance matrix elements at high, zero and low temperature cases. We further examine the temperature dependence of the covariance matrix elements and the validity of the relevant approximations in Section 5. In Sec. 6 we investigate  the separability criterion at different temperature regimes in detail and  point out  its non-monotonic behavior.  Because of this  we adopt instead negativity as a valid measure of  entanglement for quantitative analysis of quantum systems at finite temperature.  We derive some relations between the critical temperature and various coupling constants. In Sec. 7, we then offer a more intuitive viewpoint to understand how all sorts of interactions can affect entanglement between oscillators. Finally we summarize our results and compare them with the case of the shared bath in Sec. 8.

\section{The Model and the Covariance Matrix of the System}

\subsection{The Model}


Consider two coupled harmonic oscillators of equal mass $m$ and (bare) natural frequency $\omega_b$ coupled to each other with strength $\sigma$, each of which interacting with its own thermal bath with coupling constant $e$. We refer to the two oscillators together as the system, and the two baths together as the environment. This setup is a prototype used often for the investigation of nonequilibrium steady state (NESS), the existence of which is shown in a recent paper \cite{HHNESS} (see also the references therein). In the Langevin equation approach the two oscillators' amplitude $\chi_1,\chi_2$ satisfy the following equations of motion:
\begin{align}
	 \ddot{\chi}_{1}+2\gamma\,\dot{\chi}_{1}+\omega^{2}\chi_{1}+\sigma\,\chi_{2}&=\frac{1}{m}\,\xi_{1}\,,\label{E:derhs1}\\
	 \ddot{\chi}_{2}+2\gamma\,\dot{\chi}_{2}+\omega^{2}\chi_{2}+\sigma\,\chi_{1}&=\frac{1}{m}\,\xi_{2}\,,\label{E:derhs2}
\end{align}
where $\gamma$ is the damping constant related to $e$ by $\gamma=e^{2}/(8\pi m)$, and $\omega$ is the renormalized frequency (wherein the correction from the interaction with the environment has been considered before), and $\xi_1$, $\xi_2$ are the stochastic forces acting on {Oscillators 1, 2 ($O_{1,2}$)} respectively. Note they are not specified by hand but determined self-consistently. An overdot denotes taking the time derivative of a variable. The initial state of the oscillator is described by a Gaussian wavepacket and both oscillators are prepared in the same initial configuration. The two private baths {($B_{1,2}$)} are modeled by massless scalar fields at different temperatures $\beta^{-1}_{i}$.

In the matrix notation, these two Langevin equations are condensed into one, namely,
\begin{equation}\label{E:dhewca}
	\ddot{\pmb{\chi}}+2\gamma\,\dot{\pmb{\chi}}+\pmb{\Omega}^{2}\cdot\pmb{\chi}=\frac{1}{m}\,\pmb{\xi}\,,
\end{equation}
where
\begin{align}
	\pmb{\chi}&=\begin{pmatrix}\chi_{1}\\ \chi_{2}\end{pmatrix}\,,&\pmb{\Omega}^{2}&=\begin{pmatrix}\omega^{2}&\sigma\\\sigma&\omega^{2}\end{pmatrix}\,,&\pmb{\xi}&=\begin{pmatrix}\xi_{1}\\ \xi_{2}\end{pmatrix}\,.
\end{align}
The solutions to this equation are given by,
\begin{equation}\label{E:wiejwks}
	 \pmb{\chi}(t)=\mathbf{D}_{1}(t)\cdot\pmb{\chi}(0)+\mathbf{D}_{2}(t)\cdot\dot{\pmb{\chi}}(0)+\frac{1}{m}\int^{t}_{0}\!ds\;\mathbf{D}_{2}(t-s)\cdot\pmb{\xi}(s)\,.
\end{equation}
where {$\pmb{\chi}(0)$, $\dot{\pmb{\chi}}(0)$ represent the initial configuration of the oscillators. The fundamental solution matrices} $\mathbf{D}_{1}$, $\mathbf{D}_{2}$ are a special set of homogeneous solutions to the Langevin equation \eqref{E:dhewca},
\begin{align}
	 \mathbf{D}_{1}(0)&=\mathbf{1}\,,&\dot{\mathbf{D}}_{1}(0)&=\mathbf{0}\,,&\mathbf{D}_{2}(0)&=\mathbf{0}\,,&\dot{\mathbf{D}}_{2}(0)&=\mathbf{1}\,.
\end{align}
In particular,  the Fourier transformation of
\begin{equation}
	\bigl(-\kappa^{2}\mathbf{I}+\pmb{\Omega}^{2}-i\,2\kappa\,\mathbf{I}\bigr)^{-1}
\end{equation}
is equal to $\theta(\tau)\,\mathbf{D}_{2}(\tau)$, that is,
\begin{equation}
	 \theta(\tau)\,\mathbf{D}_{2}(\tau)=\int_{-\infty}^{\infty}\!\frac{d\kappa}{2\pi}\,\frac{e^{-i\,\kappa\tau}}{-\kappa^{2}\mathbf{I}+\pmb{\Omega}^{2}-i\,2\kappa\,\mathbf{I}}\,.
\end{equation}
The function $\theta(\tau)$ is the unit-step function. Unless mentioned otherwise, we will not distinguish $\theta(\tau)\,\mathbf{D}_{2}(\tau)$ from $\mathbf{D}_{2}(\tau)$ for all practical purposes, and denote $\bigl(-\kappa^{2}\mathbf{I}+\pmb{\Omega}^{2}-i\,2\kappa\,\mathbf{I}\bigr)^{-1}$ by $\widetilde{\mathbf{D}}_{2}(\kappa)$.

The force term $\xi_{i}(t)$ is a stochastic $c$-number with the statistical properties
\begin{align}
	 \langle\pmb{\xi}(t)\rangle&=0\,,&\langle\pmb{\xi}(t)\,\pmb{\xi}^{T}(t')\rangle=&e^{2}\,\mathbf{G}_{H}(t-t')=e^{2}\begin{pmatrix} G^{11}_{H}(t-t')&0\\0&G^{22}_{H}(t-t')\end{pmatrix}\,,
\end{align}
where $G^{ii}_{H}(t-t')$ is the Hadamard function of the bath scalar field, associated with the $i$th {oscillator~\cite{HHNESS}}. This stochastic force in essence represents the quantum fluctuations of the private bath at a finite temperature.

\subsection{Entanglement Measures}\label{S:ekjbek}
For continuous-variable systems, the entanglement measure based on the density matrix is not conveniently calculable because the density matrix in this case is infinite-dimensional. However, it has been shown~\cite{PPTSimon} that in the case of continuous \textit{Gaussian} variables, the Peres-Horodecki separability criterion~\cite{Peres,Horod} can be reformulated in terms of the covariance matrix of the bipartite system,
\begin{align}\label{E:eriuhks}
	 \zeta_{+}&=\det\mathbf{A}\,\det\mathbf{B}-\operatorname{Tr}\bigl\{\mathbf{A}\cdot\mathbf{J}\cdot\mathbf{C}\cdot\mathbf{J}\cdot\mathbf{B}\cdot\mathbf{J}\cdot\mathbf{C}^{T}\cdot\mathbf{J}\bigr\}+\bigl(\det\mathbf{C}+\frac{1}{4}\bigr)^{2}\notag\\
	 &\qquad\qquad\qquad\qquad\qquad\qquad\qquad\qquad\qquad\qquad\qquad-\frac{1}{4}\bigl(\det\mathbf{A}+\det\mathbf{B}\bigr)\geq0\,,
\end{align}
with
\begin{equation*}
\mathbf{J}=\begin{pmatrix}0&+1\\-1&0\end{pmatrix}\,.
\end{equation*}
Here the matrices $\mathbf{A}$, $\mathbf{B}$, $\mathbf{C}$ are the block matrices in the covariance matrix $\mathbf{V}$,
\begin{align}
	\mathbf{V}=\begin{pmatrix}\mathbf{A} &\mathbf{C} \\
					\mathbf{C}^{T} &\mathbf{B}
				\end{pmatrix}\,,
\end{align}
while the covariance matrix $\mathbf{V}$ itself is defined by the canonical variables of the two subsystems
\begin{equation}
	\mathbf{V}=\frac{1}{2}\,\operatorname{Tr}\Bigl[\rho\bigl\{\mathbf{R},\mathbf{R}^{T}\bigr\}\Bigr]=\frac{1}{2}\,\langle\bigl\{\mathbf{R},\mathbf{R}^{T}\bigr\}\rangle\,,
\end{equation}
where $\rho$ is the density matrix of the state we are interested in. We have assumed $\langle\mathbf{R}\rangle=0$. The column matrix $\mathbf{R}$ takes the form $\mathbf{R}^{T}=(\chi_{1},p_{1},\chi_{2},p_{2})$, and $p_{i}$ is the canonical momentum conjugate to $\chi_{i}$ associated with the subsystem $i$. The angular brackets denote taking the quantum expectation value. In our case, once we have the covariance matrix for the coupled harmonic oscillators in the NESS configuration, we may construct $\zeta_{+}$ according to \eqref{E:eriuhks}. A negative value of $\zeta_{+}$ thus implies the existence of quantum entanglement.

Although \eqref{E:eriuhks} constitutes only the second moments of the canonical variables, it offers a complete description of the Gaussian system since for a Gaussian system, all higher moments can be expressed in terms of the second moments. Oftentimes it is instructive to write the Peres-Horodecki separability criterion in terms of the symplectic eigenvalues of the partially transposed covariance matrix.  {Let $\eta_{\gtrless}$ stand for the symplectic eigenvalues of $\mathbf{V}^{pt}$, the partial transposition of $\mathbf{V}$. Without loss of generality we assume $\eta_{>}$ is greater than $\eta_{<}$. In fact they can be found by solving the eigenvalues of the matrix $i\,\pmb{\Omega}\cdot\mathbf{V}^{pt}$, with $\pmb{\Omega}=\bigoplus_{k=1}^{2}\mathbf{J}$. The resulting eigenvalues will appear in pairs by the form $\pm\eta_{>}$, $\pm\eta_{<}$, so the symplectic eigenvalues of $\mathbf{V}^{pt}$ are given by the absolute value of the eigenvalues of $i\,\pmb{\Omega}\cdot\mathbf{V}^{pt}$. When we write $\mathbf{V}^{pt}$ into the Williamson's form, the separability criterion $\mathbf{V}^{pt}+i\,\pmb{\Omega}/2\geq0$ becomes
\begin{align}\label{E:anekrhe}
	\begin{pmatrix}\eta_{>}&0&0&0\\0&\eta_{>}&0&0\\0&0&\eta_{<}&0\\0&0&0&\eta_{<}\end{pmatrix}+\frac{i}{2}\begin{pmatrix}0&+1&0&0\\-1&0&0&0\\0&0&0&+1\\0&0&-1&0\end{pmatrix}&\geq0\,,&&\Rightarrow&\bigl(\eta_{>}^{2}-\frac{1}{4}\bigr)\bigl(\eta_{<}^{2}-\frac{1}{4}\bigr)&\geq0\,.
\end{align}
When $\eta_{<}<1/2$, entanglement occurs. Notice that $\eta_{>}$ is assumed to be larger than $\eta_{<}$, so $\eta_{>}$ is always greater than $1/2$. We observe that although a violation of the Peres-Horodecki-Simon separability criterion signals the \textit{existence} of entanglement,  it is not a good \textsl{measure} for a quantitative description of entanglement, in that the criterion includes a unwelcome factor $(\eta_{>}-1/2)$, which does not affect the identification of  the existence of entanglement,  it messes up the correct evaluation of entanglement. This can be understood if we examine the behavior of the symplectic eigenvalues $\eta_{\gtrless}$ about $\eta_{<}\sim1/2$. For definiteness, we assume that the symplectic eigenvalues are similar monotonic functions of the parameters of the entangled system. We can easily see that if $\eta_{>}$ changes too fast in the vicinity of $\eta_{<}=1/2$, the product $\bigl(\eta_{>}^{2}-1/4\bigr)\bigl(\eta_{<}^{2}-1/4\bigr)$ will not be monotonic there.}

As is perhaps better known, a simple calculable measure of entanglement which also provides quantifiable accuracy  is negativity~\cite{Vidal}, denoted by $\mathcal{N}$ or its {logarithm (strictly speaking logarithmic negativity is not merely the logarithm of negativity, although it is related to)}~\cite{Plenio}, the logarithmic negativity $E_{\mathcal{N}}$. For the Gaussian states under study they can be respectively defined by
\begin{align}\label{E:dfkefjdkw}
	\mathcal{N}(\rho)&=\max\bigl\{0,\frac{1-2\eta_{<}}{2\eta_{<}}\bigr\}\,,&E_{\mathcal{N}}(\rho)&=\max\bigl\{0,-\ln 2\eta_{<}\bigr\}\,,
\end{align}
in terms of the symplectic eigenvalue $\eta_{<}$ of the partially transposed covariance matrix. When $\eta_{<}<1/2$, the Gaussian state $\rho$ is entangled and both measures take nonzero values between $0^{+}$ to $+\infty$. In addition, the logarithmic negativity has a convenient feature of being additive.

{Comparing the negativity~\eqref{E:dfkefjdkw} with the Peres-Horodecki-Simon criterion \eqref{E:anekrhe}, we observe that they are all based on the smaller symplectic eigenvalue $\eta_{<}$ of the partially transposed covariance matrix $\mathbf{V}^{pt}$, so they will give the same prediction on the existence of entanglement. However, the separability criterion carries an additional undesired factor $(\eta_{>}-1/2)$, which may inadvertently scale $(\eta_{<}-1/2)$. Thus the separability criterion is not suitable for quantifying entanglement.}

{Finally, we remark on a subtlety of the entanglement measure. It has been pointed out~\cite{Eisert98,Virmani,Adesso05} that different measures may give different ordering of density operators with respect to the amount of entanglement. To be more specific, given two density matrix $\rho_{1}$ and $\rho_{2}$, we can have $E_{1}(\rho_{1})\leq E_{1}(\rho_{2})$ for one entanglement measure, while $E_{2}(\rho_{1})\geq E_{2}(\rho_{2})$ for another. In particular, negativity and Gaussian entanglement of formation, the latter forming an upper bound to the true  entanglement of formation, have been found to be inequivalent for asymmetric Gaussian states~\cite{Adesso05}. For symmetric states, the predictions from both measures coincide.}

The next few sections will be dedicated to the calculation of elements of the covariance matrix.

\subsection{Elements of the Covariance Matrix}

We use \eqref{E:wiejwks} to find the elements of the covariance matrix $\mathbf{V}$.  Assume that the initial state of each oscillator is depicted by a Gaussian wave packet of the same shape, at rest initially at the bottom of the harmonic potential associated with each oscillator, such that
\begin{align}
	\langle\chi_{i}(0)\rangle&=\langle p_{i}(0)\rangle=0\,, &\langle\{\chi_{i}(0),p_{j}(0)\}\rangle&=0\,,\\
	 \langle\{\chi_{i}(0),\chi_{j}(0)\}\rangle&=\langle\chi_{i}^{2}(0)\rangle\,\delta_{ij}\,,&\langle\{p_{i}(0),p_{j}(0)\}\rangle&=\langle p_{i}^{2}(0)\rangle\,\delta_{ij}\,,
\end{align}
with $p_{i}=m\dot{\chi}_{i}$. Thus these two oscillators are initially in a separable state.  From the solutions \eqref{E:wiejwks} one can identify the role of the interaction, either between the oscillators or between the oscillator and its private bath, in creating or sustaining the quantum entanglement in the system.

To calculate the elements of the covariance matrix $\mathbf{V}$ one can show, for example, that
\begin{align}
				 \frac{1}{2}\langle\bigl\{\chi_{i}(t),\chi_{j}(t)\bigr\}\rangle&=\mathbf{D}^{ik}_{1}(t)\mathbf{D}^{jk}_{1}(t)\,\langle\chi_{k}^{2}(0)\rangle+\frac{1}{m^{2}}\,\mathbf{D}^{ik}_{2}(t)\mathbf{D}^{jk}_{2}(t)\,\langle p_{k}^{2}(0)\rangle\notag\\
				 &\qquad\qquad+\frac{e^{2}}{m^{2}}\int^{t}_{0}\!ds\,ds'\;\mathbf{D}^{ik}_{2}(t-s)\mathbf{D}^{jk}_{2}(t-s')\,\mathbf{G}^{kk}_{H}(s-s')\,.
\end{align}
When the dynamics of the system evolves into relaxation as $t\to\infty$, the first two terms on the righthand side will be exponentially small if the coupling constant between the oscillator and the bath is not vanishing. Thus at late time $\langle\bigl\{\chi_{i}(t),\chi_{j}(t)\bigr\}\rangle/2$ simplifies to
\begin{align}
			 \lim_{t\to\infty}\frac{1}{2}\langle\bigl\{\chi_{i}(t),\chi_{j}(t)\bigr\}\rangle&=\frac{e^{2}}{m^{2}}\int^{\infty}_{-\infty}\!ds\,ds'\;\mathbf{D}^{ik}_{2}(s)\mathbf{D}^{jk}_{2}(s')\,\mathbf{G}^{kk}_{H}(s-s')\notag\\
			 &=\frac{e^{2}}{m^{2}}\int^{\infty}_{-\infty}\!\frac{d\kappa}{2\pi}\;\widetilde{\mathbf{D}}^{ik\,*}_{2}(\kappa)\widetilde{\mathbf{D}}^{jk}_{2}(\kappa)\,\widetilde{\mathbf{G}}^{kk}_{H}(\kappa)\,,\label{E:nvkrw}
\end{align}
where we have used the fact that $\mathbf{D}_{2}(\tau)=0$ if $\tau<0$. Since the Fourier transform of $\mathbf{D}_{2}(s)$ is defined by
\begin{equation}			
	\widetilde{\mathbf{D}}_{2}(\kappa)=\frac{1}{-\kappa^{2}\mathbf{I}+\pmb{\Omega}^{2}-i\,2\kappa\,\mathbf{I}}\,,
\end{equation}
we use the property $\widetilde{\mathbf{D}}_{2}(-\kappa)=\widetilde{\mathbf{D}}^{*}_{2}(\kappa)$ to arrive at \eqref{E:nvkrw}.

At this point, let us look at a more specific example: the element
$V_{11}(t)=\langle\bigl\{\chi_{1}(t),\chi_{1}(t)\bigr\}\rangle/2=\langle\chi_{1}^{2}(t)\rangle$. At late time it takes on the value $\mathcal{V}_{11}$,
\begin{align}\label{E:ejrhes}
			 \mathcal{V}_{11}=\lim_{t\to\infty}V_{11}(t)&=\frac{e^{2}}{m^{2}}\int^{\infty}_{-\infty}\!\frac{d\kappa}{2\pi}\;\biggl[\lvert\widetilde{\mathbf{D}}^{11}_{2}(\kappa)\rvert^{2}\,\widetilde{\mathbf{G}}^{11}_{H}(\kappa)+\lvert\widetilde{\mathbf{D}}^{12}_{2}(\kappa)\rvert^{2}\,\widetilde{\mathbf{G}}^{22}_{H}(\kappa)\biggr]\,,
\end{align}
and
\begin{align}
			 \lvert\widetilde{\mathbf{D}}^{11}_{2}(\kappa)\rvert^{2}&=\frac{(\kappa^{2}-\omega^{2})^{2}+4\gamma^{2}\kappa^{2}}{\bigl[(\kappa^{2}-\omega_{+}^{2})^{2}+4\gamma^{2}\kappa^{2}\bigr]\bigl[(\kappa^{2}-\omega_{-}^{2})^{2}+4\gamma^{2}\kappa^{2}\bigr]}\,,\\
			 \lvert\widetilde{\mathbf{D}}^{12}_{2}(\kappa)\rvert^{2}&=\frac{\sigma^{2}}{\bigl[(\kappa^{2}-\omega_{+}^{2})^{2}+4\gamma^{2}\kappa^{2}\bigr]\bigl[(\kappa^{2}-\omega_{-}^{2})^{2}+4\gamma^{2}\kappa^{2}\bigr]}\,,
\end{align}
with $\gamma=e^{2}/(8\pi m)$. The frequencies $\omega^{2}_{\pm}=\omega^{2}\pm\sigma$ are the oscillating frequencies of the normal modes, which can be constructed from the superpositions of \eqref{E:derhs1} and \eqref{E:derhs2}. The Fourier transformation of the Hadamard function takes the form
\begin{equation}\label{E:defkjd}
	 \widetilde{\mathbf{G}}^{kk}_{H}(\kappa)=\frac{\kappa}{4\pi}\,\coth\frac{\beta_{k}\kappa}{2}=\begin{cases}\dfrac{\kappa}{4\pi}+\dfrac{\kappa}{2\pi}\,e^{-\beta_{k}\kappa}\,,&\beta_{k}\kappa\gg1\,,\vphantom{\biggl|}\\ \dfrac{1}{2\pi\beta_{k}}\,,&\beta_{k}\kappa\ll1\,.\end{cases}
\end{equation}
The term $\kappa/4\pi$ represents the vacuum zero-point contribution. The off-diagonal terms of $\widetilde{\mathbf{G}}_{H}$ are zero because both private baths are not correlated.

From the late-time value $\mathcal{V}_{11}$ of the amplitude uncertainty of $O_{1}$, we observe the following distinct features: (1) it approaches a constant independent of time, (2) its integral expression \eqref{E:ejrhes} takes a form similar to the Landauer formula, where $\lvert\widetilde{\mathbf{D}}^{11}_{2}(\kappa)\rvert^{2}$ plays a role of the transmission coefficient, and (3) it depends on both thermal baths even though $O_{1}$ does not have a direct contact with $B_{2}$. The last property would not be unexpected because the coupling between the oscillators will bring in correlations between $O_{1}$ and $B_{2}$, and vice versa, between $O_{2}$ and $B_{1}$. In fact, these features hold quite generally for the all elements of the covariance matrix in the current NESS configuration.

The definition of the covariance matrix $\mathbf{V}$ and the expressions for its elements, and their corresponding values at late times are derived in Appendix A--D.

In the next sections we will explicitly evaluate the elements of the covariance matrix for three situations: (1) high temperature limit, (2) zero temperature case and (3) low temperature regime.

\section{The Covariance Matrix at High Temperatures}

We consider the high temperature limit $\beta\omega\ll1$ of the elements of the covariance matrix. In this limit, the Hadamard function of the bath \eqref{E:defkjd} is approximately given by
\begin{equation}
	\widetilde{\mathbf{G}}^{kk}_{H}(\kappa)\simeq\frac{1}{2\pi\beta_{k}}\,,
\end{equation}
We see that its vacuum contribution is relatively negligible, and can be neglected for most cases. However, extra discretion is advised for the evaluation of the momentum uncertainty where {the vacuum contribution of the bath can be significant when the coupling between the oscillator and the bath is sufficiently strong.} Thus the result can  {depend on the cutoff scale of the environment field)} (see, e.g., ~\cite{Ludwig}).

Here we merely highlight the calculation for the element $V_{11}$ at late time. To obtain the high temperature limit of $V_{11}(\infty)$, that is, $\mathcal{V}_{11}$, essentially we evaluate the following two integrals
\begin{align}
	 I_{1}&=\int_{-\infty}^{\infty}\!d\kappa\;\frac{(\kappa^{2}-\omega^{2})^{2}+4\gamma^{2}\kappa^{2}}{\bigl[(\kappa^{2}-\omega_{+}^{2})^{2}+4\gamma^{2}\kappa^{2}\bigr]\bigl[(\kappa^{2}-\omega_{-}^{2})^{2}+4\gamma^{2}\kappa^{2}\bigr]}=\frac{\pi}{4\gamma}\left(\frac{\omega^{2}}{\omega^{4}-\sigma^{2}}+\frac{4\gamma^{2}}{4\omega^{2}\gamma^{2}+\sigma^{2}}\right)\,,\\
	 I_{2}&=\int_{-\infty}^{\infty}\!d\kappa\;\frac{\sigma^{2}}{\bigl[(\kappa^{2}-\omega_{+}^{2})^{2}+4\gamma^{2}\kappa^{2}\bigr]\bigl[(\kappa^{2}-\omega_{-}^{2})^{2}+4\gamma^{2}\kappa^{2}\bigr]}=\frac{\pi\sigma^{2}}{4\omega^{2}\gamma}\left(\frac{1}{\omega^{4}-\sigma^{2}}+\frac{1}{4\omega^{2}\gamma^{2}+\sigma^{2}}\right)\,.
\end{align}
In terms of $I_{1}$ and $I_{2}$, we see from \eqref{E:ejrhes} that the high temperature limit of the element $\mathcal{V}_{11}$ at late time is given by
\begin{align}
	\mathcal{V}_{11}=\langle\chi_{1}^{2}(\infty)\rangle&=\frac{2\gamma}{\pi m}\biggl[\frac{I_{1}}{\beta_{1}}+\frac{I_{2}}{\beta_{2}}\biggr]\notag\\
	 &=\frac{1}{2m}\left\{\frac{8\omega^{4}\gamma^{2}+\omega^{2}\sigma^{2}-4\gamma^{2}\sigma^{2}}{(\omega^{4}-\sigma^{2})(4\omega^{2}\gamma^{2}+\sigma^{2})}\frac{1}{\beta_{1}}+\frac{\sigma^{2}(\omega^{2}+4\gamma^{2})}{(\omega^{4}-\sigma^{2})(4\omega^{2}\gamma^{2}+\sigma^{2})}\frac{1}{\beta_{2}}\right\}\,.\label{E:dkjfnekne}
\end{align}
{Here we would like to point out that when the mutual interaction $\sigma$ is large, in particular when $\sigma\to\omega$, the fluctuations of the oscillator grow significantly. This will be traced back to the small values of $\omega_{-}$. We will come back to this feature in due course.}

Derivations of the high temperature forms of $V_{13}$, $V_{14}$, $V_{22}$, $V_{24}$ are given in Appendix B.   Nonetheless for the following discussions we will bring forward the results for $V_{22}$ and $V_{13}$ here. When both private baths have the same temperature $\beta^{-1}$, we have from \eqref{E:fkdfb1}, \eqref{E:fkdfb3} and \eqref{E:fkdfb5}
\begin{align}
	 \mathcal{V}_{11}&=\langle\chi_{1}^{2}(\infty)\rangle=\frac{1}{m\beta}\frac{\omega^{2}}{\omega^{4}-\sigma^{2}}\,,\\
	\mathcal{V}_{22}&=\langle p_{1}^{2}(\infty)\rangle=\frac{m}{\beta}\,,\label{E:derjje}\\
	 \mathcal{V}_{13}&=\langle\chi_{1}(\infty)\chi_{2}(\infty)\rangle=-\frac{1}{m\beta}\frac{\sigma}{\omega^{4}-\sigma^{2}}\,,
\end{align}
in the weak oscillator-bath coupling limit. This implies that the average harmonic potential energy of Oscillator 1 ($O_{1}$) will be
\begin{align}
	 E_{s_{1}}=\frac{m}{2}\,\omega^{2}\langle\chi_{1}^{2}(\infty)\rangle=\frac{1}{2\beta}\frac{\omega^{4}}{\omega^{4}-\sigma^{2}}\,.
\end{align}
It is a bit off from the value $1/2\beta$ one would expect from the equipartition theorem for a free harmonic oscillator in the high temperature limit. This difference will disappear when the mutual coupling $\sigma$ between the two oscillators are turned off.

Eq.~\eqref{E:derjje} on the other hand tells us the corresponding average value of the kinetic energy in the high temperature limit,
\begin{equation}
	E_{k_{1}}=\frac{\mathcal{V}_{22}}{2m}=\frac{1}{2\beta},
\end{equation}
is the same as the value obtained from the classical equipartition theorem. We observe that in the high temperature limit the mean kinetic energy is not equal to the mean harmonic potential energy in general, and the sum of the kinetic energy and the harmonic potential energy is not equal to $kT$:
\begin{equation}
	 E_{k_{1}}+E_{s_{1}}=\frac{1}{2\beta}+\frac{1}{2\beta}\left(1-\frac{\sigma^{2}}{\omega^{4}}\right)^{-1}\neq\frac{1}{\beta}\,.
\end{equation}
Let us compare this with the average total energy of a free harmonic oscillator in a closed system,
\begin{align}\label{E:verge}
	\langle H\rangle&=E_{k}+E_{s}=\frac{\displaystyle\sum_{n=0}^{\infty}E_{n}\,e^{-\beta E_{n}}}{\displaystyle\sum_{n=0}^{\infty}e^{-\beta E_{n}}}=-\frac{\partial\ln Z}{\partial\beta}\simeq\frac{1}{\beta}\,,&Z&=\sum_{n=0}^{\infty}e^{-\beta E_{n}}\,,
\end{align}
in the high temperature limit and $E_{n}=\bigl(n+\frac{1}{2}\bigr)\omega$.

The deviation can be accounted for by the fact that some portion of the total energy of both oscillators is stored in the mutual interaction between $O_{1}$ and $O_{2}$. Accordingly the missing piece should come from the expectation value of $m\sigma\,\chi_{1}\chi_{2}$. Its contribution to the mechanical energy is
\begin{align}
	 E_{\sigma}&=\lim_{t\to\infty}m\sigma\,\langle\chi_{1}(t)\chi_{2}(t)\rangle=m\sigma\,\mathcal{V}_{13}=-\frac{1}{\beta}\frac{\sigma^{2}}{\omega^{4}-\sigma^{2}}\,,
\end{align}
when $\beta_{1}=\beta=\beta_{2}$. Including this contribution we see the total energy for the two-oscillator system in the high-temperature limit becomes
\begin{align}\label{E:hdjhfs}
	 E=E_{k_{1}}+E_{s_{1}}+E_{k_{2}}+E_{s_{2}}+E_{\sigma}=\frac{1}{\beta}+\frac{1}{\beta}\frac{\omega^{4}}{\omega^{4}-\sigma^{2}}-\frac{1}{\beta}\frac{\sigma^{2}}{\omega^{4}-\sigma^{2}}=\frac{2}{\beta}\,.
\end{align}
which is that obtained by the classical equipartition theorem for two coupled linear oscillators. This also serves as a consistency check of our calculation.

Finally we comment on two issues. First, weak oscillator-bath coupling enables us to ignore the cutoff-dependent effect from the bath. This may not be true in the strong coupling case. Second,  despite the resemblance of  \eqref{E:hdjhfs} with  \eqref{E:verge}, they are quite different in the physical context. The former is considered in the context of  open systems while the latter under the assumption of a closed system. It has been shown~\cite{YFTH} that both results can be equivalent only in the limit of vanishingly weak oscillator-bath coupling.

\section{The Covariance Matrix at Zero and Low Temperatures} 
Here we evaluate the vacuum contributions and the low temperature correction of the covariance matrix elements. Due to the zero-point fluctuations of all bath modes, the vacuum contributions of some covariance matrix elements can be divergent.  Suitable cutoffs need be introduced to regularize them, with due consideration of the particulars of the bath the system interacts with.

Let us examine, for example, $\mathcal{V}_{11}=V_{11}(\infty)$ and work out its zero and low temperature expressions.

\subsection{$\mathcal{V}_{11}$ at zero temperature}

The vacuum contribution of $\widetilde{\mathbf{G}}^{kk}_{H}(\kappa)$ is
\begin{equation}
	\widetilde{\mathbf{G}}^{kk}_{H}(\kappa)=\operatorname{sgn}(\kappa)\,\frac{\kappa}{4\pi}\,,
\end{equation}
so we need the following two integrals to evaluate the vacuum contribution of $\mathcal{V}_{11}$,
\begin{align}
	 J_{1}&=\int_{0}^{\infty}\!d\kappa\;\frac{\kappa\bigl[(\kappa^{2}-\omega^{2})^{2}+4\gamma^{2}\kappa^{2}\bigr]}{\bigl[(\kappa^{2}-\omega_{+}^{2})^{2}+4\gamma^{2}\kappa^{2}\bigr]\bigl[(\kappa^{2}-\omega_{-}^{2})^{2}+4\gamma^{2}\kappa^{2}\bigr]}\,,\\
	 J_{2}&=\int_{0}^{\infty}\!d\kappa\;\frac{\kappa\sigma^{2}}{\bigl[(\kappa^{2}-\omega_{+}^{2})^{2}+4\gamma^{2}\kappa^{2}\bigr]\bigl[(\kappa^{2}-\omega_{-}^{2})^{2}+4\gamma^{2}\kappa^{2}\bigr]}\,.
\end{align}
The sum of $J_{1}$ and $J_{2}$ can be expressed as
\begin{align}
	 J_{1}+J_{2}&=\frac{\pi}{16\gamma}\left[\frac{f(\Omega_{+})}{\Omega_{+}}+\frac{f(\Omega_{-})}{\Omega_{-}}\right]\,,&\Omega_{\pm}^{2}&=\omega_{\pm}^{2}-\gamma^{2}
\end{align}
where the dimensionless function $f(z)$ is defined by
\begin{equation}\label{E:pondkfer}
	f(z)=1+\frac{2}{\pi}\cot^{-1}\frac{2\gamma z}{z^{2}-\gamma^{2}}\,.
\end{equation}
It is clear that $\Omega_{\pm}$ are the resonance frequencies of the two normal modes. Therefore from \eqref{E:ejrhes}, the zero-temperature (vacuum) contribution of $\mathcal{V}_{11}$ is given by
\begin{align}\label{E:nvkdbfkw1}
	\mathcal{V}_{11}^{(0)}&=\frac{2\gamma}{\pi m}\biggl[J_{1}+J_{2}\biggr]=\frac{1}{8m}\left[\frac{f(\Omega_{+})}{\Omega_{+}}+\frac{f(\Omega_{-})}{\Omega_{-}}\right]\,.
\end{align}
We observe that the vacuum contribution can be clearly separated into decoupled components of two normal modes, with oscillating frequency $\omega_{\pm}$ respectively. This is another general feature of this system.

The zero-temperature expressions for $\mathcal{V}_{13}$, $\mathcal{V}_{14}$, $\mathcal{V}_{22}$, $\mathcal{V}_{24}$ are given in Appendix C.

\subsection{$\mathcal{V}_{11}$  at low temperature $\beta\omega\gg1$}

The low temperature corrections to the covariance matrix elements basically result from the corresponding correction in the Hadamard function,
\begin{equation}\label{E:hfeirhis}
	 \widetilde{\mathbf{G}}^{kk}_{H}(\kappa)\simeq\text{vac.}+\frac{\kappa}{2\pi}\sum_{n=1}^{\infty}e^{-n\beta_{k}\kappa}\,,
\end{equation}
because the fundamental solution matrix $\mathbf{D}_{1,2}$ does not depend on temperature. This is a consequence of the fact that the retarded Green's function of the scalar field, which accounts for dissipation in the Langevin equation, is state-independent.

As is seen from \eqref{E:ejrhes}, we need the following two integrals to evaluate the low temperature correction of $\mathcal{V}_{11}$,
\begin{align}
	 K_{1}&=2\int_{0}^{\infty}\!d\kappa\;\frac{\kappa\bigl[(\kappa^{2}-\omega^{2})^{2}+4\gamma^{2}\kappa^{2}\bigr]\,e^{-\beta\kappa}}{\bigl[(\kappa^{2}-\omega_{+}^{2})^{2}+4\gamma^{2}\kappa^{2}\bigr]\bigl[(\kappa^{2}-\omega_{-}^{2})^{2}+4\gamma^{2}\kappa^{2}\bigr]}=\frac{2\omega^{4}}{\omega_{+}^{4}\omega_{-}^{4}}\frac{1}{\beta^{2}}+\mathcal{O}(\beta^{-3})\,,\\
	 K_{2}&=2\int_{0}^{\infty}\!d\kappa\;\frac{\sigma^{2}\kappa\,e^{-\beta\kappa}}{\bigl[(\kappa^{2}-\omega_{+}^{2})^{2}+4\gamma^{2}\kappa^{2}\bigr]\bigl[(\kappa^{2}-\omega_{-}^{2})^{2}+4\gamma^{2}\kappa^{2}\bigr]}=\frac{2\sigma^{2}}{\omega_{+}^{4}\omega_{-}^{4}}\frac{1}{\beta^{2}}+\mathcal{O}(\beta^{-3})\,.
\end{align}
We then have the low temperature correction to $\mathcal{V}_{11}$ given by
\begin{align}\label{E:nkdfhd}
	\mathcal{V}_{11}^{(\beta)}\sim\frac{2\gamma}{\pi m}\biggl[K_{1}+K_{2}\biggr]=\frac{4\gamma}{\pi m}\biggl[\frac{\omega^{4}}{\beta_{1}^{2}}+\frac{\sigma^{2}}{\beta^{2}_{2}}\biggr]\frac{1}{\omega_{+}^{4}\omega_{-}^{4}}\,.
\end{align}
However this is merely the contribution from the first term in the summation of all finite temperature corrections in \eqref{E:hfeirhis}. Since the remaining terms $(n>1)$ will have algebraically comparable contributions, we have to take them into consideration. We note that the leading term in $K_{1,2}$ has a temperature dependence $\beta^{-2}$ in the low temperature limit. Thus we expect that the leading contribution for the remainder of the finite temperature corrections in \eqref{E:hfeirhis} should be proportional to $n^{-2}\beta^{-2}$. Their overall contributions will introduce an addition factor
\begin{equation}
	\sum_{n=1}^{\infty}\frac{1}{n^{2}\beta^{2}}=\frac{\pi^{2}}{6}\frac{1}{\beta^{2}}\,,
\end{equation}
to \eqref{E:nkdfhd}. Therefore after taking this into account, we obtain the low temperature correction to $\mathcal{V}_{11}$ as follow:
\begin{align}\label{E:nvkdbfkw2}
	 \mathcal{V}_{11}^{(\beta)}=\frac{2\pi\gamma}{3m}\biggl[\frac{\omega^{4}}{\beta_{1}^{2}}+\frac{\sigma^{2}}{\beta^{2}_{2}}\biggr]\frac{1}{\omega_{+}^{4}\omega_{-}^{4}}\,.
\end{align}
The low temperature expression of $\mathcal{V}_{11}$ is then given by the sum of \eqref{E:nvkdbfkw1} and \eqref{E:nvkdbfkw2},
\begin{align}
	 \mathcal{V}_{11}&=\mathcal{V}_{11}^{(0)}+\mathcal{V}_{11}^{(\beta)}=\frac{1}{8m}\left[\frac{f(\Omega_{+})}{\Omega_{+}}+\frac{f(\Omega_{-})}{\Omega_{-}}\right]+\frac{2\pi\gamma}{3m}\biggl[\frac{\omega^{4}}{\beta_{1}^{2}}+\frac{\sigma^{2}}{\beta^{2}_{2}}\biggr]\frac{1}{\omega_{+}^{4}\omega_{-}^{4}}+\mathcal{O}(\beta_{k}^{-3})\,.
\end{align}
We leave the derivations of the zero and the low temperature expressions for $\mathcal{V}_{13}$, $\mathcal{V}_{14}$, $\mathcal{V}_{22}$, $\mathcal{V}_{24}$ in Appendix D.

\section{Temperature Dependence of the Covariance Matrix}

Because elements of the covariance matrix at finite temperature may prove useful for more general purposes, we collect their  expressions for both high and low temperatures for the system at late times when it reaches a NESS, the existence of which for this setup is shown in our earlier paper
\cite{HHNESS}.

Here we summarize the temperature dependence of the elements of the covariance matrix.
\begin{enumerate}
	\item $\mathcal{V}_{11}=\dfrac{1}{2}\langle\{\chi_{1}(\infty),\chi_{1}(\infty)\}\rangle$:
		\begin{equation}\label{E:fkdfb1}
			\mathcal{V}_{11}=\begin{cases}
											 \dfrac{1}{8m}\left[\dfrac{f(\Omega_{+})}{\Omega_{+}}+\dfrac{f(\Omega_{-})}{\Omega_{-}}\right]+\dfrac{2\pi\gamma}{3m}\biggl[\dfrac{\omega^{4}}{\beta_{1}^{2}}+\dfrac{\sigma^{2}}{\beta^{2}_{2}}\biggr]\dfrac{1}{(\omega^{4}-\sigma^{2})^{2}}\,,&\beta_{1,2}\omega\gg1\,,\vspace{9pt}\\
											 \dfrac{1}{2m}\left[\dfrac{8\omega^{4}\gamma^{2}+\omega^{2}\sigma^{2}-4\gamma^{2}\sigma^{2}}{(\omega^{4}-\sigma^{2})(4\omega^{2}\gamma^{2}+\sigma^{2})}\dfrac{1}{\beta_{1}}+\dfrac{\sigma^{2}(\omega^{2}+4\gamma^{2})}{(\omega^{4}-\sigma^{2})(4\omega^{2}\gamma^{2}+\sigma^{2})}\dfrac{1}{\beta_{2}}\right]\,,&\beta_{1,2}\omega\ll1\,.\vspace{9pt}
										\end{cases}
		\end{equation}
	\item $\mathcal{V}_{12}=\dfrac{1}{2}\langle\{\chi_{1}(\infty),p_{1}(\infty)\}\rangle$:
		\begin{equation}
			\mathcal{V}_{12}=0\,.
		\end{equation}
	\item $\mathcal{V}_{13}=\dfrac{1}{2}\langle\{\chi_{1}(\infty),\chi_{2}(\infty)\}\rangle$:
		\begin{equation}\label{E:fkdfb3}
			\mathcal{V}_{13}=\begin{cases}
											 \dfrac{1}{8m}\left[\dfrac{f(\Omega_{+})}{\Omega_{+}}-\dfrac{f(\Omega_{-})}{\Omega_{-}}\right]-\dfrac{2\pi\gamma}{3m}\dfrac{\omega^{2}\sigma}{(\omega^{4}-\sigma^{2})^{2}}\left[\dfrac{1}{\beta_{1}^{2}}+\dfrac{1}{\beta_{2}^{2}}\right]\,,&\beta_{1,2}\omega\gg1\,,\vspace{9pt}\\
											 -\dfrac{1}{2m}\dfrac{\sigma}{\omega^{4}-\sigma^{2}}\left[\dfrac{1}{\beta_{1}}+\dfrac{1}{\beta_{2}}\right]\,,&\beta_{1,2}\omega\ll1\,.\vspace{9pt}
										\end{cases}
		\end{equation}
	\item $\mathcal{V}_{14}=\dfrac{1}{2}\langle\{\chi_{1}(\infty),p_{2}(\infty)\}\rangle$:
		\begin{equation}
			\mathcal{V}_{14}=0-\begin{cases}
								 \dfrac{8\pi^{3}}{15}\dfrac{\gamma^{2}\sigma}{(\omega^{4}-\sigma^{2})^{2}}\left[\dfrac{1}{\beta^{4}_{1}}-\dfrac{1}{\beta^{4}_{2}}\right]\,,&\beta_{1,2}\omega\gg1\,,\vspace{9pt}\\
								 \dfrac{\gamma\sigma}{4\omega^{2}\gamma^{2}+\sigma^{2}}\left[\dfrac{1}{\beta_{1}}-\dfrac{1}{\beta_{2}}\right]\,,&\beta_{1,2}\omega\ll1\,.\vspace{9pt}
							\end{cases}
		\end{equation}
	\item $\mathcal{V}_{22}=\dfrac{1}{2}\langle\{ p_{1}(\infty),p_{1}(\infty)\}\rangle$:
		\begin{align}\label{E:fkdfb5}
			 \mathcal{V}_{22}&=\dfrac{m\gamma}{2\pi}\,\ln\dfrac{\Lambda^{4}}{\omega^{4}-\sigma^{2}}+\dfrac{m}{8}\left[\dfrac{\Omega_{+}^{2}-\gamma^{2}}{\Omega_{+}}\,f(\Omega_{+})+\dfrac{\Omega_{-}^{2}-\gamma^{2}}{\Omega_{-}}\,f(\Omega_{-})\right]\notag\\
			 &\qquad\qquad\qquad\qquad\qquad+\dfrac{4\pi^{3}}{15}\dfrac{m\gamma}{(\omega^{4}-\sigma^{2})^{2}}\left[\dfrac{\omega^{4}}{\beta_{1}^{4}}+\dfrac{\sigma^{2}}{\beta_{2}^{4}}\right]\,,\qquad\quad\beta_{1,2}\omega\gg1\,,\vspace{9pt}\\
			 &=\dfrac{m\gamma}{\pi}\biggl[\displaystyle\sum_{j=1}^{2}\theta(\beta_{j}\Lambda-1)\,\ln\bigl(\beta_{j}\Lambda\bigr)\biggr]+\dfrac{m}{2}\left[\dfrac{8\omega^{2}\gamma^{2}+\sigma^{2}}{4\omega^{2}\gamma^{2}+\sigma^{2}}\dfrac{1}{\beta_{1}}+\dfrac{\sigma^{2}}{4\omega^{2}\gamma^{2}+\sigma^{2}}\dfrac{1}{\beta_{2}}\right]\,,\notag\\
			 &\qquad\qquad\qquad\qquad\qquad\qquad\qquad\qquad\qquad\qquad\qquad\qquad\qquad\quad\beta_{1,2}\omega\ll1\,.\vspace{9pt}
		\end{align}
		Mathematically speaking, the inclusion of the unit-step function $\theta(\beta\Lambda-1)$ is to account for the vacuum contribution of the bath modes in the case $\beta\Lambda>1$, because when $\beta\kappa>1$, the Hadamard function $\widetilde{G}_{H}^{kk}(\kappa)$ takes the low-temperature form as shown in \eqref{E:defkjd}. On the other hand when $\beta\Lambda<1$, the high-temperature approximation of $\widetilde{G}_{H}^{kk}(\kappa)$ is entirely valid up to the cutoff scale, the cutoff-dependent term being subdominant. However on physical grounds, since the cutoff scale by construction is the highest energy scale compatible with the model, the thermal excitation energy thus must be smaller than the cutoff scale. It then implies that even in the high temperature limit, we still have $\omega\ll\beta^{-1}<\Lambda$.
		\begin{align*}
    		&\raisebox{-0.9em}{\scalebox{0.45}{\includegraphics{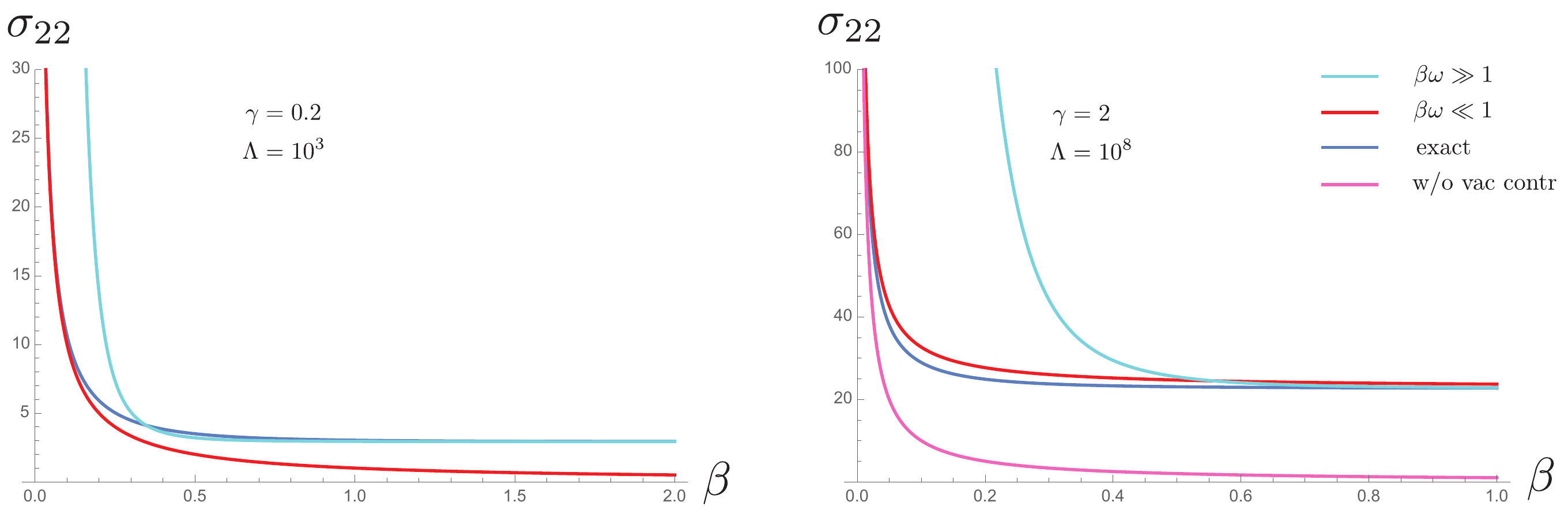}}}
		\end{align*}
		Here we show the high/low temperature approximations of $\mathcal{V}_{22}$ with a numerical calculation. In particular we explicitly highlight the role of the vacuum contribution, that is, the cutoff dependent terms, even in the high temperature approximation for strong oscillator-bath interaction.  {The pink curve in the plot on the right shows} that if the vacuum contribution of the bath is not taken into account, the {analytical high-temperature} approximation will be way off from the numerical result (the purple curve) in the region $\beta\omega\sim\mathcal{O}(1)$.  {On the other hand, the red curve, which includes the vaccuum contribution, fits nicely with the numerical result}. The parameters are chosen to be $\gamma=0.2$, $\sigma=18$, $\omega=5$, and $\Lambda=1000$. The plot on the left is drawn for weak oscillator-bath interaction $\gamma=0.2$, i.e. $\gamma/\omega\ll1$. The cutoff-dependence is seen as dispensable.
	\item $\mathcal{V}_{24}=\dfrac{1}{2}\langle\{p_{1}(\infty),p_{2}(\infty)\}\rangle$:
		\begin{align}
			 \mathcal{V}_{24}&=-\dfrac{m\gamma}{\pi}\,\ln\dfrac{\omega_{+}}{\omega_{-}}+\dfrac{m}{8}\biggl[\dfrac{\Omega_{+}^{2}-\gamma^{2}}{\Omega_{+}}\,f(\Omega_{+})-\dfrac{\Omega_{-}^{2}-\gamma^{2}}{\Omega_{-}}\,f(\Omega_{-})\biggr]\notag\\
			 &\qquad\qquad\qquad\qquad-\dfrac{4\pi^{3}}{15}\dfrac{m\omega^{2}\gamma\sigma}{(\omega^{4}-\sigma^{2})^{2}}\left[\dfrac{1}{\beta_{1}^{4}}+\dfrac{1}{\beta_{2}^{4}}\right]\,,\qquad\qquad\quad\beta_{1,2}\omega\gg1\,,\vspace{9pt}\\
			 &=\dfrac{m\sigma}{24}\Bigl[\beta_{1}+\beta_{2}\Bigr]\,,\qquad\qquad\qquad\qquad\qquad\qquad\qquad\qquad\qquad\quad\beta_{1,2}\omega\ll1\,.\vspace{9pt}
		\end{align}
		In this case, since the leading contribution of the high-temperature approximation vanishes, we have to include the next-order term.
	\item $\mathcal{V}_{23}=\dfrac{1}{2}\langle\{\chi_{2}(\infty),p_{1}(\infty)\}\rangle$: it is equal to $-\mathcal{V}_{14}$, so
		\begin{equation}
			\mathcal{V}_{23}=-\mathcal{V}_{14}=0+\begin{cases}
								 \dfrac{8\pi^{3}}{15}\dfrac{\gamma^{2}\sigma}{(\omega^{4}-\sigma^{2})^{2}}\left[\dfrac{1}{\beta^{4}_{1}}-\dfrac{1}{\beta^{4}_{2}}\right]\,,&\beta_{1,2}\omega\gg1\,,\vspace{9pt}\\
								 \dfrac{\gamma\sigma}{4\omega^{2}\gamma^{2}+\sigma^{2}}\left[\dfrac{1}{\beta_{1}}-\dfrac{1}{\beta_{2}}\right]\,,&\beta_{1,2}\omega\ll1\,.\vspace{9pt}
							\end{cases}
		\end{equation}
	\item $\mathcal{V}_{34}=\dfrac{1}{2}\langle\{\chi_{2}(\infty),p_{2}(\infty)\}\rangle$:
		\begin{equation}
			\mathcal{V}_{34}=0\,.
		\end{equation}
	\item $\mathcal{V}_{33}=\dfrac{1}{2}\langle\{\chi_{2}(\infty),\chi_{2}(\infty)\}\rangle$: it is similar to $\mathcal{V}_{11}$ except that we swap $\beta_{1}$ and $\beta_{2}$,
		\begin{equation}
			\mathcal{V}_{33}=\begin{cases}
											 \dfrac{1}{8m}\left[\dfrac{f(\Omega_{+})}{\Omega_{+}}+\dfrac{f(\Omega_{-})}{\Omega_{-}}\right]+\dfrac{2\pi\gamma}{3m}\biggl[\dfrac{\omega^{4}}{\beta_{2}^{2}}+\dfrac{\sigma^{2}}{\beta^{1}_{2}}\biggr]\dfrac{1}{(\omega^{4}-\sigma^{2})^{2}}\,,&\beta_{1,2}\omega\gg1\,,\vspace{9pt}\\
											 \dfrac{1}{2m}\left[\dfrac{8\omega^{4}\gamma^{2}+\omega^{2}\sigma^{2}-4\gamma^{2}\sigma^{2}}{(\omega^{4}-\sigma^{2})(4\omega^{2}\gamma^{2}+\sigma^{2})}\dfrac{1}{\beta_{2}}+\dfrac{\sigma^{2}(\omega^{2}+4\gamma^{2})}{(\omega^{4}-\sigma^{2})(4\omega^{2}\gamma^{2}+\sigma^{2})}\dfrac{1}{\beta_{1}}\right]\,,&\beta_{1,2}\omega\ll1\,.\vspace{9pt}
										\end{cases}
		\end{equation}
	\item $\mathcal{V}_{44}=\dfrac{1}{2}\langle\{ p_{2}(\infty),p_{2}(\infty)\}\rangle$: it is similar to $\mathcal{V}_{22}$,
		\begin{align}
			 \mathcal{V}_{44}&=\dfrac{m\gamma}{2\pi}\,\ln\dfrac{\Lambda^{4}}{\omega^{4}-\sigma^{2}}+\dfrac{m}{8}\left[\dfrac{\Omega_{+}^{2}-\gamma^{2}}{\Omega_{+}}\,f(\Omega_{+})+\dfrac{\Omega_{-}^{2}-\gamma^{2}}{\Omega_{-}}\,f(\Omega_{-})\right]\notag\\
			 &\qquad\qquad\qquad\qquad+\dfrac{4\pi^{3}}{15}\dfrac{m\gamma}{(\omega^{4}-\sigma^{2})^{2}}\left[\dfrac{\omega^{4}}{\beta_{2}^{4}}+\dfrac{\sigma^{2}}{\beta_{1}^{4}}\right]\,,\qquad\qquad\quad\beta_{1,2}\omega\gg1\,,\vspace{9pt}\\
			 &=\dfrac{m\gamma}{\pi}\biggl[\displaystyle\sum_{j=1}^{2}\theta(\beta_{j}\Lambda-1)\,\ln\bigl(\beta_{j}\Lambda\bigr)\biggr]+\dfrac{m}{2}\left[\dfrac{8\omega^{2}\gamma^{2}+\sigma^{2}}{4\omega^{2}\gamma^{2}+\sigma^{2}}\dfrac{1}{\beta_{2}}+\dfrac{\sigma^{2}}{4\omega^{2}\gamma^{2}+\sigma^{2}}\dfrac{1}{\beta_{1}}\right]\,,\notag\\
			 &\qquad\qquad\qquad\qquad\qquad\qquad\qquad\qquad\qquad\qquad\qquad\qquad\qquad\quad\beta_{1,2}\omega\ll1\,.\vspace{9pt}
		\end{align}
\end{enumerate}
Some comments are in place here:  Both oscillator are initially prepared in a state of non-overlapping Gaussian wavepackets with the same width $\varsigma$. As they come into interaction with their own private baths, the evolution of each individual oscillator is then driven by its bath and the other oscillator it is directly coupled with. Due to the dissipative self-force on the oscillator arising from its interaction with its own bath, the intrinsic information of the initial state is dispersed away exponentially fast as the system evolves in time. In the end when $t\to\infty$, the values of the dynamical variables of the oscillator are determined by its private bath and  by the other oscillator. We want to bring up this point because even when  the oscillator-bath coupling constant $\gamma$ approaches zero,  not all of the asymptotic values of the covariant matrix elements are zero.  In this limit their values are independent of the parameter $\varsigma$ characterizing the initial state, so they are not related  to the intrinsic evolution that begins with the initial configuration. Instead they are the induced components as a consequence of the interaction between the oscillator and the bath. In other words, the results of the covariance matrix in the vanishing $\gamma$ limit should be understood by the limiting procedures of taking $t\to\infty$ first and then taking $\gamma\to0$.

This is a good point to discuss in more details in what is meant by the high/low temperature approximations. We only cover the generic situation and discard some extreme cases such as $\omega_{-}$, $\Omega_{-}$, $\sigma\to0$, so we assume that $\sigma^{\frac{1}{2}}$, $\omega_{\pm}$ and $\Omega_{\pm}$ are about the same order of magnitude as the oscillating frequency $\omega$. The cutoff frequency is assumed to be much larger than $\omega$, i.e., $\Lambda\gg\omega$. The magnitude of the parameters $\gamma$ and $\beta_{1,2}$ are not restricted as long as $\Omega_{-}$ remains well-defined. We use $\mathcal{V}_{11}$ and $\mathcal{V}_{22}$ as illustrating examples,
\begin{enumerate}
	\item $\mathcal{V}_{11}$: as far as the order of magnitude is concerned, we see
		\begin{align*}
			 &\text{{vacumm:}}&\mathcal{V}_{11}^{(0)}&=\dfrac{1}{8m}\left[\dfrac{f(\Omega_{+})}{\Omega_{+}}+\dfrac{f(\Omega_{-})}{\Omega_{-}}\right]\sim\frac{1}{m\omega}\,,\\
			&\text{{low temp:}}&\mathcal{V}_{11}^{(\beta)}&=\dfrac{2\pi\gamma}{3m}\biggl[\dfrac{\omega^{4}}{\beta_{1}^{2}}+\dfrac{\sigma^{2}}{\beta^{2}_{2}}\biggr]\dfrac{1}{(\omega^{4}-\sigma^{2})^{2}}\sim\frac{1}{m\omega}\frac{\gamma}{\omega}\frac{1}{(\beta\omega)^{2}}\,,\\
			&\text{{high temp:}}&\mathcal{V}_{11}&=\dfrac{1}{2m}\left[\dfrac{8\omega^{4}\gamma^{2}+\omega^{2}\sigma^{2}-4\gamma^{2}\sigma^{2}}{(\omega^{4}-\sigma^{2})(4\omega^{2}\gamma^{2}+\sigma^{2})}\dfrac{1}{\beta_{1}}+\dfrac{\sigma^{2}(\omega^{2}+4\gamma^{2})}{(\omega^{4}-\sigma^{2})(4\omega^{2}\gamma^{2}+\sigma^{2})}\dfrac{1}{\beta_{2}}\right]\notag\\
			&&&\sim\frac{1}{m\omega}\frac{1}{\beta\omega}\,.
		\end{align*}
		Roughly speaking, the high temperature limit refers to the case $\beta\omega\ll1$; on the other hand a consistent low temperature correction requires
		\begin{equation*}
			\frac{\gamma}{\omega}\frac{1}{(\beta\omega)^{2}}\ll1\,,
		\end{equation*}
		which can be weaker than the naive low temperature limit $\beta\omega\gg1$, especially in the weak coupling limit $\gamma/\omega\ll1$. It implies that in the weak oscillator-bath coupling limit, the low temperature correction has a much wider range of validity. In the strong coupling regime $\gamma\lesssim\omega$, the low temperature correction is remains fully valid for the $\beta\omega\gg1$.
	\item $\mathcal{V}_{22}$:
		\begin{align*}
			 &\text{{vacuum:}}&\mathcal{V}_{22}^{(0)}&=\frac{m\gamma}{2\pi}\,\ln\frac{\Lambda^{4}}{\omega^{4}-\sigma^{2}}+\frac{m}{8}\left[\frac{\Omega_{+}^{2}-\gamma^{2}}{\Omega_{+}}\,f(\Omega_{+})+\frac{\Omega_{-}^{2}-\gamma^{2}}{\Omega_{-}}\,f(\Omega_{-})\right]\notag\\
			 &&&\sim\begin{cases}m\omega\,\dfrac{\gamma}{\omega}\ln\dfrac{\Lambda}{\omega}\,,\\m\omega\,,\end{cases}\\
			&\text{{low temp:}}&\mathcal{V}_{22}^{(\beta)}&=\dfrac{4\pi^{3}}{15}\dfrac{m\gamma}{(\omega^{4}-\sigma^{2})^{2}}\left[\dfrac{\omega^{4}}{\beta_{1}^{4}}+\dfrac{\sigma^{2}}{\beta_{2}^{4}}\right]\sim m\omega\,\frac{\gamma}{\omega}\frac{1}{(\beta\omega)^{4}}\,,\\
			&\text{{high temp:}}&\mathcal{V}_{22}^{(\beta)}&=\dfrac{m}{2}\left[\dfrac{8\omega^{2}\gamma^{2}+\sigma^{2}}{4\omega^{2}\gamma^{2}+\sigma^{2}}\dfrac{1}{\beta_{1}}+\dfrac{\sigma^{2}}{4\omega^{2}\gamma^{2}+\sigma^{2}}\dfrac{1}{\beta_{2}}\right]\sim m\omega\,\frac{1}{\beta\omega}\,.
		\end{align*}
		Here, additional subtlety arises due to the presence of the cutoff frequency $\Lambda$. The importance of the cutoff-dependent term relies on how the factor
		\begin{equation*}
			\dfrac{\gamma}{\omega}\ln\dfrac{\Lambda}{\omega}
		\end{equation*}
		is compared with unity. In the weak coupling limit, the cutoff dependent term is negligible, so we can safely ignore it unless the cutoff frequency is extremely high, such as
		\begin{equation*}
			\Lambda\simeq\mathcal{O}(\omega\,e^{\frac{\omega}{\gamma}})\,.
		\end{equation*}
		In the strong coupling regime $\gamma\lesssim\omega$, we see that the cutoff-dependent term still has a comparable magnitude relative to the high temperature approximation in the interval of the high-to-low temperature transition $\beta\omega\simeq\mathcal{O}(1)$. This interval has a special significance because, as we will see later, this is the region where thermal entanglement may disappear in the nonequilibrium steady state configuration.
\end{enumerate}
Thus at this point, generically speaking, the high-temperature limit refers to $\beta\omega\ll1$ while the low-temperature limit refers to $\beta\omega\gg1$. For weak oscillator-bath coupling, the low temperature correction has a wider range of validity than is implied by $\beta\omega\gg1$ due to the additional factor $\gamma/\omega$ in the corresponding expression. In addition, the cutoff is completely negligible in normal circumstances.  By contrast,  in the strong coupling regime, the cutoff-dependent contribution enters in determining the critical temperature of thermal entanglement.

\section{Entanglement of System in Nonequilibrium Steady State}

\subsection{Late Time Behavior of the Covariance Matrix}

At late time when the system reaches the steady state, the covariance matrix takes the form
\begin{equation}
	\mathbf{V}=\begin{pmatrix}
		\mathcal{V}_{11}&\mathcal{V}_{12}&\mathcal{V}_{13}&\mathcal{V}_{14}\\
		\mathcal{V}_{21}&\mathcal{V}_{22}&\mathcal{V}_{23}&\mathcal{V}_{24}\\
		\mathcal{V}_{31}&\mathcal{V}_{32}&\mathcal{V}_{33}&\mathcal{V}_{34}\\
		\mathcal{V}_{41}&\mathcal{V}_{42}&\mathcal{V}_{43}&\mathcal{V}_{44}
	\end{pmatrix}
	=\begin{pmatrix}
		\mathcal{V}_{11}&0&\mathcal{V}_{13}&\mathcal{V}_{14}\\
		0&\mathcal{V}_{22}&-\mathcal{V}_{14}&\mathcal{V}_{24}\\
		\mathcal{V}_{13}&-\mathcal{V}_{14}&\mathcal{V}_{33}&0\\
		\mathcal{V}_{14}&\mathcal{V}_{24}&0&\mathcal{V}_{44}
	\end{pmatrix}=\begin{pmatrix}\mathbf{A} &\mathbf{C} \\
					\mathbf{C}^{T} &\mathbf{B}
				\end{pmatrix}\,,
\end{equation}
with
\begin{align}
	 \mathbf{A}&=\begin{pmatrix}\mathcal{V}_{11}&0\\0&\mathcal{V}_{22}\end{pmatrix}\,,&\mathbf{B}&=\begin{pmatrix}\mathcal{V}_{33}&0\\0&\mathcal{V}_{44}\end{pmatrix}\,,\\
	 \mathbf{C}&=\begin{pmatrix}\mathcal{V}_{13}&\mathcal{V}_{14}\\-\mathcal{V}_{14}&\mathcal{V}_{24}\end{pmatrix}\,,&\mathbf{J}&=\begin{pmatrix}0&+1\\-1&0\end{pmatrix}\,.
\end{align}
The determinants of the matrices $\mathbf{A}$, $\mathbf{B}$ are related to the generalized uncertainty relation for each single subsystem, which also takes into account the correlation between canonical variables. The matrix $\mathbf{C}$ contains the cross-correlation among canonical variables between two subsystems.

As is briefly discussed in Sec.~\ref{S:ekjbek}, the knowledge of the covariance matrix enables us to use the Peres-Horodecki-Simon separability criterion to determine the quantum entanglement. In fact the separability criterion can be combined with the generalized uncertainty relation to form an unified statement
\begin{align}\label{E:dhbeh}
	 \zeta_{\pm}=\det\mathbf{A}\,\det\mathbf{B}-\operatorname{Tr}\bigl\{\mathbf{A}\cdot\mathbf{J}\cdot\mathbf{C}\cdot\mathbf{J}\cdot\mathbf{B}\cdot\mathbf{J}\cdot\mathbf{C}^{T}\cdot\mathbf{J}\bigr\}+\bigl(\det\mathbf{C}\pm\frac{1}{4}\bigr)^{2}-\frac{1}{4}\bigl(\det\mathbf{A}+\det\mathbf{B}\bigr)\geq0\,,
\end{align}
The expression containing the $-$ sign represents the uncertainty relation  while that with the $+$ sign represents the separability criterion. We immediately see that in the current case $\det\mathbf{A}$ and $\det\mathbf{B}$ are always positive definite by construction. In addition, the expression $\operatorname{Tr}\bigl\{\mathbf{A}\cdot\mathbf{J}\cdot\mathbf{C}\cdot\mathbf{J}\cdot\mathbf{B}\cdot\mathbf{J}\cdot\mathbf{C}^{T}\cdot\mathbf{J}\bigr\}$, once written explicitly in term of the covariance matrix elements,
\begin{equation}
	\operatorname{Tr}\bigl\{\mathbf{A}\cdot\mathbf{J}\cdot\mathbf{C}\cdot\mathbf{J}\cdot\mathbf{B}\cdot\mathbf{J}\cdot\mathbf{C}^{T}\cdot\mathbf{J}\bigr\}=\mathcal{V}_{22}\mathcal{V}_{44}\mathcal{V}_{13}^{2}+\mathcal{V}_{11}\mathcal{V}_{22}\mathcal{V}_{24}^{2}+\left(\mathcal{V}_{11}\mathcal{V}_{44}+\mathcal{V}_{22}\mathcal{V}_{33}\right)\mathcal{V}_{14}^{2}>0\,,
\end{equation} 
is found to be always positive. Thus when we rewrite \eqref{E:dhbeh} as,
\begin{align}
	\zeta_{\pm}=\left[\Bigl(\det\mathbf{A}-\frac{1}{4}\Bigr)\Bigl(\det\mathbf{B}-\frac{1}{4}\Bigr)+\Bigl(\det\mathbf{C}\pm\frac{1}{4}\Bigr)^{2}\right]-\left[\operatorname{Tr}\Bigl\{\mathbf{A}\cdot\mathbf{J}\cdot\mathbf{C}\cdot\mathbf{J}\cdot\mathbf{B}\cdot\mathbf{J}\cdot\mathbf{C}^{T}\cdot\mathbf{J}\Bigr\}+\frac{1}{16}\right]\,.
\end{align}
we immediately recognize that $\zeta_{\pm}$ actually contains two positive but competing components. This makes it difficult to determine the sign of $\zeta_{\pm}$. However,  we can use the following argument: Suppose that the uncertainty relation $\zeta_{-}\geq0$ always holds. Since
\begin{equation}
	\zeta_{+}=\zeta_{-}+\det\mathbf{C}\,,
\end{equation}
the condition $\zeta_{-}<0$ implies that $\det\mathbf{C}$ must be negative. Therefore  the appearance of negative values  of $\det\mathbf{C}$ may help to signify the existence of entanglement. The sign of $\det\mathbf{C}$ is less clear,
\begin{equation}
	\det\mathbf{C}=\mathcal{V}_{13}\mathcal{V}_{24}+\mathcal{V}_{14}^{2}\,,
\end{equation}
depending on how negative $\mathcal{V}_{13}\mathcal{V}_{24}$ can be allowed. Although this is not a sufficient condition, it highlights the role of cross-correlations in entanglement.

\begin{figure}
\centering
    \scalebox{0.5}{\includegraphics{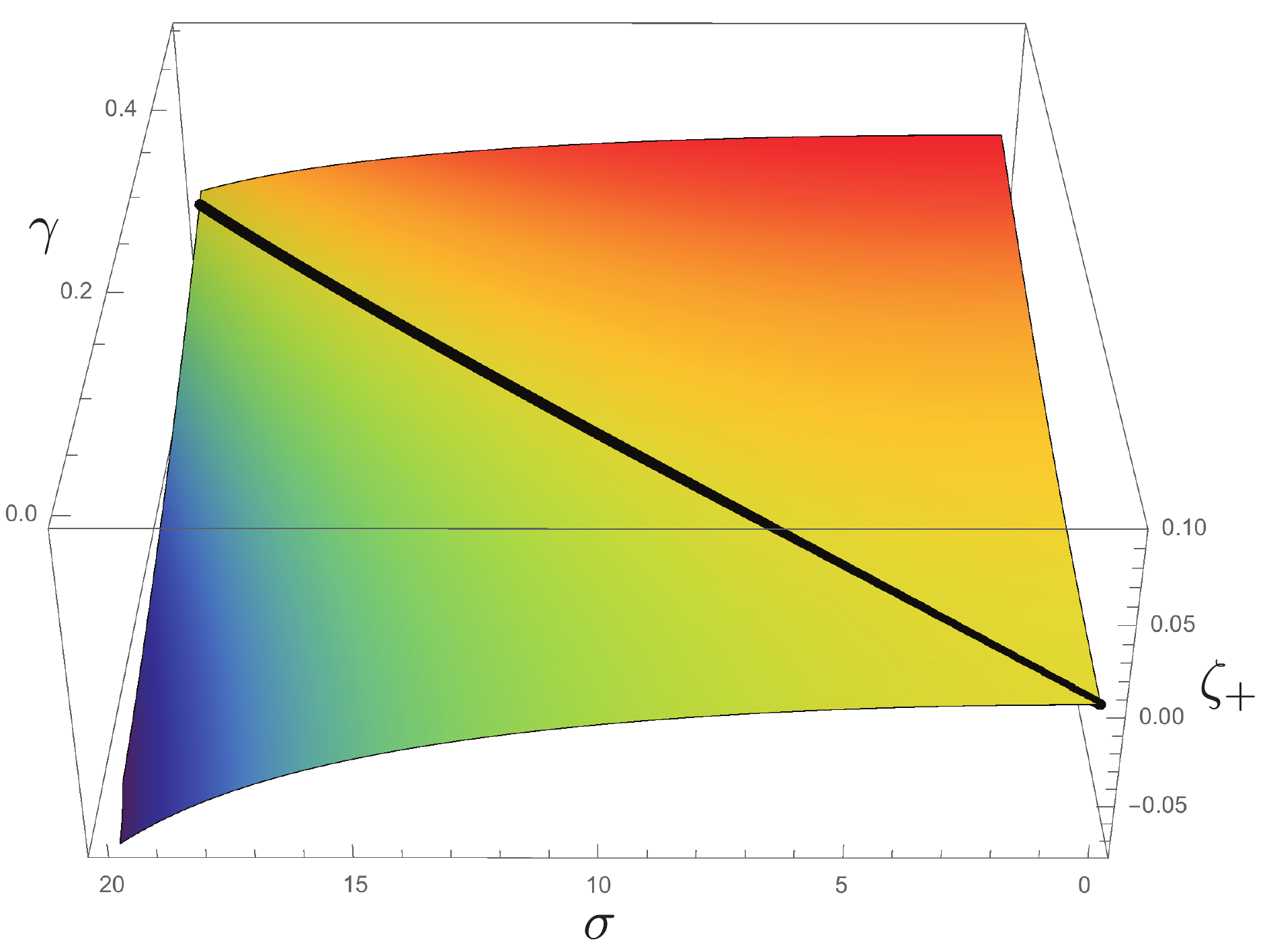}}
    \caption{The separability criterion $\zeta_{+}$ is plotted against the oscillator-bath coupling $\gamma$ and the mutual coupling strength $\sigma$ between the oscillators. The black curve demarcates the separate ($\zeta_{+}>0$) and the entangled ($\zeta_{+}<0$) regions. The choices for the parameters are $\omega=5$ and $\Lambda=10000$.}\label{Fi:vacuum ent}
\end{figure}

Before we proceed to evaluate $\zeta_{+}$, we observe that among the elements of the covariance matrix, two of them, $\mathcal{V}_{22}$ and $\mathcal{V}_{44}$, have dependence on the cutoff frequency $\Lambda$, which is the highest energy scale that is consistent with the theory. Thus we expect that $\zeta_{\pm}$,  and in particular,  the separability criterion, will depend on the cutoff scale. Since the cutoff-dependent term always has the form $\gamma\ln\Lambda$, where $\gamma$ is the system-environment coupling constant, it implies that this cutoff dependence will be suppressed in the weak coupling limit. However, when the system interacts strongly with the environment in the sense that $\gamma/\omega$ is close to unity, the contribution from the factor $\gamma\ln\Lambda$ can be significant, and can make the separability criterion ambiguous.

{Likewise in terms of the matrices $\mathbf{A}$, $\mathbf{B}$, and $\mathbf{C}$ we can construct the symplectic eigenvalues $\eta_{\gtrless}$ of the partial transpose $\mathbf{V}^{pt}$ of the covariance matrix $\mathbf{V}$~\cite{Illuminati04},
\begin{align}\label{E:bheuhs}
	\eta_{\gtrless}&=\left[\left(\frac{\det\mathbf{A}+\det\mathbf{B}}{2}-\det\mathbf{C}\right)\pm\sqrt{\left(\frac{\det\mathbf{A}+\det\mathbf{B}}{2}-\det\mathbf{C}\right)^{2}-\det\mathbf{V}}\right]^{\frac{1}{2}}\,,
\end{align}
where alternatively $\det\mathbf{V}$ can be written as $\det\mathbf{A}\,\det\mathbf{B}+(\det\mathbf{C})^{2}-\operatorname{Tr}\{\mathbf{A}\cdot\mathbf{J}\cdot\mathbf{C}\cdot\mathbf{J}\cdot\mathbf{B}\cdot\mathbf{J}\cdot\mathbf{C}^{T}\cdot\mathbf{J}\}$. This enables us to calculate the quantitative entanglement measures like negativity or logarithmic negativity for the Gaussian state.}

{In the sections that follow, we will refer to the special case when both thermal reservoirs have the same temperature. In this case, the Gaussian state becomes symmetric, so (logarithmic) negativity will give an unambiguous ordering of density matrices, in comparison with other quantitative entanglement measures. Since we have $\mathbf{A}=\mathbf{B}$, and the matrix $\mathbf{C}$ becomes diagonal, the symplectic eigenvalues $\eta_{\gtrless}$ takes a particularly neat form
\begin{equation}\label{E:dbekrheq}
	\eta_{\gtrless}=\Bigl[\bigl(\mathcal{V}_{11}\mp\mathcal{V}_{13}\bigr)\bigl(\mathcal{V}_{22}\pm\mathcal{V}_{24}\bigr)\Bigr]^{\frac{1}{2}}\,,
\end{equation}
with
\begin{align*}
	\mathcal{V}_{11}&=\langle\chi_{1}^{2}(t)\rangle\,,&\mathcal{V}_{22}&=\langle p_{1}^{2}(t)\rangle\,,&\mathcal{V}_{13}&=\frac{1}{2}\,\langle\bigl\{\chi_{1}(t),\chi_{2}(t)\bigr\}\rangle\,,&\mathcal{V}_{24}&=\frac{1}{2}\,\langle\bigl\{p_{1}(t),p_{2}(t)\bigr\}\rangle\,.
\end{align*}
We readily see that
\begin{align}
	\mathcal{V}_{11}\mp\mathcal{V}_{13}&=\frac{1}{2}\,\langle\bigl\{\chi_{1},\chi_{1}\mp\chi_{2}\bigr\}\rangle\,,\\
	\mathcal{V}_{22}\pm\mathcal{V}_{24}&=\frac{1}{2}\,\langle\bigl\{p_{1},p_{1}\pm p_{2}\bigr\}\rangle\,,
\end{align}
are associated with the dynamics of the normal modes of the joint system.  This elicits a transparent connection between the entanglement behavior and the underlying dynamics.}

\subsection{Entanglement Behavior}
As stated earlier the Peres-Simon-Horodecki criterion can be used to identity the existence of entanglement of the Gaussian states, but it  may be inadequate to provide a quantitative description of entanglement, in particular, for a quantum system at finite temperature.  We will show later that it does not vary monotonically with temperature and coupling constants. This discrepancy comes from an additional factor in the criterion. It has no effect on the identification of entanglement but it will give an unwarranted bias on the values, rendering it inappropriate for quantifying  entanglement. While the separability criterion can be used for the system under study at zero temperature,  we need a different  measure to quantify  finite-temperature entanglement, namely,  negativity.

\begin{figure}
\centering
    \scalebox{0.5}{\includegraphics{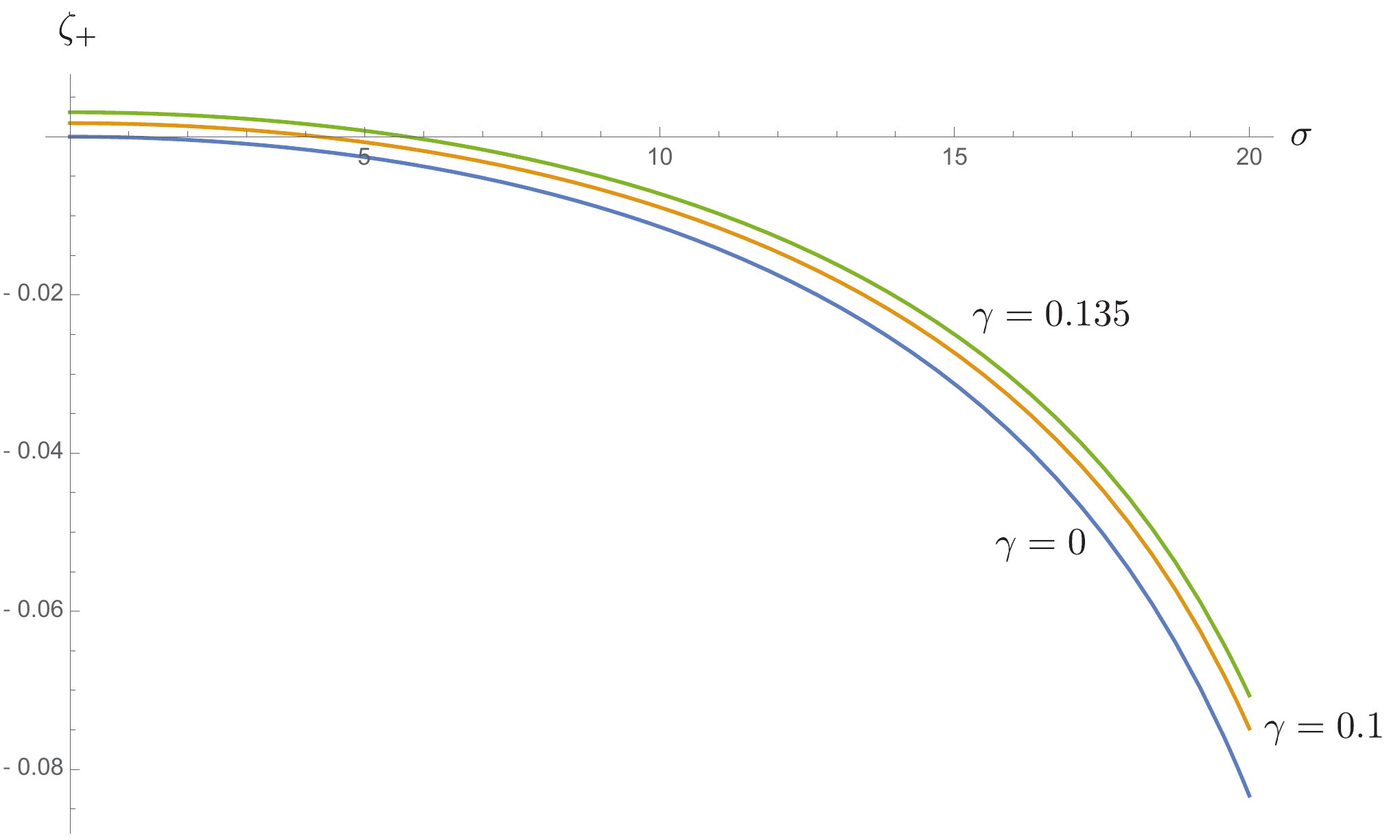}}
    \caption{The separability criterion $\zeta_{+}$ is plotted against the mutual coupling strength $\sigma$ between the oscillators at zero temperature. Larger values of the damping constant $\gamma$ will move the curve upwards and make the entanglement between the two oscillators harder to sustain. The oscillating frequency $\omega$ and the cutoff frequency $\Lambda$ are chosen to be 5 and 10000, respectively}\label{Fi:th_ent}
\end{figure}

\subsubsection{zero temperature}
We first examine the separability criterion $\zeta_{+}$ at zero temperature. The whole expression for $\zeta_{+}$ can be exactly found but it is tremendously large. Here we present the leading terms in the weak oscillator-bath coupling limit, i.e., $\gamma<\omega_{\pm}$ is the smallest parameter at hand,
\begin{align}
	 \zeta_{+}&=-\frac{(\omega_{+}-\omega_{-})^{2}}{16\omega_{+}\omega_{-}}+\gamma\,\frac{(\omega_{+}-\omega_{-})}{8\pi\omega_{+}^{2}\omega_{-}^{2}}\biggl[\omega_{+}^{2}-\omega_{-}^{2}+\omega_{+}\omega_{-}\ln\frac{\omega_{+}^{2}}{\omega_{-}^{2}}\biggr]\notag\\
	 &\qquad\qquad+\frac{\gamma^{2}}{32\pi^{2}\omega_{+}^{2}\omega_{-}^{2}}\biggl\{\pi^{2}\bigl(\omega_{+}-\omega_{-}\bigr)^{2}\bigl(\omega_{+}^{2}+4\omega_{+}\omega_{-}+\omega_{-}^{2}\bigr)-32\omega_{+}^{2}\omega_{-}^{2}\biggr.\notag\\
	 &\qquad\qquad+\biggl.16\omega_{+}\omega_{-}\biggl[\omega_{+}^{2}\ln\frac{\omega_{+}}{\omega_{-}}-2\omega_{+}\omega_{-}\biggl(\ln\frac{\omega_{+}}{\Lambda}\ln\frac{\omega_{-}}{\Lambda}+\ln\frac{\omega_{+}}{\Lambda}+\ln\frac{\omega_{-}}{\Lambda}\biggr)-\omega_{-}^{2}\ln\frac{\omega_{+}}{\omega_{-}}\biggr]\biggr\}\notag\\
	&\qquad\qquad+\mathcal{O}(\gamma \ln\Lambda)^{3}\,.\label{E:dheura}
\end{align}
Note that it is not sufficient to expand $\zeta_{+}$ to first order in $\gamma$ because they all depend on $(\omega_{+}-\omega_{-})$. This factor will make the first-order expansion of $\zeta_{+}$ vanish when $\sigma\to0$ no matter what value $\gamma$ has. In fact $\zeta_{+}$ has a finite value when $\gamma\neq0$, so we have to include terms which are at least of second order in $\gamma$.

In addition, as far as the leading contribution is concerned, we see that $(\omega_{+}-\omega_{-})^{2}$ is always positive, so $\zeta_{+}$ is negative for all nonzero mutual coupling strength $\sigma$ between the two oscillators.  This implies that the oscillators will become entangled aympotically  once they are coupled. On the other hand, when we consider contributions due to the finite value of the damping constant $\gamma$, we find that, in particular in the limit $\sigma\to0$, we have $\omega_{+}\to\omega_{-}$ and
\begin{align}
	\lim_{\sigma\to0}\zeta_{+}=\frac{\gamma^{2}}{\pi^{2}\omega^{2}}\left(\ln\frac{\Lambda}{\omega}-1\right)^{2}>0\,.
\end{align}
The separability criterion $\zeta_{+}$ is positive for $\sigma=0$ when $\gamma\neq0$. With increasing $\sigma$, the value of $\zeta_{+}$ gradually falls below zero at some critical value of $\sigma_{c}$. Thus we see that the curve of the separability criterion will move upwards with larger values of the damping constant $\gamma$, that is, with stronger interaction between the oscillator and its private bath. Furthermore, it also indicates that these oscillators are not always entangled, and they can be separable for some choices of $\gamma$ and $\sigma$. For a specific value of $\gamma$, the mutual coupling strength $\sigma$ must be greater than the critical value to render both oscillators entangled. In other words, the bonding between two oscillators has to be strong enough to overcome the incoherent disturbance from their respective baths in order to maintain their entanglement.  The larger the values of $\sigma$ the more stable the mutual entanglement is. Therefore we see that the oscillator-bath interaction and the coupling between oscillators play competing roles in sustaining the entanglement.

We now derive a relation between the critical values of different couplings. For the case of a small damping constant $\gamma$, the critical value $\sigma_{c}$ can be obtained by solving \eqref{E:dheura} with $\zeta_{+}=0$,  yielding
\begin{align}\label{E:beers}
	 \gamma=\frac{\pi\sigma}{4\omega}\frac{1}{\ln\frac{\Lambda}{\omega}-1}-\frac{\pi\sigma^{2}}{4\omega^{3}}\frac{1}{\left(\ln\frac{\Lambda}{\omega}-1\right)^{2}}+\cdots\,.
\end{align}
Inverting it leads to
\begin{equation}\label{E:dfberjh}
	 \sigma_{c}=\left[\ln\frac{\Lambda}{\omega}-1\right]\left[\frac{4\omega}{\pi}\,\gamma-\frac{16}{\pi^{2}}\,\gamma^{2}+\cdots\right]\,.
\end{equation}
Taking Fig~\ref{Fi:th_ent} as an example,  setting $\omega=6$, $\gamma=0.135$ and $\Lambda=10000$ in \eqref{E:dfberjh} gives $\sigma_{c}=5.868$. It is nicely consistent with the intersection of the green curve with the horizontal axis in Fig~\ref{Fi:th_ent}.

Therefore in the weak coupling regime where $\frac{\gamma}{\omega}\ln\frac{\Lambda}{\omega}$ is small but not vanishing, we find that if the two oscillators are initially prepared in a disentangled state, they can become entangled for sufficiently strong direct mutual coupling between the oscillators. Otherwise, they remain asymptotically in a separable state when the mutual coupling is weak. 
\begin{figure}
\centering
    \scalebox{0.6}{\includegraphics{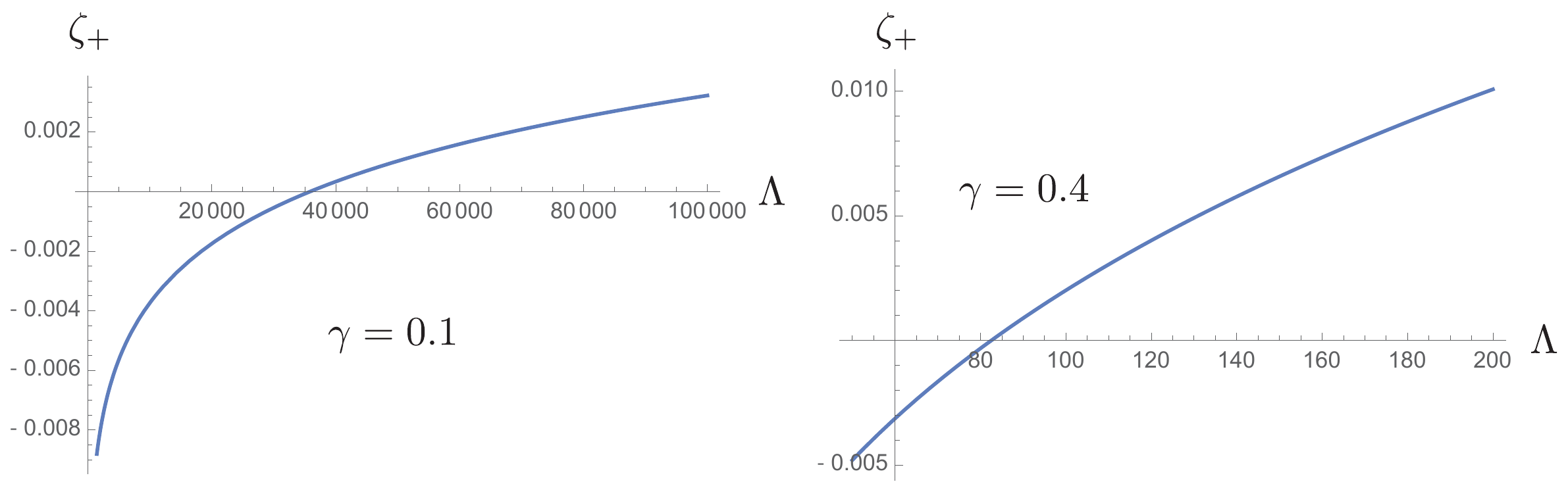}}
    \caption{The separability criterion $\zeta_{+}$ plotted against the cutoff scale $\Lambda$. The oscillating frequency $\omega$ and the inter-oscillator coupling $\sigma$ are chosen to be 5 and 21, respectively}\label{Fi:cutoff}
\end{figure}
Finally, we add some remarks on the cutoff dependence of the separability criterion. From \eqref{E:dfberjh}, we see that the dependence on the cutoff always occurs as long as $\gamma\neq0$. This implies some discretion is needed in the treatment of the cutoff scale. If we ignore this contribution, then one will encounter the following unphysical situation:  If the oscillators are initially in a separable state , prepared as Gaussian wavepackets, their final state is always, at least marginally, entangled even though the mutual interaction is turned off.  In contrast, if the cutoff contribution is accounted for, then the final state of the oscillators will not be entangled unless their mutual interaction is strong enough. Secondly, the cutoff scale always appears in the form  $\ln\Lambda$, so the separability criterion is not very sensitive to the choice of the cutoff scale unless it takes some extreme values. In Fig~\ref{Fi:cutoff}, we let the cutoff scale $\Lambda$ go up to a very high value relative to $\omega$. We see that the separability criterion $\zeta_{+}$ become always positive above a critical value $\Lambda_{c}$, and $\Lambda_{c}$ is highly sensitive to the choice of $\gamma$. Comparing the two plots in Fig~\ref{Fi:cutoff}, we see a mere change in $\gamma$ causes a dramatic shift in the value of $\Lambda_{c}$. Generally speaking, only for very weak oscillator-bath coupling will the cutoff-dependent terms play a subdominant role in the separability criterion.

So far we have presented the general features  in how the separability criterion depends on the interactions.  We now investigate the role of temperature in the criterion.

\begin{figure}
\centering
    \scalebox{0.5}{\includegraphics{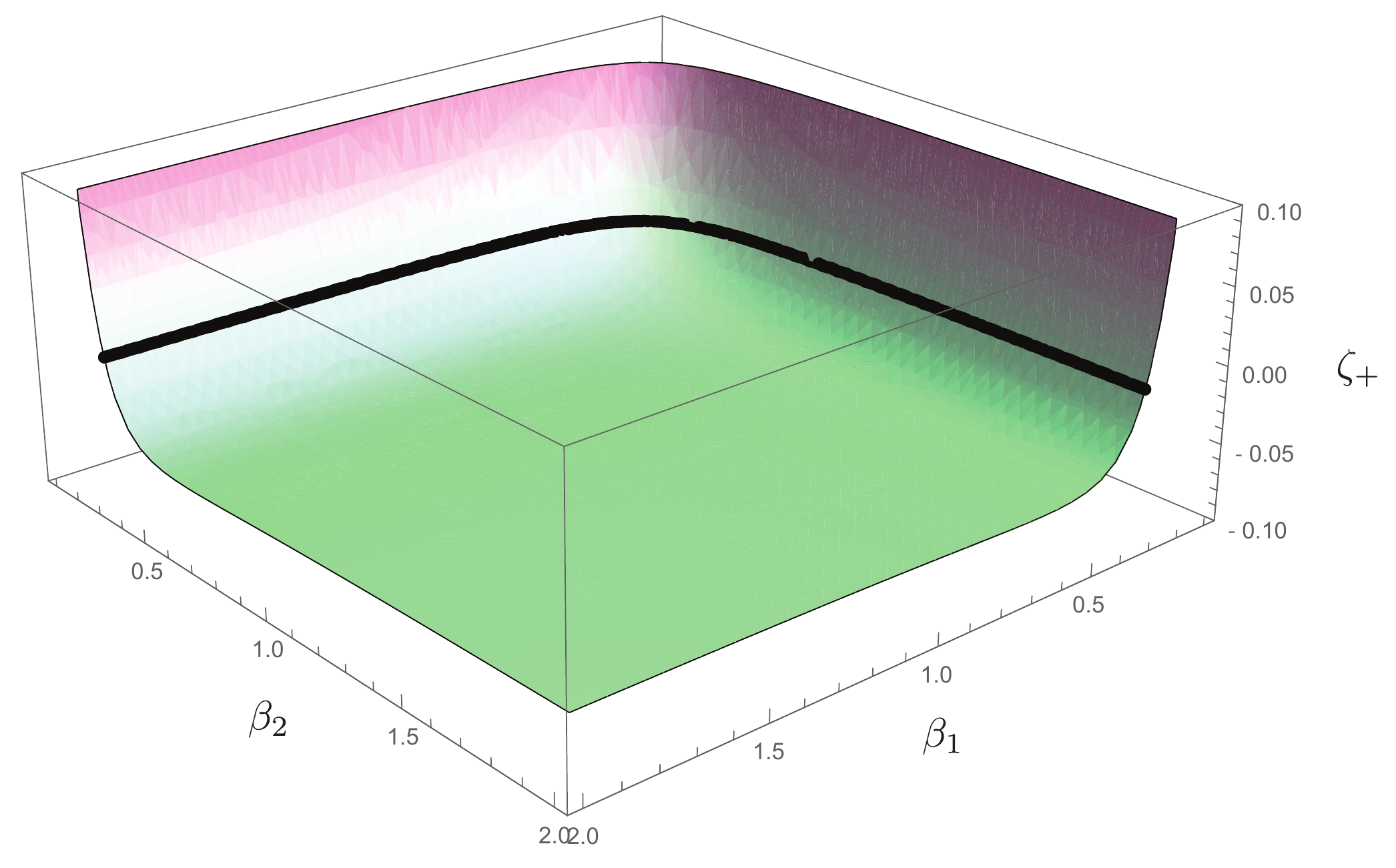}}
    \caption{The separate criterion $\zeta_{+}$ is plotted with respect to the temperatures of two private baths. The black curve $\zeta_{+}=0$ divides the separable state ($\zeta_{+}>0$, pink shade) from the entangled state ($\zeta_{+}<0$, green shade). Essentially the curve traces along the region $\beta\omega=\mathcal{O}(1)$. The oscillating frequency $\omega$ and the cutoff $\Lambda$ are chosen to be 5 and 10000, respectively. The damping constant $\gamma$ is 0.5.}\label{Fi:sepT}
\end{figure}
\subsubsection{low temperature $\beta\omega\gg1$}
Generally speaking, with increased temperature, thermal fluctuations will become increasingly important in affecting the dynamics of the oscillators from their respective baths. Quantum coherence is expected to deteriorate accordingly. We expect similar degradation may occur in entanglement. Thus it is reasonable to conjecture that once the temperatures of the baths are raised above a certain critical value, the degradation can be so severe that the oscillators become completely separable. However, the situation is more complicated for the present setup because two independent thermal baths are involved. It turns out that lowering the temperature of one of the thermal baths does not necessarily guarantee entanglement between the oscillators. Thus the concept of a universal critical temperature is less well-defined in multiple bath situations.

\begin{figure}
\centering
    \scalebox{0.6}{\includegraphics{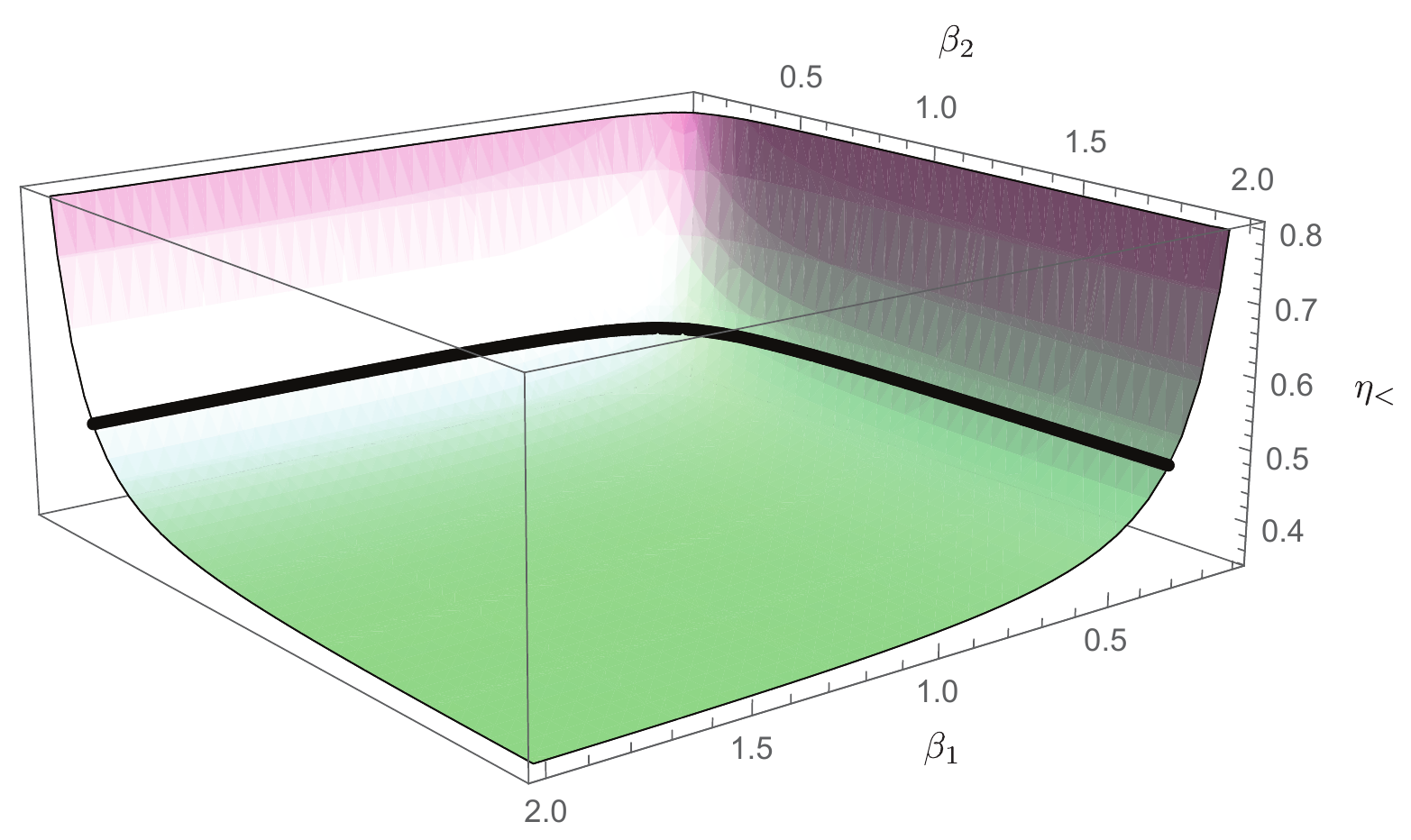}}
    \caption{The symplectic eigenvalue $\eta_{<}$ is plotted with respect to the temperatures of two private baths. The black curve $\eta_{<}=0$ divides the separable state ($\eta_{<}>0$, pink shade) from the entangled state ($\eta_{<}<0$, green shade). Essentially the curve traces along the region $\beta\omega=\mathcal{O}(1)$. This result can be easily mapped to the logarithmic negativity by $E_{\mathcal{N}}(\rho)=\max\bigl\{0,-\ln 2\eta_{<}\bigr\}$. The oscillating frequency $\omega$ and the cutoff $\Lambda$ are chosen to be 5 and 10000, respectively. The damping constant $\gamma$ is 0.1, and the inter-oscillator coupling $\sigma$ is 20.}\label{Fi:etas}
\end{figure}
Here, we discuss the functional dependence of the separability criterion $\zeta_{+}$ on temperatures. To begin with, let us suppose that it takes on a generic form $\zeta_{+}=\zeta(\beta_{1},\beta_{2})$. When a steady state is reached, the separability criterion should be invariant under the exchange of $\beta_{1}$ and $\beta_{2}$ because the configuration of the total system is designed to be symmetric when we swap one oscillator and  its private bath with the  other oscillator and its private bath. This implies that $\zeta_{+}(\beta_{1},\beta_{2})=\zeta_{+}(\beta_{2},\beta_{1})$. However, it is unlikely that the temperature dependence of the separability criterion can be reduced to a function of $\lvert\beta_{1}-\beta_{2}\rvert$ solely. If $\zeta_{+}$ were a function of $\lvert\beta_{1}-\beta_{2}\rvert$, it would imply that the separability criterion could be independent of temperature for the case $\beta_{1}=\beta_{2}$ where it would further suggest that both oscillators should be either separable or entangled for all temperatures. We have shown that at least they can not always be separable because in the zero temperature case, we found that both oscillators can be entangled for certain choices of parameters. On the other hand, it is hard to believe that both oscillators remain entangled even at very high temperature. Thus we rule out the possibility that the separability criterion may depend on $\lvert\beta_{1}-\beta_{2}\rvert$. {The same features are also shared by the symplectic eigenvalue $\eta_{<}$, as can be seen in Fig~\ref{Fi:etas}, but there are two differences:  $\eta_{<}$ is monotonic with respect to the parameters of the joint system and it does not rise up as steeply as the separability criterion in the high temperature regime. The latter is related to the extra factor $(\eta_{>}^{2}-1/4)$ in the criterion. Furthermore we observe that even for $\beta_{1}\neq\beta_{2}$ where the reduced system is described @by?@ asymmetric Gaussian states, the symplectic eigenvalue $\eta_{<}$, thus negativity, still gives a consistent and physical picture with respective to the ordering of the density matrix in terms of the relevant parameters in question. 
}

Thus, to define more precisely a critical temperature $\beta_{c}$, we will look at the special case of $\beta_{1}=\beta_{2}$. In this case both private reservoirs have the same temperature,  yet they are totally uncorrelated. This setup is still distinct from the case that two oscillators share a common bath. In the shared bath case,  the oscillators can influence  each other indirectly through their interaction with the same bath, whereby non-Markovian effects enter in their dynamics,  with dependence on their spatial separation (see, e.g., \cite{LinHu09}) . This type of effects are absent in the private bath configuration; nonetheless, other than the direct influence from its own bath, each oscillator can still experience, by means of direct coupling between the two oscillators, the action of the  other bath associated with the other oscillator. Therefore the equal-temperature private baths  and the single common bath configurations are not the same,  but, as we shall see, there are some similar features. {Moreover, in this special case the two-mode Gaussian state becomes symmetric so the negativity can give an unambiguous comparison of entanglement between states.}

In the low temperature limit, we find the finite temperature correction to the separability criterion is given by
\begin{align}\label{E:droeroe}
	 \zeta_{+}&=\zeta_{+}^{(0)}+\frac{1}{3\beta^2\omega_+^4\omega_-^4}\biggl\{-\frac{\pi\gamma}{4}\,\Bigl[\left(\omega_+-\omega_-\right)^2\left(\omega_++\omega_-\right) \left(\omega_+^2+\omega_-^2\right)\Bigr]\biggr.\\
	 &\qquad\qquad\qquad\quad+\biggl.\gamma^{2}\Bigl[\omega_{+}^{4}\ln\frac{\omega_{+}}{\omega_{-}}+\omega_{+}^{3}\omega_{-}\bigl(\ln\frac{\Lambda}{\omega_{+}}-1\bigr)+\omega_{+}\omega_{-}^{3}\bigl(\ln\frac{\Lambda}{\omega_{-}}-1\bigr)+\omega_{-}^{4}\ln\frac{\omega_{-}}{\omega_{+}}\Bigr]\biggr\}\,,\notag
\end{align}
where $\zeta_{+}^{(0)}$ is the zero-temperature result in \eqref{E:dheura}. It is interesting to note that the correction may change  sign as the inter-oscillator coupling $\sigma$ varies from zero to its upper limit. The upper limit of $\sigma$ is constrained by the condition $\Omega_{-}=\sqrt{\omega_{-}^{2}-\gamma^{2}}=0$, so $\sigma_{\max}\sim\mathcal{O}(\omega^{2})$. When $\sigma=0$, the term linear in $\gamma$ vanishes due to $\omega_{+}=\omega_{-}$ there, but the term quadratic in $\gamma$ is positive. Hence the correction starts off with a positive value. On the other hand in the limit $\omega_{-}\to\gamma$  {(i.e., $\sigma\to\sigma_{\max}$)}, we find the finite temperature correction gradually becomes negative
\begin{align}
	 &\quad\lim_{\omega_{-}\to\gamma}\frac{1}{3\beta^2\omega_+^4\omega_-^4}\biggl\{-\frac{\pi\gamma}{4}\,\Bigl[\left(\omega_+-\omega_-\right)^2\left(\omega_++\omega_-\right) \left(\omega_+^2+\omega_-^2\right)\Bigr]\biggr.\notag\\
	 &\qquad\qquad\qquad\qquad\qquad+\biggl.\gamma^{2}\Bigl[\omega_{+}^{4}\ln\frac{\omega_{+}}{\omega_{-}}+\omega_{+}^{3}\omega_{-}\bigl(\ln\frac{\Lambda}{\omega_{+}}-1\bigr)+\omega_{+}\omega_{-}^{3}\bigl(\ln\frac{\Lambda}{\omega_{-}}-1\bigr)+\omega_{-}^{4}\ln\frac{\omega_{-}}{\omega_{+}}\Bigr]\biggr\}\notag\\
	 &=\frac{\omega_{+}\gamma}{3\beta^2\omega_-^4}\biggl\{-\frac{\pi}{4}+\frac{\gamma}{\omega_{+}}\ln\frac{\omega_{+}}{\gamma}\biggr\}\sim-\frac{\pi\omega_{+}\gamma}{12\beta^2\omega_-^4}<0\,,
\end{align}
where we have used the L'H\^opital's rule to evaluate the limit of such an expression
\begin{equation}
	\lim_{x\to0}x\ln\frac{1}{x}=-\lim_{x\to0}\frac{\ln x}{\dfrac{1}{x}}=-\lim_{x\to0}\frac{\dfrac{1}{x}}{-\dfrac{1}{x^{2}}}=\lim_{x\to0}x=0\,.
\end{equation}
{This feature  reveals the non-monotonicity of the separability criterion at finite temperature. We stress that this errant behavior does not affect us from reading off the critical values of the parameters.}
\begin{figure}
\centering
    \scalebox{0.6}{\includegraphics{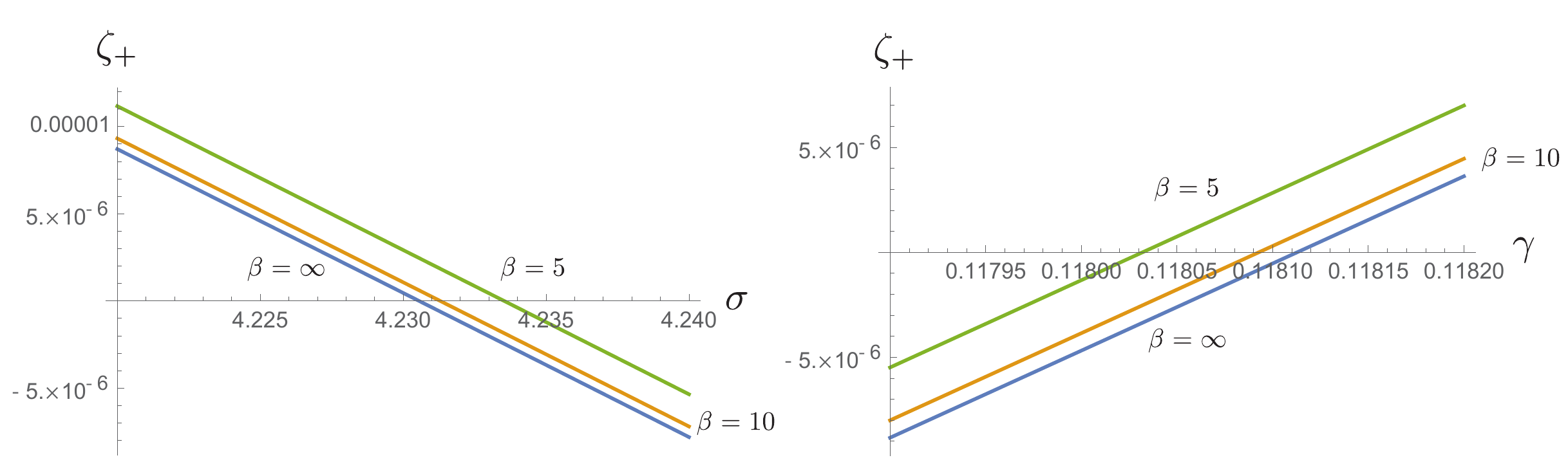}}
    \caption{The separability criterion $\zeta_{+}$ is plotted against the inter-oscillator coupling $\sigma$ and the oscillator-bath coupling $\gamma$ at low temperature. In each plot, we show the $\zeta_{+}$ curve for three different bath temperatures. We see that the critical temperpareue is higher for stronger $\sigma$ but weaker $\gamma$. The oscillating frequency $\omega$ and the cutoff $\Lambda$ are chosen to be 5 and 10000, respectively}\label{Fi:sigmaT}
\end{figure}

Next we look into the effect of finite temperature correction on the critical value of $\sigma_{c}$ where $\zeta_{+}$ transits from a positive to a negative value. In the zero temperature case, we have found this critical value in \eqref{E:dfberjh}, now denoted by $\sigma_{c}^{(0)}$. Generally speaking the finite temperature correction of $\zeta_{+}$ does not necessarily vanish at $\sigma=\sigma_{c}^{(0)}$ as $\zeta_{+}^{(0)}$ does. Instead we find at $\sigma=\sigma_{c}^{(0)}$ the finite temperature correction of $\zeta_{+}$ is
\begin{equation}
	\frac{2\gamma^{2}}{3\beta^{2}\omega^{4}}\Bigl[\ln\frac{\Lambda}{\omega}-1\Bigr]+\cdots\,,
\end{equation}
which is always positive. It means that this correction shifts the curve $\zeta_{+}$ upwards by about the order of magnitude $\mathcal{O}(\gamma^{2})$. It thus implies that the critical values of $\sigma$ will increase because in general $\zeta_{+}$ decreases with $\sigma$, as seen in Fig~\ref{Fi:sigmaT}. In addition, a higher bath temperature results in a larger correction, and in turn causes $\sigma_{c}$ to be even greater. Therefore thermal fluctuations from the baths make entanglement harder to maintain. The higher the bath temperature, the more severely the entanglement will deteriorate. This is consistent with our expectation. However, we may be concerned with a possible loophole related to what we found earlier that the finite temperature correction to $\zeta_{+}$ may change sign with increasing $\sigma$. If it occurred before $\sigma_{c}$, we may encounter the opposite conclusion that the lower bath temperature will instead do more harm to the quantum entanglement in the system.  We will argue that this is not the case. Let $\zeta_{+}^{(\beta)}$ be the low-temperature correction, so that $\zeta_{+}=\zeta_{+}^{(0)}+\zeta_{+}^{(\beta)}$. Since we have shown that when $\zeta_{+}^{(0)}=0$, we have $\zeta_{+}^{(\beta)}>0$, it implies that when $\zeta_{+}=0$, we have $\zeta_{+}^{(0)}<0$ but $\zeta_{+}^{(\beta)}$ remains positive. On the other hand, since $\zeta_{+}^{(\beta)}$ monotonically decreases, $\zeta_{+}^{(\beta)}=0$ will not occur until $\sigma>\sigma_{c}$.  {Thus the separability criterion still offers consistent predictions on the behaviors of the critical parameters.}

\begin{figure}
\centering
    \scalebox{0.5}{\includegraphics{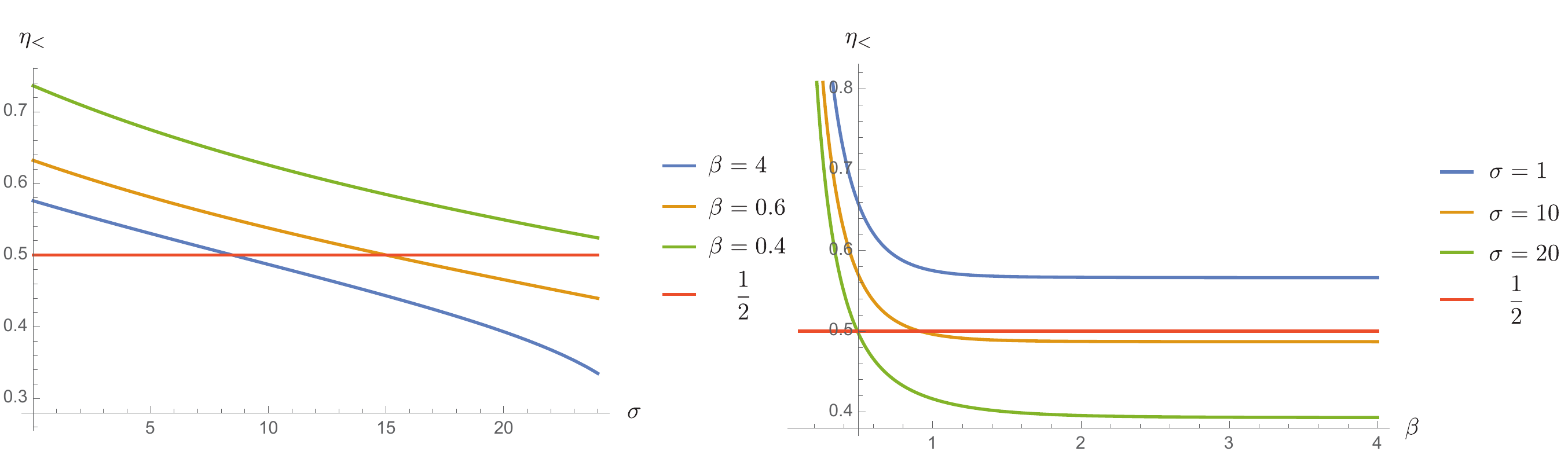}}
    \caption{The trend of $\eta_{<}$ with respect to the inter-oscillator coupling $\sigma$ and the inverse temperature $\beta$ when both private baths have the same temperature $\beta^{-1}$. It can be translated into the logarithmic negativity by $E_{\mathcal{N}}(\rho)=\max\bigl\{0,-\ln 2\eta_{<}\bigr\}$. We also draw a reference line $\eta_{<}=1/2$, the region below which represents the entangled final state of the joint system. In addition, all these curves are monotonic with respect to the parameters in discussion. The oscillating frequency $\omega$ and the cutoff $\Lambda$ are chosen to be 5 and 10000, respectively. The damping constant $\gamma$ is 0.5.}\label{Fi:eta}
\end{figure}


From \eqref{E:droeroe} we can derive a relation among the critical values of $\gamma$, $\sigma$ and $\beta$ for the small $\gamma$ cases. Similar to \eqref{E:beers}, we can show in the low temperature limit that
\begin{align}\label{E:ncuwe}
	 \gamma&=\left[\frac{\pi\sigma}{4\omega}\frac{1}{\ln\frac{\Lambda}{\omega}-1}-\frac{\pi\sigma^{2}}{4\omega^{3}}\frac{1}{\left(\ln\frac{\Lambda}{\omega}-1\right)^{2}}+\cdots\right]\notag\\
	 &\qquad\qquad+\frac{1}{\beta^{2}}\left[-\frac{\pi^{3}\sigma}{12\omega^{3}}\frac{1}{\left(\ln\frac{\Lambda}{\omega}-1\right)^{2}}+\frac{\pi^{3}\sigma^{2}}{6\omega^{5}}\frac{\ln\frac{\Lambda}{\omega}}{\left(\ln\frac{\Lambda}{\omega}-1\right)^{3}}+\cdots\right]+\mathcal{O}(\beta^{-4})\,.
\end{align}
Note that the expression in the second pair of square brackets is negative in the low temperature case. We have argued earlier that at low temperature we don't need a strong inter-oscillator coupling to safeguard quantum entanglement, so the curve $\zeta_{+}$ can vanish for the small values of $\sigma$. Furthermore if $\Lambda\gg\omega$, we have $\ln\dfrac{\Lambda}{\omega}\simeq(\ln\dfrac{\Lambda}{\omega}-1)$. Thus we may safely conclude that
\begin{equation*}
	 \frac{\dfrac{\pi^{3}\sigma^{2}}{6\omega^{5}}\dfrac{\ln\frac{\Lambda}{\omega}}{\left(\ln\frac{\Lambda}{\omega}-1\right)^{3}}}{\dfrac{\pi^{3}\sigma}{12\omega^{3}}\dfrac{1}{\left(\ln\frac{\Lambda}{\omega}-1\right)^{2}}}=\frac{2\sigma}{\omega^{2}}\frac{\ln\frac{\Lambda}{\omega}}{\ln\frac{\Lambda}{\omega}-1}<1\,,
\end{equation*}
for small $\sigma$ and $\Lambda\gg\omega$, and that the second pair of square brackets in \eqref{E:ncuwe} is negative in the low temperature case. Alternatively we may roughly see this based on the arguments that for the expansion to be valid we need $\sigma/\omega^{2}<1$ so the second term should be smaller in magnitude than the first term in the second pair of square brackets. The physical implication of \eqref{E:ncuwe} is that the critical temperature is lowered when the oscillator-bath interaction gets stronger and vice versa.

{Presently we have seen that the low-temperature correction can change sign for sufficiently large mutual coupling; however, this does not affect its usefulness to identify of the existence of entanglement. This unwelcome feature only makes murky the quantitative description of entanglement based on the separability criterion, and it can be traced back to the fact that the separability criterion contains not only $\eta_{<}$ but also $\eta_{>}$, whose existence distorts the information about entanglement, delivered by $\eta_{<}$.  By comparison negativity is freed from this  nonintuitive, unphysical behavior. Let us analyze the finite temperature correction of $\eta_{<}$. In the same configuration, it takes a much simpler form
\begin{align}
	\eta_{<}^{(\beta)}&=\frac{5\pi\beta^{2}\omega_{+}^{\frac{3}{2}}\omega_{-}^{\frac{11}{2}}+2\pi^{3}\omega_{+}^{\frac{9}{2}}\omega_{-}^{\frac{1}{2}}}{15\beta^{4}\omega_{+}^{5}\omega_{-}^{5}}\,\gamma+\mathcal{O}(\gamma^{2})\,,
\end{align}
with $\omega_{\pm}^{2}=\omega^{2}\pm\sigma$. We immediately see that it is always positive and monotonically increasing for all permissible values of $\sigma$. Moreover, the finite temperature correction of $\eta_{<}$ is a monotonically decreasing function of $\beta$. This, unlike the separability criterion, give a plausible and physically intuitive description of the extent the state is entangled. Furthermore, since the analytical expression of $\eta_{<}$ is much simpler than that of the separability criterion, it simplifies the analysis on the critical parameters.   Expand  out $\sigma_{c}=\sigma_{c}^{(0)}+\gamma\,\sigma_{c}^{(1)}+\cdots$  and plug this expression back into the symplectic eigenvalue $\eta_{<}=\eta_{<}^{(0)}+\eta_{<}^{(\beta)}+\cdots=1/2$ where $\sigma_{c}^{(0)}$ satisfies $\eta_{<}^{(0)}=1/2$ at zero temperature. We find
\begin{equation}
	\sigma_{c}^{(1)}=\frac{4\pi\bigl(\omega^2-\sigma_c^{(0)}\bigr)}{3\beta^2_{c}\omega^{2}\bigl(\omega^2+\sigma_c^{(0)}\bigr)^{\frac{1}{2}}}\,,
\end{equation}
which is positive-definite for all permissible ranges of the coupling constant $\sigma$ and temperature $\beta^{-1}$. Note that the dependence on the cutoff scale is hidden in the expression of $\sigma_{c}^{(0)}$. 
}

{Finally we calculate the critical temperature via the criterion $\eta_{<}=1/2$ in the low temperature regime. Expanding $\eta_{<}$ with respect to large $\beta$ gives
\begin{equation}
	\eta_{<}=\eta_{<}^{(0)}\biggl[1+\frac{4\pi\gamma\Omega_{+}}{3(\Omega_{+}^{2}+\gamma^{2})^{2}\,f(\Omega_{+})\,\beta^{2}}+\mathcal{O}(\beta^{-4})\biggr]\,,
\end{equation}
where $f(z)$ is defined in \eqref{E:pondkfer}. Solving $\eta_{<}=1/2$ leads to
\begin{equation}
	\beta_{c}=\left(\frac{8\pi}{3}\right)^{\frac{1}{2}}\frac{(\gamma\Omega_{+}\eta_{<}^{(0)})^{\frac{1}{2}}}{[(1-2\eta_{<}^{(0)})f(\Omega_{+})]^{\frac{1}{2}}(\Omega_{+}^{2}+\gamma^{2})}\,.
\end{equation}
The inverse critical temperature $\beta_{c}$ grows with increasing $\gamma$ but falls off with increasing $\sigma$.}

{Therefore, we can see that} in the low temperature regime the critical temperature $\beta_{c}$ at which $\eta_{<}=1/2$ is higher for stronger inter-oscillator interaction, and for weaker oscillator-bath coupling $\gamma$. This is totally in line with our intuition that the temperature and  the oscillator-bath coupling $\gamma$ will corroborate to disrupt the quantum coherence between the oscillators and make them harder to remain entangled, while the inter-oscillator coupling will tend to enhance the coordination of both oscillators so their entanglement become more robust.

\subsubsection{high temperature $\beta\omega\ll1$}

We now turn our attention to the high temperature regime and ask if entanglement at high temperatures is at all possible.

{From the plot of the symplectic eigenvalue $\eta_{<}$ against the bath temperatures $\beta_{1}$ and $\beta_{2}$ in Fig~\ref{Fi:etas} we see that the surface $\eta_{<}$ forms a very flat basin which is symmetric with respect to $\beta_{1}$ and $\beta_{2}$. The surface $\eta_{<}$ will mildly rise up in the vicinity of $\beta_{1,2}\omega=\mathcal{O}(1)$ when we approach from the low temperature end. Next \sout{we see} it sharply climbs up, crossing the dividing curve $\eta_{<}=1/2$ in the region $\beta_{1,2}\omega=\mathcal{O}(1)$, and enters the high temperature regime.  Thus we can make a first observation that, roughly speaking,  the dividing curve of $\eta_{<}=1/2$ follows $\beta_{1}\omega=\mathcal{O}(1)$ and then turns to $\beta_{2}\omega=\mathcal{O}(1)$. Secondly it implies that separability is determined by the temperature of the warmer bath, instead of the temperature difference, as was mentioned in the previous section. Thirdly, since from earlier discussion we know entanglement tends to survive at higher temperature if the mutual coupling between oscillators is stronger, we use the high temperature approximation to find the critical temperature in the strong $\sigma$ regime. As shown in Fig~\ref{Fi:etaT1}, we compare the numerical calculation of $\eta_{<}$ with its low and high temperature approximations, and see that in the large $\sigma$ case the high-temperature approximation yields a very consistent behavior of $\eta_{<}$ in the vicinity of $\beta\omega\sim\mathcal{O}(1)$, in comparison with the numerical results.
\begin{figure}
\centering
    \scalebox{0.65}{\includegraphics{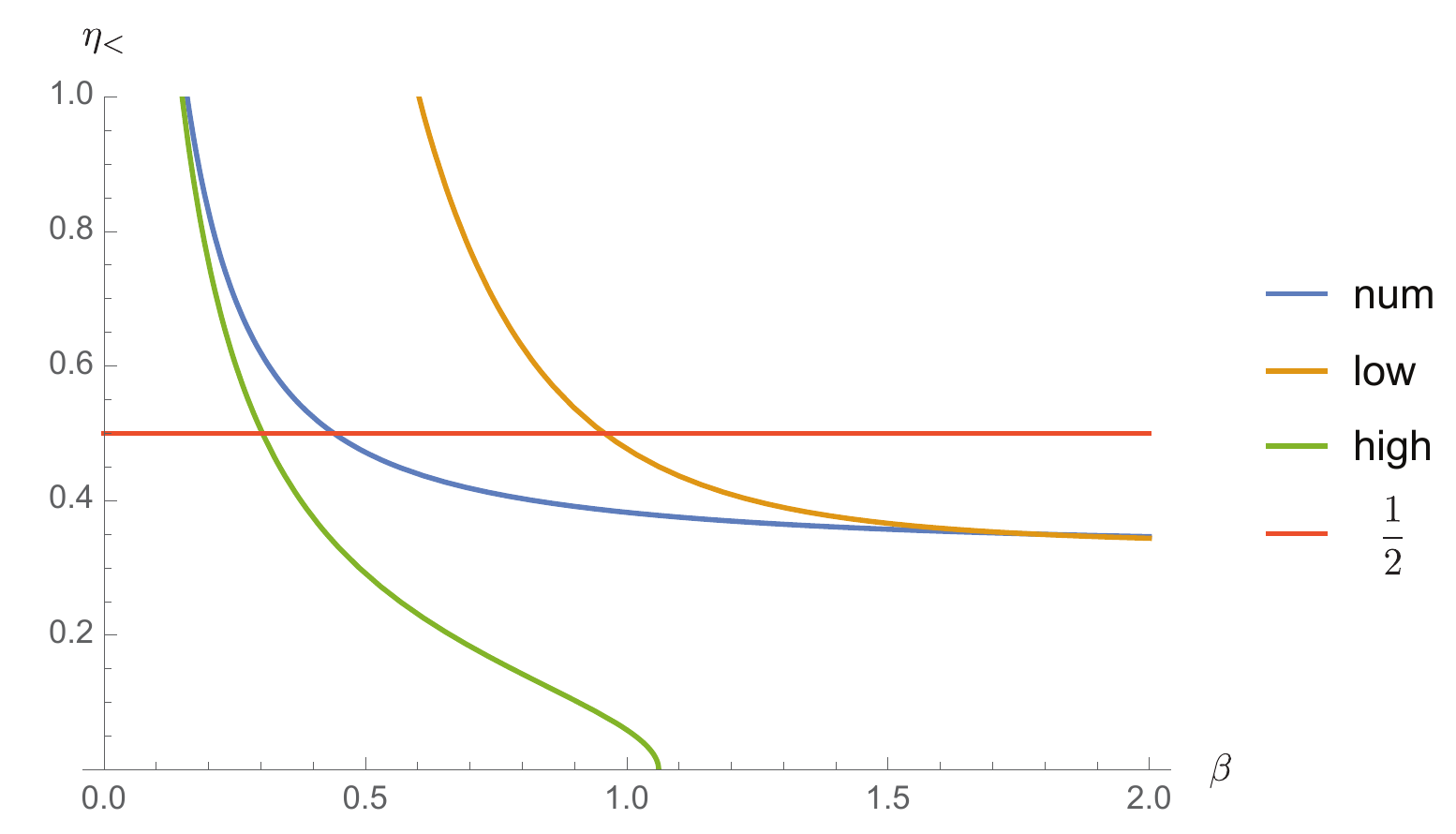}}
    \caption{The symplectic $\eta_{+}$ is plotted against temperature. We show the numerical result and the low-, high-temperature approximations of $\eta_{<}$. For stronger inter-oscillator coupling, the high-temperature approximation yields a very satisfactory result in the region where the transition occurs, in comparison with the numerical calculations. The oscillating frequency $\omega$ and the cutoff $\Lambda$ are chosen to be 5 and 10000, respectively. The damping constant $\gamma$ and the inter-oscillator coupling $\sigma$ are 0.2 and 24 respectively.}\label{Fi:etaT1}
\end{figure}
}

{In the high temperature approximation, the symplectic eigenvalue $\eta_{<}$ is given by
\begin{equation}\label{E:ndkkejrs}
	\eta_{<}\simeq\frac{1}{2\sqrt{3}}\sqrt{\frac{12-\beta^{2}\sigma}{\omega^{2}+\sigma}}+\frac{\gamma\, \ln\beta^2 \Lambda^2}{\pi(12-\beta ^2 \sigma)}\sqrt{\frac{3(12-\beta ^2 \sigma)}{\omega ^2+\sigma}}+\mathcal{O}(\gamma^{2})\,.
\end{equation}
The cutoff-dependent factor in those higher order expressions is less important in the weak $\gamma$ limit because it always appears with the small parameter $\gamma/\omega$. The critical temperature occurs at $\eta_{<}=\frac{1}{2}$. Directly solving a transcendental equation like \eqref{E:ndkkejrs} for $\beta_{c}$ is not possible. Nonetheless since $\ln\beta\Lambda$ always pairs up with $\gamma$, we can use the iteration method to derive $\beta_{c}$. If we first ignore terms of the order $\mathcal{O}(\gamma)$ and higher, we find $\beta_{c}$ is given by $2\sqrt{3}/\sqrt{3\omega^{2}+4\sigma}$.  Substituting it back to seek a correction of order $\mathcal{O}(\gamma)$  we obtain
\begin{equation}
	\beta_{c}=\frac{2\sqrt{3}}{\sqrt{3\omega^{2}+4\sigma}}+\frac{6 \gamma  }{\pi  \left(3 \omega ^2+4 \sigma\right)}\,\ln\frac{12 \Lambda^2}{3 \omega ^2+4 \sigma}+\mathcal{O}(\gamma^{2})\,.
\end{equation}
It is indeed consistent with the statement that $\beta_{c}\omega=\mathcal{O}(1)$, and it rules out the possibility of the existence of entanglement in the regime $\beta\omega\ll1$. Again it reveals the fact that with larger inter-oscillator coupling $\sigma$ we see a higher critical temperature; on the other hand,  stronger oscillator-bath interaction $\gamma$ will cause the critical temperature to decrease.
}

\begin{figure}
\centering
    \scalebox{0.6}{\includegraphics{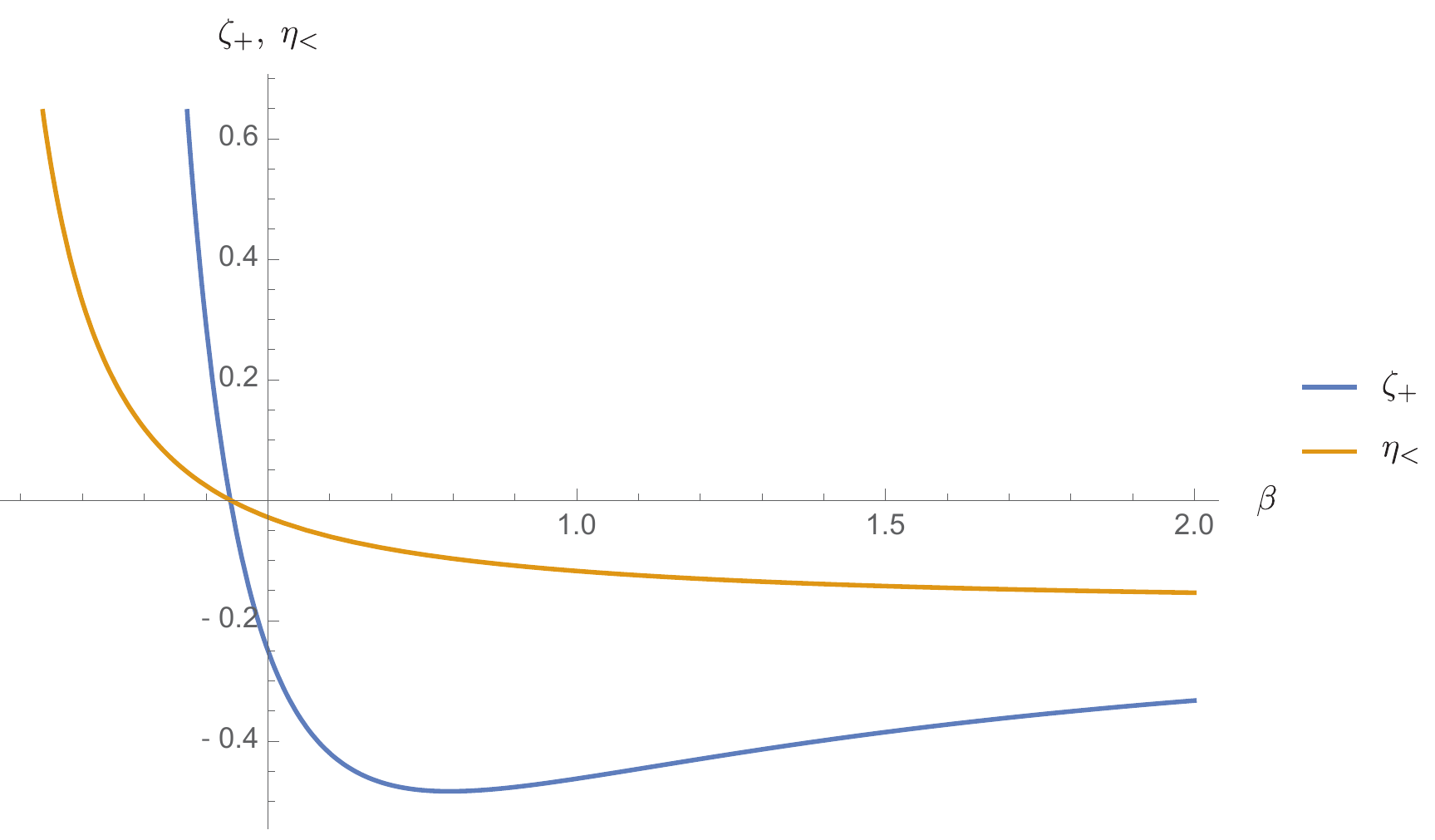}}
    \caption{We plot the separability criterion $\zeta_{+}$ and the symplectic eigenvalue $\eta_{<}$ together. They crisscross at the critical temperature, and therefor give the same information about the existence of entanglement. However, separability criterion falls off and rises up with increasing $\beta$. Note that we shift the values of $\zeta_{<}$ downward by $1/2$, That is, what we plot in fact is $\eta_{<}-1/2$. The oscillating frequency $\omega$, the cutoff $\Lambda$, the damping constant $\gamma$ and the inter-oscillator coupling $\sigma$ are chosen to be 5, 10000, 0.2 and 24 respectively.}\label{Fi:etasep}
\end{figure}

{The same results can be found if we investigate the high temperature approximation of the separability criterion. This is no surprise since we have previously discussed that the separability criterion is perfectly valid for identification of entanglement except for a quantitative measure of entanglement. For example,  as shown in Fig.~\ref{Fi:etasep}, the separability criterion $\zeta_{+}$ and the symplectic eigenvalue $\eta_{<}$ give the same prediction about the location of the critical temperature, but the separability criterion is not a monotonic function of the temperature, which  makes it inappropriate as an entanglement measure.
}

{With temperature measured in ratio to the oscillator's natural frequency $\beta \omega$ we can conclude that \textit{there is no high temperature entanglement in Case C1}, namely, between two oscillators each interacting with its own bath.}\\\\


\section{Intuitive Understanding of Entanglement Behavior}

So far we have taken quite some labor to show that asymptotic entanglement between oscillators are easier to sustain for stronger inter-oscillator coupling but weaker oscillator-bath interaction. 

{Here} we would like to offer a physically more transparent illustration as to the competing roles between these two kinds of interactions in terms of normal modes of the oscillator. The Langevin equations \eqref{E:derhs1} and \eqref{E:derhs2} can be easily decoupled by forming two normal modes
\begin{align}
	\chi_{+}&=\frac{\chi_{1}+\chi_{2}}{2}\,,&\chi_{-}=\chi_{1}-\chi_{2}\,,
\end{align}
and the corresponding equations of motion are
\begin{align}
	\ddot{\chi}_{+}+2\gamma\,\dot{\chi}_{+}+\omega_{+}^{2}\,\chi_{+}&=\frac{1}{2m}\bigl(\xi_{1}+\xi_{2}\bigr)\,,\\
	\ddot{\chi}_{-}+2\gamma\,\dot{\chi}_{-}+\omega_{-}^{2}\,\chi_{-}&=\frac{1}{m}\bigl(\xi_{1}-\xi_{2}\bigr)\,.
\end{align}
Since we are interested in the late-time dynamics, we will not write down the homogeneous solutions to the Langevin equations. Following the earlier discussions we find that the inhomogeneous solutions are given by
\begin{align}
	\chi_{+}(t)&=\frac{1}{2m}\int^{t}_{0}\!ds\;d_{2}^{(+)}(t-s)\Bigl[\xi_{1}(s)+\xi_{2}(s)\Bigr]\,,\\
	\chi_{-}(t)&=\frac{1}{m}\int^{t}_{0}\!ds\;d_{2}^{(-)}(t-s)\Bigl[\xi_{1}(s)-\xi_{2}(s)\Bigr]\,,
\end{align}
with
\begin{align}
	 d_{2}^{(\pm)}(\tau)&=\frac{\gamma}{\Omega_{\pm}}\,e^{-\gamma\tau}\sin\Omega_{\pm}\tau\,,&\Omega_{\pm}^{2}&=\omega_{\pm}^{2}-\gamma^{2}\,.
\end{align}
The frequency $\Omega_{\pm}$ is the resonance frequency of the normal modes $\chi_{\pm}$. Hence the stronger inter-oscillator coupling $\sigma$ implies smaller values of $\omega_{-}$ but larger values of $\omega_{+}$, and in turn smaller $\Omega_{-}$ and larger $\Omega_{+}$. Since the amplitude of the normal modes $\chi_{\pm}$ is related to the ratio $\gamma/\Omega_{\pm}$,  stronger inter-oscillator interaction will {induce a larger amplitude of the mode $\chi_{-}$, which will grow with increasing values of $\sigma$, meanwhile it causes the mode $\chi_{+}$ to oscillate subdominantly and its amplitude decreases with $\sigma$. This is intuitively plausible since, e.g., for a very soft spring, or for a particle in a very shallow harmonic potential, a small disturbance could easily induce a large displacement in its motion. Thus in these circumstances it tends to have a large position uncertainty. Furthermore, the consequence from the normal-mode dynamics} hints at the fact that when we form the displacements of two oscillators by superposing the normal modes
\begin{align}\label{E:ncvdhj}
	\chi_{1}&=\chi_{+}+\frac{1}{2}\,\chi_{-}\,,&\chi_{2}&=\chi_{+}-\frac{1}{2}\,\chi_{-}\,,
\end{align}
the mode $\chi_{+}$ can be overshadowed by $\chi_{-}$. The original displacements $\chi_{1}$, $\chi_{2}$ are more or less determined solely by the mode $\chi_{-}$, in particular in the strong mutual coupling limit $\Omega_{-}\to0$. Furthermore, in this limit, $\chi_{1}$ and $\chi_{2}$ will be out of phase by $\pi$.  Likewise, following similar arguments and taking care of contributions from the resonance, we see that in the case of the conjugate momentum $p_{1}$, $p_{2}$ of the two oscillators, the contribution of $p_{+}$ can dominate over that of $p_{-}$ for strong mutual coupling between the oscillators. 
Furthermore, the phase difference between $\chi_{1}$ and $\chi_{2}$ is reflected by the fact that in this limit we should have $\mathcal{V}_{11}\sim-\mathcal{V}_{13}$. It is particularly easy to see this for the special case $\beta_{1}=\beta_{2}$. The formal late-time expressions of $\mathcal{V}_{11}$ and $\mathcal{V}_{13}$ in this case are
\begin{align*}
	\mathcal{V}_{11}&=\frac{e^{2}}{2m^{2}}\int_{-\infty}^{\infty}\!\frac{d\kappa}{2\pi}\;\Bigl\{\lvert\widetilde{d}_{2}^{(+)}(\kappa)\rvert^{2}+\lvert\widetilde{d}_{2}^{(-)}(\kappa)\rvert^{2}\Bigr\}\,\widetilde{G}_{H}(\kappa)\simeq+\frac{e^{2}}{2m^{2}}\int_{-\infty}^{\infty}\!\frac{d\kappa}{2\pi}\;\lvert\widetilde{d}_{2}^{(-)}(\kappa)\rvert^{2}\,\widetilde{G}_{H}(\kappa)\,,\\
	\mathcal{V}_{13}&=\frac{e^{2}}{2m^{2}}\int_{-\infty}^{\infty}\!\frac{d\kappa}{2\pi}\;\Bigl\{\lvert\widetilde{d}_{2}^{(+)}(\kappa)\rvert^{2}-\lvert\widetilde{d}_{2}^{(-)}(\kappa)\rvert^{2}\Bigr\}\,\widetilde{G}_{H}(\kappa)\simeq-\frac{e^{2}}{2m^{2}}\int_{-\infty}^{\infty}\!\frac{d\kappa}{2\pi}\;\lvert\widetilde{d}_{2}^{(-)}(\kappa)\rvert^{2}\,\widetilde{G}_{H}(\kappa)\,,
\end{align*}
in the limit $\sigma\to\omega^{2}$ where $\lvert\widetilde{d}_{2}^{(-)}(\kappa)\rvert^{2}>\lvert\widetilde{d}_{2}^{(+)}(\kappa)\rvert^{2}$. In addition Eqs.~\eqref{E:fkdfb1} and~\eqref{E:fkdfb3} also explicitly demonstrate the same relation.  Similarly $p_{+}$ is the dominant mode in the conjugate momenta $p_{1}$, $p_{2}$, so we may expect $\mathcal{V}_{22}\sim+\mathcal{V}_{24}$ and this is clear from
\begin{align*}
	\mathcal{V}_{22}&=\frac{e^{2}}{2}\int_{-\infty}^{\infty}\!\frac{d\kappa}{2\pi}\;\kappa^{2}\Bigl\{\lvert\widetilde{d}_{2}^{(+)}(\kappa)\rvert^{2}+\lvert\widetilde{d}_{2}^{(-)}(\kappa)\rvert^{2}\Bigr\}\,\widetilde{G}_{H}(\kappa)\simeq\frac{e^{2}}{2}\int_{-\infty}^{\infty}\!\frac{d\kappa}{2\pi}\;\kappa^{2}\lvert\widetilde{d}_{2}^{(+)}(\kappa)\rvert^{2}\,\widetilde{G}_{H}(\kappa)\,,\\
	\mathcal{V}_{24}&=\frac{e^{2}}{2}\int_{-\infty}^{\infty}\!\frac{d\kappa}{2\pi}\;\kappa^{2}\Bigl\{\lvert\widetilde{d}_{2}^{(+)}(\kappa)\rvert^{2}-\lvert\widetilde{d}_{2}^{(-)}(\kappa)\rvert^{2}\Bigr\}\,\widetilde{G}_{H}(\kappa)\simeq\frac{e^{2}}{2}\int_{-\infty}^{\infty}\!\frac{d\kappa}{2\pi}\;\kappa^{2}\lvert\widetilde{d}_{2}^{(+)}(\kappa)\rvert^{2}\,\widetilde{G}_{H}(\kappa)\,,
\end{align*}
for the case $\beta_{1}=\beta_{2}$.

{Now let us take a look at the formal expression of the symplectic eigenvalue $\eta_{<}$. From \eqref{E:dbekrheq}, we see
\begin{align}
	\eta_{<}^{2}=\langle\bigl\{\chi_{1},\chi_{+}\bigr\}\rangle\langle\bigl\{p_{1},p_{-}\bigr\}\rangle=\frac{1}{4}\langle\bigl\{\chi_{+},\chi_{+}\bigr\}\rangle\langle\bigl\{p_{-},p_{-}\bigr\}\rangle\,,
\end{align}
due to the fact that  there is no cross-correlation between two normal modes in the case $\beta_{1}=\beta_{2}$. It is clearly seen that $\eta_{<}$ is composed of subdominant contributions only, which all have smaller uncertainty with larger $\sigma$. Moreover they decrease with increasing values of the inter-oscillator coupling $\sigma$. This makes possible that the symplectic eigenvalue $\eta_{<}$ can fall off with strong inter-oscillator coupling, thus that entanglement can be sustained at higher temperature.}

In summary we show that by  analyzing the behaviors of the normal mode frequencies with respect to various couplings and parameters of interest, we may get an intuitive understanding of the general features of the entanglement between two oscillators in relation to these parameters.

\section{Summary of Results and Comparisons}

\subsection{Summary of Results for Entanglement in Systems in NESS}

Having shown the quantitative details in the last section we now provide a summary of the qualitative features of entanglement dynamics in the case (\textbf{Case C1}) studied here for quantum systems in NESS. For two bilinearly coupled oscillators each interacting with its own bath, we find:
\begin{enumerate}
	\item Quantum entanglement in systems of this setup is harder to sustain at finite temperatures. Thermal fluctuations from the baths disrrupt the coherence between the oscillators.
	\item {Both the separability criterion and the negativity are perfectly good indicators for the existence of entanglement. However, the former is not necessarily a monotonic function of the parameters in the configuration, so it does not qualify as an entanglement measure. It cannot give a consistent, quantitative comparison between different entangled configurations.}
	\item The entanglement criterion $\zeta_{+}$  {or the symplectic eigenvalue $\eta_{<}$} in general is not a function of the bath temperature difference, in contrast to thermal transport in the same setting \cite{HHNESS}.
		\begin{itemize}
			\item Lowering the temperature of one of the thermal baths does not necessarily help to keep the entanglement between the oscillators.
			\item The notion of a critical temperature, where $\zeta_{+}=0$ or {$\eta_{<}=1/2$}, is better defined when two private baths have the same temperature.
		\end{itemize}
	\item Entanglement between the two oscillators is reduced for stronger oscillator-bath interaction, but enhanced for larger inter-oscillator coupling. They play competing roles as far as their influence on entanglement is concerned.
		\begin{itemize}
			\item strong inter-oscillator coupling better links the dynamics of the two oscillators, and thus improves the coherence between them.
			\item uncorrelated environment fluctuations corrupts the correlations between the oscillators;  stronger oscillator-bath interaction will compound this effect.
		\end{itemize}
	\item For weak oscillator-bath coupling the critical temperature satisfies $\beta_{c}\omega\sim2\bigl(1+4\sigma/3\omega^{2}\bigr)^{-1}$.  This supports the rough estimate condition $\beta_{c}\omega\sim\mathcal{O}(1)$.
		\begin{itemize}
			\item For strong oscillator-bath coupling the critical temperature depends on the damping constant $\gamma$ and the environment cutoff frequency $\Lambda$.
			\item The effect of environment cutoff cannot be ignored in the low temperature and the strong oscillator-bath coupling cases.
		\end{itemize}
	\item Asymptotic quantum entanglement disappears in the high temperature regime $\beta\omega\ll1$. There is no hot entanglement in  systems (with bilinear constant coupling) under NESS conditions.
\end{enumerate}

\subsection{Comparison:  System in a Private Bath vs in a Common Bath}

It is useful to make a comparison of the case studied here (Case C1) with what we have studied in Paper I (\textbf{Case B}), namely,  a system of two bilinearly coupled oscillators interacting with one common bath.

\paragraph {Case B:  common bath.}   In Paper I we have studied the case of two coupled oscillators at a finite spatial separation, both interacting with a common thermal field bath, which is a finite temperature generalization of the work  ~\cite{LinHu09,Goan}.  For comparison with Case C1 studied here we only need to consider the limiting case when the two oscillators are placed next to each other \footnote{See discussions in \cite{LinHu09} in how close the two oscillators can be placed to avoid possible singular retardation effect and non-Markovian behavior.}, thus ignoring the spatial variation of entanglement.  The action for this setup takes the form
\begin{align}
	 S&=\int\!dt\;\left\{\sum_{i=1}^{2}\left[\frac{m}{2}\,\dot{\chi}_{i}^{2}(t)-\frac{m\omega^{2}}{2}\,\chi_{i}^{2}(t)\right]-m\sigma\,\chi_{1}(t)\chi_{2}(t)\right\}+\sum_{i=1}^{2}e\int\!d^{4}x\;\chi_{i}(t)\delta(x-z(t))\,\phi(x)\notag\\
	&\qquad\qquad\qquad\qquad\qquad\qquad+\int\!d^{4}x\;\frac{1}{2}\Bigl[\partial\phi(x)\Bigr]^{2}\,.
\end{align}
Since the two oscillators share the same bath, we can decompose the two harmonic oscillator variables into the fast mode (or center of mass)  and the slow mode (or relative coordinate)  variables, $\chi_+ = \ha (\chi_1+\chi_2 ),  \;\; \chi_-=\chi_1 - \chi_2 $  whence the action becomes
\begin{align}
	 S&=\int\!dt\;\frac{1}{2}\biggl[\frac{m}{2}\,\dot{\chi}_{-}^{2}-\frac{m\omega_{-}^{2}}{2}\,\chi_{-}^{2}\biggr]+2\int\!dt\;\biggl[\frac{m}{2}\,\dot{\chi}_{+}^{2}-\frac{m\omega_{+}^{2}}{2}\,\chi_{+}^{2}\biggr]\notag\\
	 &\qquad\qquad\qquad\qquad\qquad\qquad+2e\int\!d^{4}x\;\chi_{+}(t)\delta(x-z(t))\,\phi(x)+\int\!d^{4}x\;\frac{1}{2}\Bigl[\partial\phi(x)\Bigr]^{2}\,,
\end{align}
where $\omega_{\pm}^{2}=\omega^{2}\pm\sigma$.

We see that  the slow mode $\chi_{-}$ is decoupled from the bath, while the fast mode $\chi_{+}$  now interacts with an effective bath, described by the same scalar field but with the reduced amplitude, $\phi/\sqrt{2}$ and with an effective coupling strength enhanced to $\sqrt{2}\,e$. The Langevin equations for the fast and slow  variables are
\begin{align}
	\ddot{\chi}_{+}+2\gamma\,\dot{\chi_{+}}+\omega_{+}^{2}&=\frac{1}{m}\,\xi\,,\label{E:ndekwi1}\\
	\ddot{\chi}_{-}+\omega_{-}^{2}&=0\,.\label{E:ndekwi2}
\end{align}
Note there is a subtle issue in this pair of Langevin equations. Although we still write the oscillating frequencies of the fast and the slow mode as $\omega_{\pm}$, now they have quite different physical contents. We observe that the fast mode couples with the bath, so its oscillating frequency $\omega_{+}$ will acquire a correction due to its interaction with the bath. This correction is absent for the slow mode. Nonetheless because the correction depends on the environment cutoff and it is of the order $\delta\omega^{2}/\omega^{2}\sim\mathcal{O}(\gamma\Lambda/\omega^{2})$, we expect that in the weak oscillator-bath coupling limit, this correction is moderate. In fact it has been shown in~\cite{LinHu09} that the oscillator-bath coupling constant  should be reasonably small or else in may induce instability due to the Coulomb-like interaction at short distances.

The stochastic force $\xi$ in this case still possesses the statistical properties
\begin{align}
	\langle\xi(t)\rangle&=0\,,&\langle\xi(t)\xi(t')\rangle&=e^{2}\,\,G_{H}(t,t')\,.
\end{align}
The solutions to \eqref{E:ndekwi1} and \eqref{E:ndekwi2} are
\begin{align}
	 \chi_{+}(t)&=\mathfrak{d}_{1}^{(+)}(t)\,\chi_{+}(0)+\mathfrak{d}_{2}^{(+)}(t)\,\dot{\chi}_{+}(0)+\frac{1}{m}\int_{0}^{t}\!ds\;\mathfrak{d}_{2}^{(+)}(t-s)\xi(s)\,,\\
	\chi_{-}(t)&=\mathfrak{d}_{1}^{(-)}(t)\,\chi_{-}(0)+\mathfrak{d}_{2}^{(-)}(t)\,\dot{\chi}_{-}(0)\,,
\end{align}
with
\begin{align}
	\mathfrak{d}_{1}^{(+)}(t)&=e^{-\gamma t}\left[\cos W_{+}t+\frac{\gamma}{W_{+}}\,\sin W_{+}t\right]\,,&\mathfrak{d}_{2}^{(+)}(t)&=\frac{1}{W_{+}}\,e^{-\gamma t}\sin W_{+}t\,,&W_{+}^{2}&=\omega_{+}^{2}-\gamma^{2}\,.\\
	\mathfrak{d}_{1}^{(-)}(t)&=\cos W_{-}t\,,&\mathfrak{d}_{2}^{(-)}(t)&=\frac{1}{W_{-}}\,\sin W_{-}t\,,&W_{-}^{2}&=\omega_{-}^{2}\,.
\end{align}
Here we note that the slow mode in the common bath case is non-decaying, so some of the initial information of the system can be kept to the very end of the evolution. This is in strong contrast with the private bath case studied here. The corresponding component of the fast mode decays with time. Thus at late times the fast mode only responds to the environment.

The original oscillator variables thus evolve according to
\begin{align}
	 \chi_{1}(t)&\simeq+\frac{1}{2}\left[\mathfrak{d}_{1}^{(-)}(t)\,\chi_{-}(0)+\mathfrak{d}_{2}^{(-)}(t)\,\dot{\chi}_{-}(0)\right]+\frac{1}{m}\int_{0}^{t}\!ds\;\mathfrak{d}_{2}^{(+)}(t-s)\xi(s)\,,\\
	 \chi_{2}(t)&\simeq-\frac{1}{2}\left[\mathfrak{d}_{1}^{(-)}(t)\,\chi_{-}(0)+\mathfrak{d}_{2}^{(-)}(t)\,\dot{\chi}_{-}(0)\right]+\frac{1}{m}\int_{0}^{t}\!ds\;\mathfrak{d}_{2}^{(+)}(t-s)\xi(s)\,,
\end{align}
at late time $t\gg\gamma^{-1}$. We find the elements of the covariance matrix in this case are given by
\begin{align}
	\langle\chi_{1}^{2}(t)\rangle &=+\lambda\,\frac{1}{4}\left[\mathfrak{d}_{1}^{(-)\,2}(t)\,\langle\chi_{-}^{2}(0)\rangle+\mathfrak{d}_{2}^{(-)\,2}(t)\,\langle\dot{\chi}_{-}^{2}(0)\rangle\right]\notag\\
	 &\qquad\qquad\qquad\qquad\quad+\frac{e^{2}}{m^{2}}\int_{0}^{t}\!ds\,ds'\;\mathfrak{d}_{2}^{(+)}(s)\mathfrak{d}_{2}^{(+)}(s')G_{H}(s-s')\,,\\
	\langle\chi_{2}^{2}(t)\rangle &=\langle\chi_{1}^{2}(t)\rangle\,,\\
	 \frac{1}{2}\langle\{\chi_{1}(t),\chi_{2}(t)\}\rangle&=-\lambda\,\frac{1}{4}\left[\mathfrak{d}_{1}^{(-)\,2}(t)\,\langle\chi_{-}^{2}(0)\rangle+\mathfrak{d}_{2}^{(-)\,2}(t)\,\langle\dot{\chi}_{-}^{2}(0)\rangle\right]\notag\\
	 &\qquad\qquad\qquad\qquad\quad+\frac{e^{2}}{m^{2}}\int_{0}^{t}\!ds\,ds'\;\mathfrak{d}_{2}^{(+)}(s)\mathfrak{d}_{2}^{(+)}(s')G_{H}(s-s')\,,\label{E:neriwns}\\
	\langle p_{1}^{2}(t)\rangle &=+\lambda\,\frac{m^{2}}{4}\left[\dot{\mathfrak{d}}_{1}^{(-)\,2}(t)\,\langle\chi_{-}^{2}(0)\rangle+\dot{\mathfrak{d}}_{2}^{(-)\,2}(t)\,\langle\dot{\chi}_{-}^{2}(0)\rangle\right]\notag\\
	& \qquad\qquad\qquad\qquad\quad+e^{2}\int_{0}^{t}\!ds\,ds'\;\dot{\mathfrak{d}}_{2}^{(+)}(s)\dot{\mathfrak{d}}_{2}^{(+)}(s')G_{H}(s-s')\,,\\
	\langle p_{2}^{2}(t)\rangle &=\langle p_{1}^{2}(t)\rangle\,,\\
	 \frac{1}{2}\langle\{p_{1}(t),p_{2}(t)\}\rangle&=-\lambda\,\frac{m^{2}}{4}\left[\dot{\mathfrak{d}}_{1}^{(-)\,2}(t)\,\langle\chi_{-}^{2}(0)\rangle+\dot{\mathfrak{d}}_{2}^{(-)\,2}(t)\,\langle\dot{\chi}_{-}^{2}(0)\rangle\right]\notag\\
	 &\qquad\qquad\qquad\qquad\quad+e^{2}\int_{0}^{t}\!ds\,ds'\;\dot{\mathfrak{d}}_{2}^{(+)}(s)\dot{\mathfrak{d}}_{2}^{(+)}(s')G_{H}(s-s')\,.
\end{align}
and
\begin{align}
	 \frac{1}{2}\langle\{\chi_{1}(t),p_{1}(t)\}\rangle&=+\lambda\,\frac{m}{4}\left[\mathfrak{d}_{1}^{(-)}(t)\dot{\mathfrak{d}}^{(-)}_{1}(t)\,\langle\chi_{-}^{2}(0)\rangle+\mathfrak{d}_{2}^{(-)\,2}(t)\dot{\mathfrak{d}}^{(-)}_{2}(t)\,\langle\dot{\chi}_{-}^{2}(0)\rangle\right]\,,\\
	 \frac{1}{2}\langle\{\chi_{1}(t),p_{2}(t)\}\rangle&=-\lambda\,\frac{m}{4}\left[\mathfrak{d}_{1}^{(-)}(t)\dot{\mathfrak{d}}^{(-)}_{1}(t)\,\langle\chi_{-}^{2}(0)\rangle+\mathfrak{d}_{2}^{(-)\,2}(t)\dot{\mathfrak{d}}^{(-)}_{2}(t)\,\langle\dot{\chi}_{-}^{2}(0)\rangle\right]\,,\\
	\frac{1}{2}\langle\{\chi_{2}(t),p_{2}(t)\}\rangle&=\frac{1}{2}\langle\{\chi_{1}(t),p_{1}(t)\}\rangle\,,\\
	\frac{1}{2}\langle\{\chi_{2}(t),p_{1}(t)\}\rangle&=\frac{1}{2}\langle\{\chi_{1}(t),p_{2}(t)\}\rangle\,.
\end{align}
For the last four elements, the term caused by the environment vanishes in the limit $t\to\infty$. The parameter $\lambda$ is used as a marker for the intrinsic components, and can be set to unity  with no consequence.  This is in contrast to the components induced by the environment which have  $e^{2}$ dependence.

One feature  in the common bath case noticeably different from the private bath case is that the elements of the covariance matrix still contain the information about the initial conditions even though the system has evolved to late time. This is a consequence of the fact that one of the normal modes is completely decoupled from the bath such that part of the initial information is retained in the system and is not lost into the environment. On the contrary, for the private bath case, both the slow and the fast modes are coupled to the environment, and it causes the dispersion of the initial information into the environment. We also note that the components induced by the environment are typically smaller by an order $\mathcal{O}(\gamma)$ than the components intrinsic to the quantum evolution of the oscillators.

Moreover, we have shown that in the private bath case, stronger inter-oscillator coupling renders the oscillating frequency of the slow mode  smaller than that of the fast mode. It implies that when the interaction between the oscillators are comparable with the original oscillating frequency  $\omega$ the slow mode will dominate over the fast mode. From \eqref{E:ncvdhj}, we see that the late-time correlation between $\chi_{1}$ and $\chi_{2}$ is prone to be negative, meaning that $\chi_{1}$ tends to be \textit{anti-correlated} with $\chi_{2}$.  This is not the case for the common bath case. When both oscillators share a common bath, only the fast mode couples with the bath. The coupling between the oscillators plays a minor role because $W_{+}$ is always of the order $\omega$. At late time $t\gg\gamma^{-1}$, we see that both $\chi_{1}$ and $\chi_{2}$ are more or less led by the fast mode, apart from the intrinsic quantum evolution of the system inherited in the slow mode. Hence the bath tends to drive two neighboring oscillators into correlation, meaning that the correlation between $\chi_{1}$ and $\chi_{2}$ \textit{induced by the shared bath} tends to be positive.

At this point, some discretion is advised. First, we observe from \eqref{E:neriwns} that the correlation caused by the intrinsic quantum evolution of the system carries a negative sign, and will counteract with the correlation induced by the environment, so the total correlation between $\chi_{1}$ and $\chi_{2}$ may not always be positive definite at late time. This will also introduce additional complexity in the entanglement for the shared bath case. 
Secondly, unlike the private bath case where the elements of the covariance matrix approach a time-independent constant when the NESS is reached, the corresponding elements in the shared bath case remain oscillatory in time reflective of the intrinsic quantum dynamics of the system.

We end with a summary of the qualitative behavior of entanglement in a system of two coupled oscillators comparing between the two cases: in one case this system interacts with a common bath, and in the other case, with their own private baths, but kept at the same temperature.
\begin{enumerate}
	\item From the structure of the normal modes, we find that
		\begin{itemize}
			\item private bath: both degrees of freedom are coupled to the bath, so they behave like a pair of damped driven oscillators with different oscillating frequency.
			\item common bath: only the  fast mode is coupled to the bath,  the slow mode is totally decoupled from the bath and it acts like a free oscillator.
		\end{itemize}
	\item If we separate the elements of the covariance matrix into the intrinsic and the induced components; the former depends on the initial conditions of the oscillators but the latter is entirely driven by the environment, independent of the initial conditions of the oscillators. We see that
		\begin{itemize}
			\item for the mode coupled to the bath, the intrinsic component will be exponentially small at late time, and only the induced component survives.
			\item for the free mode, the intrinsic component oscillates all the time and there is no induced component.
		\end{itemize}
	\item This implies that the initial Gaussian conditions will be washed out for the mode coupled to the bath, but they will survive at late times for the uncoupled mode.
		\begin{itemize}
			\item private bath: the initial Gaussian conditions will be irrelevant to the asymptotic entanglement,
			\item common bath: they remain significant, so the final state of entanglement depends on the choice of the initial conditions.
		\end{itemize}
	\item At late times the entanglement measure for the private bath case is time-independent, but for the common bath it continues oscillating in time.
	\item The amplitude of the driven mode is related to the mode frequency. The smaller the frequency is, the larger the driven amplitude will be.
		\begin{itemize}
			\item private bath: slow mode will have larger driven amplitude than that of the fast mode, so the dynamics of the original canonical variables, which are the superposition of these two modes, will be dominated by the slow mode, especially when the mutual interaction is strong.
			\item common bath: since there is one driven mode and it is the fast mode, the driven amplitude does not change too much as the mutual coupling strength varies. However, the asymptotic dynamics is determined by the relative magnitude between the slow mode (intrinsic component only) and the fast mode (induced component only).
				\begin{itemize}
					\item if the fast mode dominates, then the asymptotic elements of the covariance matrix will be more or less constant in time with small ripples.
					\item if the slow mode dominates, then they will oscillate in time.
				\end{itemize}
		\end{itemize}
	\item The inter-oscillator coupling $(\sigma>0)$ plays a more important role in the private bath case, but a minor role in the shared bath case.
	\item In the private bath case, entanglement is easier to survive for stronger inter-oscillator and weak oscillator-bath coupling, but in the shared bath case, both factors can be overshadowed by the intrinsic components, which are sensitive to the initial conditions of the oscillators.
 \item The asymptotic entanglement criterion in the common bath case can thus be broken into three components: one involving the fast mode only, one with slow mode only and the cross term.
		\begin{itemize}
			\item if the fast-mode part is subdominant, then the resulting entanglement criterion will oscillate with time, and that can cause sudden death \cite{YuEberly} and revival~\cite{Kanu,Goan,Paz} (SDR).
			\item if the fast-mode part is dominant, then there is no SDR phase.
		\end{itemize}
\end{enumerate}

\noindent {\bf Acknowledgment}  JTH thanks Shih-Yuin Lin for valuable discussions.  BLH  has discussed this problem with Rong Zhou and Yigit Subasi in August 2013 before the commencement of this work.  While engaging in this work, in the summer of 2014 and winter of 2015 he enjoyed the hospitality of the Center for Theoretical Physics at Fudan University, Shanghai, China.

\newpage

\appendix

\section{Expressions for $V_{13}(t)$, $V_{12}(t)$, $V_{14}(t)$, $V_{22}(t)$, $V_{32}(t)$, $V_{24}(t)$}

\subsection{$V_{13}(t)=\langle\bigl\{\chi_{1}(t),\chi_{2}(t)\bigr\}\rangle/2$:}
From \eqref{E:nvkrw}, we see that
\begin{align}
	 \mathcal{V}_{13}=\lim_{t\to\infty}V_{13}(t)&=\frac{e^{2}}{m^{2}}\int^{\infty}_{-\infty}\!\frac{d\kappa}{2\pi}\;\biggl[\widetilde{\mathbf{D}}^{11\,*}_{2}(\kappa)\widetilde{\mathbf{D}}^{21}_{2}(\kappa)\,\widetilde{\mathbf{G}}^{11}_{H}(\kappa)+\widetilde{\mathbf{D}}^{12\,*}_{2}(\kappa)\widetilde{\mathbf{D}}^{22}_{2}(\kappa)\,\widetilde{\mathbf{G}}^{22}_{H}(\kappa)\biggr]\notag\\
	 &=\frac{e^{2}}{m^{2}}\int^{\infty}_{-\infty}\!\frac{d\kappa}{2\pi}\;\frac{\sigma(\kappa^{2}-\omega^{2})\bigl[\widetilde{\mathbf{G}}^{11}_{H}(\kappa)+\widetilde{\mathbf{G}}^{22}_{H}(\kappa)\bigr]}{\bigl[(\kappa^{2}-\omega_{+}^{2})^{2}+4\gamma^{2}\kappa^{2}\bigr]\bigl[(\kappa^{2}-\omega_{-}^{2})^{2}+4\gamma^{2}\kappa^{2}\bigr]}\,,\label{E:yrtywc}
\end{align}
where
\begin{align}
	 \widetilde{\mathbf{D}}^{11\,*}_{2}(\kappa)\widetilde{\mathbf{D}}^{21}_{2}(\kappa)&=\frac{\sigma\bigl[(\kappa^{2}-\omega^{2})-i\,2\gamma\kappa\bigr]}{\bigl[(\kappa^{2}-\omega_{+}^{2})^{2}+4\gamma^{2}\kappa^{2}\bigr]\bigl[(\kappa^{2}-\omega_{-}^{2})^{2}+4\gamma^{2}\kappa^{2}\bigr]}\,,\\
	 \widetilde{\mathbf{D}}^{12\,*}_{2}(\kappa)\widetilde{\mathbf{D}}^{22}_{2}(\kappa)&=\frac{\sigma\bigl[(\kappa^{2}-\omega^{2})+i\,2\gamma\kappa\bigr]}{\bigl[(\kappa^{2}-\omega_{+}^{2})^{2}+4\gamma^{2}\kappa^{2}\bigr]\bigl[(\kappa^{2}-\omega_{-}^{2})^{2}+4\gamma^{2}\kappa^{2}\bigr]}\,.
\end{align}
We see that both baths contribute equally.

\subsection{$V_{12}(t)=\langle\bigl\{\chi_{1}(t),p_{1}(t)\bigr\}\rangle/2$:}
We find
\begin{align}
	 \frac{1}{2}\,\langle\bigl\{\chi_{i}(t),\,p_{j}(t)\bigr\}\rangle&=m\,\mathbf{D}^{ik}_{1}(t)\dot{\mathbf{D}}^{jk}_{1}(t)\,\langle\chi_{k}^{2}(0)\rangle+\frac{1}{m}\,\mathbf{D}^{ik}_{2}(t)\dot{\mathbf{D}}^{jk}_{2}(t)\,\langle p_{k}^{2}(0)\rangle\notag\\
				 &\qquad\qquad\qquad+\frac{e^{2}}{m}\int^{t}_{0}\!ds\,ds'\;\mathbf{D}^{ik}_{2}(t-s)\dot{\mathbf{D}}^{jk}_{2}(t-s')\,\mathbf{G}^{kk}_{H}(s-s')\,.\label{E:loerp}
\end{align}
Note that the overdot represents time derivative with respect to the argument of the variable. The late time limit of $V_{12}(t)$ is then given by
\begin{align}
	 \mathcal{V}_{12}=\lim_{t\to\infty}V_{12}(t)&=-i\,\frac{e^{2}}{m}\int^{\infty}_{-\infty}\!\frac{d\kappa}{2\pi}\;\kappa\,\widetilde{\mathbf{D}}^{1k\,*}_{2}(\kappa)\widetilde{\mathbf{D}}^{1k}_{2}(\kappa)\,\mathbf{G}^{kk}_{H}(\kappa)=0\,.
\end{align}
The result is identically zero because the integrand is odd in $\kappa$.

\subsection{$V_{14}(t)=\langle\bigl\{\chi_{1}(t),p_{2}(t)\bigr\}\rangle/2$:}
The late-time limit of $V_{14}(t)$ can be inferred from \eqref{E:loerp},
\begin{align}
	 \mathcal{V}_{14}&=-i\,\frac{e^{2}}{m}\int^{\infty}_{-\infty}\!\frac{d\kappa}{2\pi}\;\kappa\,\widetilde{\mathbf{D}}^{1k\,*}_{2}(\kappa)\widetilde{\mathbf{D}}^{2k}_{2}(\kappa)\,\mathbf{G}^{kk}_{H}(\kappa)\notag\\
	 &=-\frac{e^{2}}{m}\int^{\infty}_{-\infty}\!\frac{d\kappa}{2\pi}\;\frac{2\sigma\gamma\kappa^{2}\bigl[\widetilde{\mathbf{G}}^{11}_{H}(\kappa)-\widetilde{\mathbf{G}}^{22}_{H}(\kappa)\bigr]}{\bigl[(\kappa^{2}-\omega_{+}^{2})^{2}+4\gamma^{2}\kappa^{2}\bigr]\bigl[(\kappa^{2}-\omega_{-}^{2})^{2}+4\gamma^{2}\kappa^{2}\bigr]}\,.\label{E:erueow}
\end{align}
It is interesting to compare this result with \eqref{E:yrtywc}. The latter depends on additive contribution from both baths while the former has a subtractive contribution between baths.

\subsection{$V_{22}(t)=\langle\bigl\{p_{1}(t),p_{1}(t)\bigr\}\rangle/2=\langle p_{1}^{2}(t)\rangle$:}
By a similar derivation of $\langle\bigl\{\chi_{i}(t),\chi_{j}(t)\bigr\}\rangle/2$, we find
\begin{align}
				 \frac{1}{2}\langle\bigl\{p_{i}(t),p_{j}(t)\bigr\}\rangle&=m^{2}\dot{\mathbf{D}}^{ik}_{1}(t)\dot{\mathbf{D}}^{jk}_{1}(t)\,\langle\chi_{k}^{2}(0)\rangle+\dot{\mathbf{D}}^{ik}_{2}(t)\dot{\mathbf{D}}^{jk}_{2}(t)\,\langle p_{k}^{2}(0)\rangle\notag\\
				 &\qquad\qquad\qquad+e^{2}\int^{t}_{0}\!ds\,ds'\;\dot{\mathbf{D}}^{ik}_{2}(t-s)\dot{\mathbf{D}}^{jk}_{2}(t-s')\,\mathbf{G}^{kk}_{H}(s-s')\,.\label{E:ywbsdn}
\end{align}
Thus we have the late-time limit of $V_{22}(t)$ given by
\begin{align}
	 \mathcal{V}_{22}=\lim_{t\to\infty}V_{22}(t)&=e^{2}\int^{\infty}_{-\infty}\!ds\,ds'\;\dot{\mathbf{D}}^{ik}_{2}(s)\dot{\mathbf{D}}^{jk}_{2}(s')\,\mathbf{G}^{kk}_{H}(s-s')\notag\\
	 &=e^{2}\int^{\infty}_{-\infty}\!\frac{d\kappa}{2\pi}\;\kappa^{2}\biggl[\lvert\widetilde{\mathbf{D}}^{11}_{2}(\kappa)\rvert^{2}\,\widetilde{\mathbf{G}}^{11}_{H}(\kappa)+\lvert\widetilde{\mathbf{D}}^{12}_{2}(\kappa)\rvert^{2}\,\widetilde{\mathbf{G}}^{22}_{H}(\kappa)\biggr]\,.
\end{align}
This is similar to $\mathcal{V}_{11}$.

\subsection{$V_{32}(t)=\langle\bigl\{\chi_{2}(t),p_{1}(t)\bigr\}\rangle/2$:}
We show that $V_{32}(t)=-V_{14}(t)$, that is
\begin{align}
	 \mathcal{V}_{32}&=-i\,\frac{e^{2}}{m}\int^{\infty}_{-\infty}\!\frac{d\kappa}{2\pi}\;\kappa\biggl[\widetilde{\mathbf{D}}^{21\,*}_{2}(\kappa)\widetilde{\mathbf{D}}^{11}_{2}(\kappa)\,\mathbf{G}^{11}_{H}(\kappa)+\widetilde{\mathbf{D}}^{22\,*}_{2}(\kappa)\widetilde{\mathbf{D}}^{12}_{2}(\kappa)\,\mathbf{G}^{kk}_{H}(\kappa)\biggr]\notag\\
	 &=+\frac{e^{2}}{m}\int^{\infty}_{-\infty}\!\frac{d\kappa}{2\pi}\;\frac{2\sigma\gamma\kappa^{2}\bigl[\widetilde{\mathbf{G}}^{11}_{H}(\kappa)-\widetilde{\mathbf{G}}^{22}_{H}(\kappa)\bigr]}{\bigl[(\kappa^{2}-\omega_{+}^{2})^{2}+4\gamma^{2}\kappa^{2}\bigr]\bigl[(\kappa^{2}-\omega_{-}^{2})^{2}+4\gamma^{2}\kappa^{2}\bigr]}=-\mathcal{V}_{14}\,.
\end{align}
In the context of nonequilibrium transport, $V_{14}(t)$ is related to the power done by the oscillator 1 to oscillator 2 by means of the mutual coupling, while $V_{32}(t)$ is the other way around. At late time when the steady energy current is established, both should be equal in magnitude but opposite in sign. See. e.g., \cite{HHNESS}

\subsection{$V_{24}(t)=\langle\bigl\{p_{1}(t),p_{2}(t)\bigr\}\rangle/2$:}
Eq.~\eqref{E:ywbsdn} tells us that at late time $t\to\infty$, the element $V_{24}(t)$ becomes
\begin{align}\label{E:oierjfsj}
	 \mathcal{V}_{24}&=e^{2}\int^{\infty}_{-\infty}\!\frac{d\kappa}{2\pi}\;\kappa^{2}\widetilde{\mathbf{D}}^{1k\,*}_{2}(\kappa)\widetilde{\mathbf{D}}^{2k}_{2}(\kappa)\,\widetilde{\mathbf{G}}^{kk}_{H}(\kappa)\notag\\
	 &=e^{2}\int^{\infty}_{-\infty}\!\frac{d\kappa}{2\pi}\;\frac{\sigma\kappa^{2}(\kappa^{2}-\omega^{2})\bigl[\widetilde{\mathbf{G}}^{11}_{H}(\kappa)+\widetilde{\mathbf{G}}^{22}_{H}(\kappa)\bigr]}{\bigl[(\kappa^{2}-\omega_{+}^{2})^{2}+4\gamma^{2}\kappa^{2}\bigr]\bigl[(\kappa^{2}-\omega_{-}^{2})^{2}+4\gamma^{2}\kappa^{2}\bigr]}\,.
\end{align}
Again it has a similar structure to $\mathcal{V}_{13}$ in \eqref{E:yrtywc}.

\section{High Temperature forms of  $\mathcal{V}_{13}$, $\mathcal{V}_{14}$, $\mathcal{V}_{22}$, $\mathcal{V}_{24}$ }
At high temperature the Hadamard function $\widetilde{\mathbf{G}}^{ij}_{H}(\kappa)$ is approximately given by
\begin{equation}\label{E:erdkkww}
	\widetilde{\mathbf{G}}^{ij}_{H}(\kappa)=\frac{1}{2\pi\beta_{i}}\,\delta_{ij}\,.
\end{equation}

From \eqref{E:yrtywc} we need the integral
\begin{equation}
	 I_{3}=\int_{-\infty}^{\infty}\!d\kappa\;\frac{\kappa^{2}-\omega^{2}}{\bigl[(\kappa^{2}-\omega_{+}^{2})^{2}+4\gamma^{2}\kappa^{2}\bigr]\bigl[(\kappa^{2}-\omega_{-}^{2})^{2}+4\gamma^{2}\kappa^{2}\bigr]}=-\frac{\pi}{4\gamma}\frac{1}{\omega^{4}-\sigma^{2}}\,,
\end{equation}
to evaluate $\mathcal{V}_{13}$ at high temperature. Thus we have the high temperature limit of $\mathcal{V}_{13}$ given by
\begin{align}
	\mathcal{V}_{13}=\frac{2\gamma}{\pi m}\,I_{3}\biggl[\frac{1}{\beta_{1}}+\frac{1}{\beta_{2}}\biggr]=-\frac{1}{2m}\frac{\sigma}{\omega^{4}-\sigma^{2}}\left(\frac{1}{\beta_{1}}+\frac{1}{\beta_{2}}\right)\,.
\end{align}
It means that in this configuration, $\chi_{1}$ and $\chi_{2}$ anti-correlated and this anti-correlation grows with the mutual coupling strength $\sigma$.

To calculate {$\mathcal{V}_{14}$ we need the integral
\begin{equation}
	 I_{4}=\int_{-\infty}^{\infty}\!d\kappa\;\frac{\kappa^{2}}{\bigl[(\kappa^{2}-\omega_{+}^{2})^{2}+4\gamma^{2}\kappa^{2}\bigr]\bigl[(\kappa^{2}-\omega_{-}^{2})^{2}+4\gamma^{2}\kappa^{2}\bigr]}=\frac{\pi}{4\gamma}\frac{1}{4\omega^{2}\gamma^{2}+\sigma^{2}}\,,
\end{equation}
so that from \eqref{E:erueow} in the high temperature limit, $\mathcal{V}_{14}$ is given by
\begin{equation}
	 \mathcal{V}_{14}=-\frac{4\gamma^{2}\sigma}{\pi}\,I_{4}\biggl[\frac{1}{\beta_{1}}-\frac{1}{\beta_{2}}\biggr]=-\frac{\gamma\sigma}{4\omega^{2}\gamma^{2}+\sigma^{2}}\left(\frac{1}{\beta_{1}}-\frac{1}{\beta_{2}}\right)\,.
\end{equation}
The correlation between $\chi_{1}$ and $p_{2}$ diminishes with increasing mutual coupling. Moreover, the correlation disappear when both thermal baths have the same temperature.

Similar to those in evaluating $\mathcal{V}_{11}$, the following two integrals are needed for evaluation of $\mathcal{V}_{22}$, 
\begin{align}
	 I_{5}&=\int_{-\infty}^{\infty}\!d\kappa\;\frac{\kappa^{2}(\kappa^{2}-\omega^{2})^{2}+4\gamma^{2}\kappa^{2}}{\bigl[(\kappa^{2}-\omega_{+}^{2})^{2}+4\gamma^{2}\kappa^{2}\bigr]\bigl[(\kappa^{2}-\omega_{-}^{2})^{2}+4\gamma^{2}\kappa^{2}\bigr]}=\frac{\pi}{4\gamma}\frac{8\omega^{2}\gamma^{2}+\sigma^{2}}{4\omega^{2}\gamma^{2}+\sigma^{2}}\,,\\
	 I_{6}=\sigma^{2}I_{4}&=\int_{-\infty}^{\infty}\!d\kappa\;\frac{\kappa^{2}\sigma^{2}}{\bigl[(\kappa^{2}-\omega_{+}^{2})^{2}+4\gamma^{2}\kappa^{2}\bigr]\bigl[(\kappa^{2}-\omega_{-}^{2})^{2}+4\gamma^{2}\kappa^{2}\bigr]}=\frac{\pi}{4\gamma}\frac{\sigma^{2}}{4\omega^{2}\gamma^{2}+\sigma^{2}}\,.
\end{align}
The high temperature form of  $\mathcal{V}_{22}=\langle p_{1}^{2}(\infty)\rangle$ is given by
\begin{align}
	\mathcal{V}_{22}=\frac{2\gamma m}{\pi}\biggl[\frac{I_{5}}{\beta_{1}}+\frac{I_{6}}{\beta_{2}}\biggr]&=\frac{m}{2}\left\{\frac{8\omega^{2}\gamma^{2}+\sigma^{2}}{4\omega^{2}\gamma^{2}+\sigma^{2}}\frac{1}{\beta_{1}}+\frac{\sigma^{2}}{4\omega^{2}\gamma^{2}+\sigma^{2}}\frac{1}{\beta_{2\,.}}\right\}\,.
\end{align}
In the case $\beta_{1}=\beta=\beta_{2}$, it reduces to
\begin{equation}
	\mathcal{V}_{22}=\frac{m}{\beta}\,,
\end{equation}
which is independent of both coupling strengths $\gamma$ and $\sigma$.

Finally for $\mathcal{V}_{24}$ the integral
\begin{equation}
	 I_{7}=\int_{-\infty}^{\infty}\!d\kappa\;\frac{\kappa^{2}(\kappa^{2}-\omega^{2})}{\bigl[(\kappa^{2}-\omega_{+}^{2})^{2}+4\gamma^{2}\kappa^{2}\bigr]\bigl[(\kappa^{2}-\omega_{-}^{2})^{2}+4\gamma^{2}\kappa^{2}\bigr]}
\end{equation}
vanishes identically, so $\mathcal{V}_{24}=0$ in the high temperature limit. Thus we need the next order contribution. If we expand the Hadamard function in \eqref{E:erdkkww} one more order in $\beta$, we find
\begin{equation}\label{E:erdkkwy}
	\widetilde{\mathbf{G}}^{ij}_{H}(\kappa)=\left[\frac{1}{2\pi\beta_{i}}+\frac{\kappa^{2}}{24\pi}\,\beta_{i}+\cdots\right]\,\delta_{ij}\,.
\end{equation}
Then we need the integral
\begin{equation}
	 I_{8}=\int_{-\infty}^{\infty}\!d\kappa\;\frac{\kappa^{4}(\kappa^{2}-\omega^{2})}{\bigl[(\kappa^{2}-\omega_{+}^{2})^{2}+4\gamma^{2}\kappa^{2}\bigr]\bigl[(\kappa^{2}-\omega_{-}^{2})^{2}+4\gamma^{2}\kappa^{2}\bigr]}=\frac{\pi}{4\gamma}\,,
\end{equation}
and from \eqref{E:oierjfsj} we obtain
\begin{align}
	\mathcal{V}_{24}&=\frac{e^{2}\sigma}{48\pi^{2}}\,I_{8}\,\Bigl[\beta_{1}+\beta_{2}\Bigr]=\frac{m\sigma}{24}\,\Bigl[\beta_{1}+\beta_{2}\Bigr]\,.
\end{align}
This contribution is relatively small in the high temperature limit $\beta\omega\to0$.

\section{Zero-temperature expressions for $\mathcal{V}_{13}$, $\mathcal{V}_{14}$, $\mathcal{V}_{22}$, $\mathcal{V}_{24}$}

\subsection{$\mathcal{V}_{13}$}
We first evaluate the integral
\begin{align}
	 J_{3}&=\int_{0}^{\infty}\!d\kappa\;\frac{\kappa(\kappa^{2}-\omega^{2})}{\bigl[(\kappa^{2}-\omega_{+}^{2})^{2}+4\gamma^{2}\kappa^{2}\bigr]\bigl[(\kappa^{2}-\omega_{-}^{2})^{2}+4\gamma^{2}\kappa^{2}\bigr]}\notag\\
	&=\frac{\pi}{32\gamma\sigma}\left[\frac{f(\Omega_{+})}{\Omega_{+}}-\frac{f(\Omega_{-})}{\Omega_{-}}\right]\,.
\end{align}
We thus obtain $\mathcal{V}_{13}$
\begin{align}
	\mathcal{V}_{13}^{(0)}&=\frac{4\gamma\sigma}{\pi m}\,J_{3}=\frac{1}{8m}\left[\frac{f(\Omega_{+})}{\Omega_{+}}-\frac{f(\Omega_{-})}{\Omega_{-}}\right]\,.
\end{align}

\subsection{$\mathcal{V}_{14}$}
The elements $\mathcal{V}_{14}^{(0)}$ vanishes because the contributions from both thermal baths cancel.

\subsection{$\mathcal{V}_{22}$}
Here comes the tricky part because divergence emerges when we evaluate the vacuum component of $\mathcal{V}_{22}$. We first calculate the following two integrals
\begin{align}
	 J_{5}&=\int_{0}^{\infty}\!d\kappa\;\frac{\kappa^{3}\bigl[(\kappa^{2}-\omega^{2})^{2}+4\gamma^{2}\kappa^{2}\bigr]}{\bigl[(\kappa^{2}-\omega_{+}^{2})^{2}+4\gamma^{2}\kappa^{2}\bigr]\bigl[(\kappa^{2}-\omega_{-}^{2})^{2}+4\gamma^{2}\kappa^{2}\bigr]}\,,\\
	 J_{6}&=\int_{0}^{\infty}\!d\kappa\;\frac{\kappa^{3}\sigma^{2}}{\bigl[(\kappa^{2}-\omega_{+}^{2})^{2}+4\gamma^{2}\kappa^{2}\bigr]\bigl[(\kappa^{2}-\omega_{-}^{2})^{2}+4\gamma^{2}\kappa^{2}\bigr]}\,.
\end{align}
Apparently $J_{5}$ diverges since its integrand behaves like $\kappa^{-1}$. The sum of $J_{5}+J_{6}$ is given by
\begin{align}
	 J_{5}+J_{6}&=\frac{1}{2}\,\ln\frac{\Lambda^{2}}{\omega_{+}\omega_{-}}+\frac{\pi}{16\gamma}\left[\frac{\Omega_{+}^{2}-\gamma^{2}}{\Omega_{+}}\,f(\Omega_{+})+\frac{\Omega_{-}^{2}-\gamma^{2}}{\Omega_{-}}\,f(\Omega_{-})\right]\,.
\end{align}
The logarithmic divergence is regularized by a frequency cutoff $\Lambda$. Therefore the vacuum contribution of $\mathcal{V}_{22}$ is
\begin{align}\label{E:ueyress}
	 \mathcal{V}_{22}^{(0)}&=\frac{2m\gamma}{\pi}\biggl[J_{5}+J_{6}\biggr]=\frac{m\gamma}{\pi}\,\ln\frac{\Lambda^{2}}{\omega_{+}\omega_{-}}+\frac{m}{8}\left[\frac{\Omega_{+}^{2}-\gamma^{2}}{\Omega_{+}}\,f(\Omega_{+})+\frac{\Omega_{-}^{2}-\gamma^{2}}{\Omega_{-}}\,f(\Omega_{-})\right]\,.
\end{align}
Now let us check some limiting cases of \eqref{E:ueyress},
\begin{enumerate}
	\item $\sigma\to0$: when the mutual coupling is vanishingly small, the momentum uncertainty of Oscillator 1 becomes
		\begin{equation}
			 \lim_{\sigma\to0}\mathcal{V}_{22}^{(0)}=\frac{2m\gamma}{\pi}\,\ln\frac{\Lambda}{\omega}+\frac{m}{4}\frac{\Omega^{2}-\gamma^{2}}{\Omega}\,f(\Omega)\,,
		\end{equation}
		where $\Omega^{2}=\omega^{2}-\gamma^{2}$ is the resonance frequency of the oscillator. This is the momentum uncertainty of the uncoupled oscillator when it couples to the vacuum fluctuations of the bath.
	\item $\gamma\to0$: the leading contribution of the momentum uncertainty at late time in this case is
		\begin{equation}
			 \lim_{\gamma\to0}\mathcal{V}_{22}^{(0)}=\frac{m}{4}\Bigl(\omega_{+}+\omega_{-}\Bigr)+\mathcal{O}(\gamma)=\frac{m\omega}{2}+\mathcal{O}(\gamma)\,.
		\end{equation}
\end{enumerate}

\subsection{$\mathcal{V}_{24}$}
We need the integral
\begin{align}
	 J_{7}&=\int_{0}^{\infty}\!d\kappa\;\frac{\kappa^{3}(\kappa^{2}-\omega^{2})}{\bigl[(\kappa^{2}-\omega_{+}^{2})^{2}+4\gamma^{2}\kappa^{2}\bigr]\bigl[(\kappa^{2}-\omega_{-}^{2})^{2}+4\gamma^{2}\kappa^{2}\bigr]}\notag\\
	 &=-\frac{1}{4\sigma}\,\ln\frac{\omega_{+}}{\omega_{-}}+\frac{\pi}{32\gamma\sigma}\biggl[\frac{\Omega_{+}^{2}-\gamma^{2}}{\Omega_{+}}\,f(\Omega_{+})-\frac{\Omega_{-}^{2}-\gamma^{2}}{\Omega_{-}}\,f(\Omega_{-})\biggr]\,.
\end{align}
Thus we have the vacuum component of $\mathcal{V}_{24}$ given by
\begin{align}
	\mathcal{V}_{24}^{(0)}=\frac{4m\gamma\sigma}{\pi}\, J_{7}=-\frac{m\gamma}{\pi}\,\ln\frac{\omega_{+}}{\omega_{-}}+\frac{m}{8}\biggl[\frac{\Omega_{+}^{2}-\gamma^{2}}{\Omega_{+}}\,f(\Omega_{+})-\frac{\Omega_{-}^{2}-\gamma^{2}}{\Omega_{-}}\,f(\Omega_{-})\biggr]\,.
\end{align}

\section{Low temperature correction expressions for $\mathcal{V}_{13}$, $\mathcal{V}_{14}$, $\mathcal{V}_{22}$, $\mathcal{V}_{24}$}\label{S:eknkdw}

\subsection{$\mathcal{V}_{13}$}
We first evaluate the integral
\begin{align}
	 K_{3}&=2\int_{0}^{\infty}\!d\kappa\;\frac{\sigma\,\kappa(\kappa^{2}-\omega^{2})\,e^{-\beta\kappa}}{\bigl[(\kappa^{2}-\omega_{+}^{2})^{2}+4\gamma^{2}\kappa^{2}\bigr]\bigl[(\kappa^{2}-\omega_{-}^{2})^{2}+4\gamma^{2}\kappa^{2}\bigr]}=-\frac{2\sigma\,\omega^{2}}{\omega_{+}^{4}\omega_{-}^{4}}\frac{1}{\beta^{2}}+\cdots\,.
\end{align}
This implies that the low temperature correction to $\mathcal{V}_{13}$ is given by
\begin{equation}
	\mathcal{V}_{13}^{(\beta)}=\frac{\pi^{2}}{6}\frac{2\gamma}{\pi m}\Bigl[K_{3}(\beta_{1})+K_{3}(\beta_{2})\Bigr]=-\frac{2\pi\gamma}{3m}\frac{\omega^{2}\sigma}{(\omega^{4}-\sigma^{2})^{2}}\left[\frac{1}{\beta_{1}^{2}}+\frac{1}{\beta_{2}^{2}}\right]\,.
\end{equation}

\subsection{$\mathcal{V}_{14}$}
For $\mathcal{V}_{14}$ we need the integral
\begin{align}
	 J_{4}&=2\int_{0}^{\infty}\!d\kappa\;\frac{\gamma\sigma\,\kappa^{3}e^{-\beta\kappa}}{\bigl[(\kappa^{2}-\omega_{+}^{2})^{2}+4\gamma^{2}\kappa^{2}\bigr]\bigl[(\kappa^{2}-\omega_{-}^{2})^{2}+4\gamma^{2}\kappa^{2}\bigr]}=\frac{12\gamma\sigma}{\omega_{+}^{4}\omega_{-}^{4}}\frac{1}{\beta^{4}}+\cdots\,.
\end{align}
Here the finite temperature correction behaves like $\beta^{-4}$, so we will acquire a factor
\begin{equation}
	\sum_{n=1}^{\infty}\frac{1}{n^{4}}=\frac{\pi^{4}}{90}\,,
\end{equation}
once we consider all algebraically equivalent contributions in \eqref{E:hfeirhis}. Thus the correction to $\mathcal{V}_{14}$ is given by
\begin{equation}
	 \mathcal{V}_{14}^{(\beta)}=-\frac{\pi^{4}}{90}\frac{4\gamma}{\pi}\,K_{4}=-\frac{8\pi^{3}}{15}\frac{\gamma^{2}\sigma}{(\omega^{4}-\sigma^{2})^{2}}\left[\frac{1}{\beta^{4}_{!}}-\frac{1}{\beta^{4}_{2}}\right]+\cdots\,.
\end{equation}

\subsection{$\mathcal{V}_{22}$}
Before evaluating $\mathcal{V}_{22}$, we first evaluate the following two integrals
\begin{align}
	 K_{5}&=2\int_{0}^{\infty}\!d\kappa\;\frac{\kappa^{3}\bigl[(\kappa^{2}-\omega^{2})^{2}+4\gamma^{2}\kappa^{2}\bigr]\,e^{-\beta\kappa}}{\bigl[(\kappa^{2}-\omega_{+}^{2})^{2}+4\gamma^{2}\kappa^{2}\bigr]\bigl[(\kappa^{2}-\omega_{-}^{2})^{2}+4\gamma^{2}\kappa^{2}\bigr]}=\frac{12\omega^{4}}{\omega_{+}^{4}\omega_{-}^{4}}\frac{1}{\beta^{4}}+\mathcal{O}(\beta^{-3})\,,\\
	 K_{6}&=2\int_{0}^{\infty}\!d\kappa\;\frac{\sigma^{2}\kappa^{3}\,e^{-\beta\kappa}}{\bigl[(\kappa^{2}-\omega_{+}^{2})^{2}+4\gamma^{2}\kappa^{2}\bigr]\bigl[(\kappa^{2}-\omega_{-}^{2})^{2}+4\gamma^{2}\kappa^{2}\bigr]}=\frac{12\sigma^{2}}{\omega_{+}^{4}\omega_{-}^{4}}\frac{1}{\beta^{4}}+\mathcal{O}(\beta^{-3})\,.
\end{align}
We have the low temperature correction to $\mathcal{V}_{22}$ given by
\begin{align}
	 \mathcal{V}_{22}^{(\beta)}=\frac{\pi^{4}}{90}\frac{2m\gamma}{\pi}\left[K_{5}+K_{6}\right]=\frac{4\pi^{3}}{15}\frac{m\gamma}{(\omega^{4}-\sigma^{2})^{2}}\left[\frac{\omega^{4}}{\beta_{1}^{4}}+\frac{\sigma^{2}}{\beta_{2}^{4}}\right]+\cdots\,.
\end{align}
The corresponding finite temperature correction to the kinetic energy of Oscillator 1 is
\begin{equation}
	 E_{k_{1}}=\frac{\mathcal{V}_{22}^{(\beta)}}{2m}=\frac{2\pi^{3}}{15}\frac{\gamma}{(\omega^{4}-\sigma^{2})^{2}}\left[\frac{\omega^{4}}{\beta_{1}^{4}}+\frac{\sigma^{2}}{\beta_{2}^{4}}\right]+\cdots\,.
\end{equation}

\subsection{$\mathcal{V}_{24}$}
Here we need the integral
\begin{align}
	 K_{7}&=2\int_{0}^{\infty}\!d\kappa\;\frac{\sigma\kappa^{3}(\kappa^{2}-\omega^{2})\,e^{-\beta\kappa}}{\bigl[(\kappa^{2}-\omega_{+}^{2})^{2}+4\gamma^{2}\kappa^{2}\bigr]\bigl[(\kappa^{2}-\omega_{-}^{2})^{2}+4\gamma^{2}\kappa^{2}\bigr]}=-\frac{12\omega^{2}\sigma}{\omega_{+}^{4}\omega_{-}^{4}}\frac{1}{\beta^{4}}+\mathcal{O}(\beta^{-3})\,.
\end{align}
Therefore $\mathcal{V}_{24}$ becomes
\begin{align}
	 \mathcal{V}_{24}^{(\beta)}=\frac{\pi^{4}}{90}\frac{2m\gamma}{\pi}\,\Bigl[K_{7}(\beta_{1})+K_{7}(\beta_{2})\Bigr]=-\frac{4\pi^{3}}{15}\frac{m\omega^{2}\gamma\sigma}{(\omega^{4}-\sigma^{2})^{2}}\left[\frac{1}{\beta_{1}^{4}}+\frac{1}{\beta_{2}^{4}}\right]\,.
\end{align}

\end{document}